\pdfoutput=1
\documentclass[12pt]{article} 
\usepackage[sectionbib]{natbib}
\usepackage{array,epsfig,fancyheadings,rotating}
\usepackage[margin = 1in]{geometry}
\usepackage{sectsty, secdot}
\sectionfont{\fontsize{12}{14pt plus.8pt minus .6pt}\selectfont}
\renewcommand{\theequation}{\thesection\arabic{equation}}
\subsectionfont{\fontsize{12}{14pt plus.8pt minus .6pt}\selectfont}

\usepackage{amsmath}
\usepackage{amssymb}
\usepackage{amsfonts}
 \usepackage{booktabs, longtable, amsmath}
\usepackage{multirow}
\usepackage{booktabs,multirow}
\usepackage{amsthm}
\usepackage[ruled,vlined]{algorithm2e} 
\newtheorem{theorem}{Theorem}
\newtheorem{lemma}{Lemma}
\newtheorem{corollary}{Corollary}

\theoremstyle{definition}

\usepackage{amsmath}
\usepackage{hyperref}
\usepackage{amssymb}
\usepackage{amsfonts}
\usepackage{dsfont}
\usepackage{cleveref}
\usepackage{enumitem}
\usepackage{bm}
\usepackage{bbm}
\usepackage{multirow}
\usepackage{multicol}

\def\R{\mathbb{R}}

\def\bY{\bm{Y}}
\def\bX{\bm{X}}
\def\bC{\bm{C}}
\def\bZ{\bm{Z}}

\renewcommand{\vec}[1]{\mathbf{1}}
\DeclareMathOperator*{\argmin}{arg\min}
\newcommand{\E}{\mathbb{E}}
\newcommand{\prob}{\mathbb{P}}
\newcommand{\btheta}{\bm{\theta}}
\newcommand{\boldeta}{\bm\eta}
\newcommand{\normal}{\mathcal{N}}
\numberwithin{equation}{section}

\theoremstyle{plain}

\newtheorem{assumption}{Assumption}[section]
\theoremstyle{remark}

\begin{document}


\renewcommand{\baselinestretch}{2}

\markright{ \hbox{\footnotesize\rm Statistica Sinica
}\hfill\\[-13pt]
\hbox{\footnotesize\rm
}\hfill }

\markboth{\hfill{\footnotesize\rm SAMHITA PAL AND SUBHASHIS GHOSAL} \hfill}
{\hfill {\footnotesize\rm Sparse Bayesian Grouped-regression} \hfill}

\renewcommand{\thefootnote}{}
$\ $\par


\fontsize{12}{14pt plus.8pt minus .6pt}\selectfont \vspace{0.8pc}
\centerline{\large\bf Bayesian High-dimensional Grouped-regression}
\vspace{2pt} 
\centerline{\large\bf using Sparse Projection-posterior}
\vspace{.4cm} 
\centerline{Samhita Pal and Subhashis Ghosal} 
\vspace{.4cm} 
\centerline{\it North Carolina State University}
 \vspace{.55cm} \fontsize{9}{11.5pt plus.8pt minus.6pt}\selectfont


\begin{quotation}
\noindent {\it Abstract:}
{\bf We present a novel Bayesian approach for high-dimensional grouped regression under sparsity. We leverage a sparse projection method that uses a sparsity-inducing map to induce a posterior on a lower-dimensional parameter space. Our method introduces three distinct projection maps based on popular penalty functions: the Group LASSO projection-posterior, the Group SCAD projection-posterior, and the Adaptive Group LASSO projection-posterior. Each projection map is constructed to immerse posterior samples into a structured, sparse space, allowing for effective group selection and estimation in high-dimensional settings. We derive optimal posterior contraction rates for estimation and prediction, thereby proving that the methods are model-selection consistent. We also propose a Debiased Group LASSO Projection Map that ensures correct asymptotic coverage of credible sets. Our methodology is particularly suited for applications in nonparametric additive models, where we use B-spline expansions to capture complex relationships between covariates and the response. Extensive simulations validate our theoretical findings, demonstrating the robustness of our approach across different settings. Finally, we illustrate the practical utility of our method with an application to brain MRI volume data from the Alzheimer's Disease Neuroimaging Initiative (ADNI), where our model identifies key brain regions associated with Alzheimer’s Disease severity. }\\

\vspace{9pt}
\par
\end{quotation}\par

\def\thefigure{\arabic{figure}}
\def\thetable{\arabic{table}}

\renewcommand{\theequation}{\thesection.\arabic{equation}}

\fontsize{12}{14pt plus.8pt minus .6pt}\selectfont

\section{Introduction}
Grouped sparse regression is a powerful tool in statistical modeling that provides a structured approach for selecting and estimating subsets of variables in high-dimensional datasets. In scenarios where very high-dimensional predictors naturally group, such as genes in genomics or pixels in imaging, grouped sparse regression techniques help select relevant groups, thereby improving predictive accuracy. The group-selection approach is also instrumental for high-dimensional additive models by using basis expansions or by representing multilevel categorical variables as a set of dummy variables. There can be another approach in which, along with a subset of groups, it is further assumed that only a minor portion of each group's constituent variables is active, leading to within-group sparsity. However, we do not consider this setup in the current paper.

Assuming we have prior information on the group structure among the covariates, it is interesting to examine the effects of the most relevant groups on the response variable. This model assumes that the underlying sparsity pattern lies within the groups, while keeping all variables within each group. The pioneering work in this direction, the group LASSO, was developed by \cite{yuan2006model}. This was later extended to the sparse additive models by \cite{ravikumar2009sparse}. A very broad class of penalties, specifically, a composite absolute penalty for group selection, was developed by \cite{zhao2009composite}, of which the group LASSO is a special case. Following these initial studies, several other penalization methods of grouped regression under sparsity were proposed in the literature, including but not limited to the group SCAD \citep{zeng2014group}, group MCP \citep{huang2009group}, and the adaptive group LASSO \citep{wang2008note}. A complete review of frequentist properties of various group selection and estimation methods can be found in \cite{huang2012selective}.

In the Bayesian literature, the grouped regression method was first studied by \cite{farcomeni2010bayesian}, who extended the SSVS method of \cite{george1995stochastic} to the group-sparse setting. Similarly, \cite{raman2009bayesian} extended the Bayesian LASSO \citep{park2008bayesian} to the group setting and proposed the Bayesian group LASSO. \cite{casella2010penalized} also suggested a multivariate Laplacian prior for the group coefficients along the same line. \cite{xu2015bayesian} proposed the Bayesian group lasso model with spike and slab priors and demonstrated the oracle property of the posterior median estimator for group variable selection and estimation under orthogonal designs. \cite{xu2016bayesian} proposed the Bayesian grouped horseshoe regression, which extends the Bayesian horseshoe estimator to handle grouped variables through a grouped Bayesian horseshoe and a hierarchical Bayesian grouped horseshoe, demonstrating their effectiveness in additive models with both simulated and real data. \cite{yang2020consistent} proposed a Bayesian hierarchical model with a spike-and-slab prior to perform group selection in high-dimensional linear regression, providing theoretical consistency results. \cite{lai2021review} thoroughly reviewed and compared Bayesian approaches to group selection under the linear regression model.



When the variables in a high-dimensional linear regression are not individually meaningful but naturally form groups, a statistical method that correctly identifies these groups and accurately estimates the regression coefficients is proper. Our paper solves this problem using a novel Bayesian projection technique and provides an in-depth study of its theoretical properties. The approach used in this paper follows the concept of mapping non-sparse samples from the full posterior obtained by putting a convenient conjugate prior on the vector of regression coefficients. Here, we have proposed three maps. They closely resemble minimizers of the objective functions for the group LASSO, group SCAD, and adaptive group LASSO procedures. This idea, also called `immersion posterior' in the literature, has previously been studied in the context of sparse high-dimensional linear regression in \cite{pal2024bayesian} and in several other contexts in \cite{bhaumik2015bayesian,chakraborty2021coverage,wang2023coverage} and the references therein. This approach bypasses the need for MCMC, leading to more efficient computations, especially in large-scale problems. The method's flexibility allows different tuning parameters for varying objectives, with smaller values suited for estimation and larger values for consistent model selection. Additionally, this sparse projection-posterior method can be implemented using a distributed computing setup, enabling efficient processing of large datasets and addressing privacy concerns by avoiding the need to store data on a single machine. Although substantial time gains from distributed implementation become prominent only for very large sample sizes (Supplement A), the framework offers a more immediate and practically important benefit in smaller settings: it eliminates the need to share raw data across sites. Since our projection-posterior method operates entirely on aggregated summary statistics \((\bX^\top\bX, \bX^\top\bY, \bY^\top\bY)\) for design matrix $\bX$ and response vector $\bY$, it naturally accommodates data-federated environments where privacy, regulatory, or logistical constraints prohibit central data pooling. This feature distinguishes it from existing high-dimensional Bayesian regression approaches, which typically require full data access for MCMC sampling or posterior updates, making our approach particularly suitable for multi-institutional studies and privacy-preserving inference.
 
 Another interpretation is to view the sparse parameter as a projection of a ``super-parameter" without sparsity, using a misspecified likelihood for inference. This dual perspective makes the sparse projection-posterior approach computationally efficient and theoretically appealing for high-dimensional inference. For high-dimensional sparse group-regression, variational inference (VI), as discussed in \cite{subrahmanya2013variational, babacan2014bayesian, komodromos2023group}, iteratively obtains the Kullback-Leibler projection of the posterior distribution to a tractable class of distributions, giving a random measure that mimics the posterior distribution. The immersion posterior gives a different random measure by transforming posterior samples from an unrestricted posterior distribution via a typically non-iterative optimization. The immersion posterior usually has a more explicit description that helps establish convergence, model selection, and coverage properties. Further discussion of similarities and differences between VI and the immersion posterior is given in \cite{pal2024bayesian}. Group selection methods can address the high-dimensional additive regression problem via a basis-expansion technique. In this paper, we also develop a Bayesian method for high-dimensional additive regression models using group selection.

The rest of the paper is organized as follows. We delineate the proposed procedure in \Cref{Method}, describing the steps leading up to the three sparse projection maps. Following the detailed description, we formally state the asymptotic optimality properties of the three proposed projection-posteriors in \Cref{Results}. We show that the sparse projection-posteriors concentrate around the true parameter at rates comparable to those of their frequentist counterparts and that model selection procedure is consistent for all three maps. We also guarantee coverage of credible sets obtained from a specific debiased map of the three group sparse projection-posteriors. In \Cref{sec:additive_model}, we describe the application of grouped regression under sparsity to the nonparametric additive regression and show optimal convergence properties. We conduct extensive simulation studies under both linear and additive models with varying choices of the number of groups, the number of active groups, and sparsity. We also apply our method to MRI brain volume data from the Alzheimer's Disease Neuroimaging Initiative (ADNI). These numerical results are displayed in \Cref{Numericals}.

\section{Methodology}\label{Method}

\subsection{Model and Notation}
Consider the linear model $\bY = \bX\bm{\beta} + \bm{\varepsilon}$, with noise $\bm{\varepsilon} \sim \normal(\textbf{0},\sigma^2\bm{I}_n)$, where the response variable $\bY \in \R^n$, the design matrix $\bX \in \R^{n \times p}$, and the coefficient vector $\bm{\beta} \in \R^p$. Let the true value of the regression coefficient parameter vector be $\bm{\beta}^0$. Suppose there are $K$ disjoint groups in total, namely, $G_1,G_2,\dots,G_K$. The group sizes are $|G_k| = p_k, k = 1,2,\dots,K$, such that $\sum_{k = 1}^K p_k = p$. Without loss of generality, we can assume a few adjacent variables to be grouped, that is, $\bX = [\bX_{G_1},\bX_{G_2}\dots \bX_{G_K}]$, where $\bX_{G_k} \in \R^{n \times p_k}$ for $k = 1,2,\dots,K$. Accordingly, the coefficients can also be grouped leading to $\bm{\beta} = (\bm{\beta}_1^\mathrm{T},\dots,\bm{\beta}_K^\mathrm{T})^\mathrm{T}$, and $\bm{\beta}_k \in \R^{p_k}$. The linear model can then be re-written as $\bY = \sum_{k = 1}^K \bX_{G_k} \bm{\beta}_k + \bm\varepsilon.$ The assumption here is that the setup is group sparse, meaning that there can be more than one group that does not influence the response variable. As a result, all variables within such unimportant groups are removed from the model. Suppose only $s_0 \leq K$ important groups exist, where $s_0$ is unknown. We seek to identify the active groups and accurately recover the regression parameter. Also, let $p_0 < p$ denote the number of important variables. Let $S_0$ be the set of group indices corresponding to the signal groups, that is, $S_0 = \{k : 1 \leq k \leq K, \bm{\beta}^0_k \neq \textbf{0}_{p_k}\}$, where $\textbf{0}_{p_k}$ is the $p_k$-dimensional vector of zeroes. Consequently, the cardinality of $S_0$ is $s_0$ and moreover, we can write $p_0 = \sum_{k \in S_0} p_k$. 

Throughout this paper, we denote the minimum and maximum group sizes by $p_{\textnormal{min}}$ and $p_{\text{max}}$ respectively. The $j$-th column of the design matrix is denoted by $\bX^{(j)}$, whereas the design matrix without the $j$-th column is denoted by $\bX^{(-j)}$. Similarly, the design matrix without the $g$-th group is denoted by $\bX_{-g}$. That is, $\bX_{-g} \in \R^{n \times \sum_{k \neq g}p_k}$ does not contain any  variable belonging to the $g$-th group. The $l$-th variable within the $g$-th group is denoted by $\bX_{G_g}^{(l)}$. Define the gram matrix as $\hat{\bm\Sigma} = n^{-1}\bX^\mathrm{T}\bX$, so that the inner-product between the $j$-th and $k$-th variables forms the ($j,k$)-th element of this matrix. We define $\hat\Sigma_{j,k}$ as the $(j,k)$-th element of the Gram matrix. Moreover, let $\hat{\bm\Sigma}_{j,-j} = \hat{\bm\Sigma}_{-j,j}^\mathrm{T}$ and $\hat{\bm\Sigma}_{-j,-j}$ is the Gram matrix of $\bX^{(-j)}$. Various norms have been used across the paper. For a $d$-dimensional vector $\bm{x} \in \R^d$, we write $\|\bm{x}\|_q = (\sum_{i = 1}^{d} |a_i|^q)^{1/q}$, $\|\bm{x}\|_\infty = \max_{1 \leq i \leq d} |x_i|$ and drop the subscript $q$ if $q = 2$. We use the usual $O$ and $o$ notations, and write $a_n\asymp b_n$ if both $a_n=O(b_n)$ and $b_n=O(a_n)$.

\subsection{Prior Specification and the Unrestricted Posterior}\label{subsec:prior-posterior}
The central idea here is to initially disregard the group's sparsity, inherent to the regression model, while specifying the prior distribution on the coefficient vector. Instead, the goal here is to specify a prior distribution so that the posterior can be computed efficiently and in a relatively simple form. We thus use the conjugate normal prior for the parameter vector, assuming the error standard deviation $\sigma > 0$ to be given as $\bm{\beta}|\sigma \sim \normal_p(\textbf{0}, \sigma^2a_n^{-1} \bm{I}_p).$ Consequently, defining $\hat{\bm{\beta}}^\mathrm{R} = \big(\bX^{\mathrm{T}} \bX + a_n\bm{I}_{p}\big)^{-1} \bX^{\mathrm{T}} \bY$, the ridge regression estimator with penalty $a_n^{-1}$, 
the posterior is obtained as $\bm{\beta}|(\bX, \bY, \sigma) \sim \normal_p\big(\hat{\bm{\beta}}^\mathrm{R}, \sigma^2 \big(\bX^{\mathrm{T}} \bX + a_n\bm{I}_{p}\big)^{-1}\big).$ The posterior distribution of $\bX\bm{\beta}$ given $\sigma$ is then given by $\bX\bm{\beta}|(\sigma,\bY) \sim \normal_n \big(\bX \hat{\boldsymbol\beta}^\mathrm{R} , \sigma^2 \bm{H}(a_n)\big),$ where $\bm{H}(a_n) = \bX \big(\bX^{\mathrm{T}} \bX + a_n \bm{I}_{p}\big)^{-1} \bX^{\mathrm{T}}.$ The posterior distribution of the quantity $\bX\bm{\beta} - \bX\bm{\beta}^0 = \bm{\eta}$, given the noise variance $\sigma^2$, is given by $\normal_n\left(\bm\mu, \sigma^2 \bm{H}(a_n)\right)$, where $ \bm\mu=\bX (\hat{\boldsymbol\beta}^\mathrm{R} - \bm\beta^0)$. However, this posterior is supported on non-sparse vectors, and it does not account for the fact that all groups are irrelevant to the response. To incorporate group-level sparsity in the posterior, we rely on a transformation, immersing a posterior sample without sparsity into the fuzzy set of sparse vectors with a lower effective dimension. The induced {\em Immersion Posterior}, a term coined by \cite{wang2023coverage}, will be generically referred to as the sparse projection-posterior in our context. In the next section, we discuss the strategy we adopted in this paper to obtain the group-sparse induced posterior. Specific names will be used for the sparse projection-posterior corresponding to specific group sparsity-generating maps. 

For the sparse projection-posterior approach, conditional posterior draws are made given \(\sigma^2\), requiring the posterior distribution of \(\sigma^2\) to be consistent to ensure the desired asymptotic frequentist properties. However, when \(p\) grows at a rate comparable to or faster than \(n\), the marginal vanilla posterior for \(\sigma^2\), obtained using a standard inverse-gamma prior in the regression model, becomes inconsistent. Specifically, $\text{with the prior for }\sigma^{2}\propto 1/\sigma^2\ (\textnormal{or equivalently, }p(\upsilon)\propto 1/\upsilon),$ the posterior distribution is $\upsilon\mid \bY\sim\mathrm{Gamma}(n/2,\ n\hat\sigma_n^2/2),$ where $ \hat\sigma_n^2=n^{-1}\bY^\mathrm{T}(\bm I_n-\bm H(a_n))\bY.\ \text{Hence, } \mathbb{E}[\upsilon\mid \bY]=n/\{\bY^\mathrm{T}(\bm I_n-\bm H(a_n))\bY\},$ which diverges as $n\to\infty.$
 This issue can be addressed by applying a data-dependent map to the precision \(\upsilon\). Using a non-informative Gamma prior on $\upsilon$ and the map \( \iota: \upsilon \mapsto \kappa \upsilon = \upsilon^*\), we obtain the induced posterior \(\upsilon^* | \bY \sim \text{Gamma}({n}/{2}, \kappa^{-1} {n \hat{\sigma}_n^2}/{2})\), with mean \(\kappa / \hat{\sigma}_n^2\), where $\hat{\sigma}_n^2 = n^{-1} \bY^{\mathrm{T}} ( \bm{I}_n - \bm{H}(a_n) ) \bY$. Now, choosing \(\kappa = \hat{\sigma}_n^2 / \tilde{\sigma}^2\), where \(\tilde{\sigma}^2\) is a consistent estimator of \(\sigma^2\) (for example, from \cite{sun2012scaled}), ensures that the posterior mean of the induced posterior becomes a consistent for the true precision. {This is because, the posterior variance of this induced posterior distribution is ${2}/({n \tilde{\sigma}^4}) \to 0$ as $n \to \infty,$ since $\tilde{\sigma}^2$ converges to the finite positive constant $\sigma_0^2$ by the consistency of $\tilde{\sigma}^2$ for $\sigma^2$.} By the smooth one-to-one correspondence between precision $\upsilon$ and noise variance $\sigma^2$, we can also ensure that the induced posterior ${\sigma^*}^2 = 1/\upsilon^*$ is consistent for the true error variance ${\sigma^0}^2$.

\subsection{Sparse Projection Maps}
To transform non-sparse posterior samples into a space that reflects the desired sparsity, for $\bm{u} \in \R^p$, where $\bm{u} = (\bm{u}_1,\dots,\bm{u}_K)^\mathrm{T}$, we apply a mapping with $L$ as a loss function and $\mathcal{P}_{\lambda_n}(\cdot)$ as a penalty function, given by:
\begin{align}
\label{proj_map}
    \iota:\bm\beta \mapsto \bm\beta^*= \argmin_{u\in \R^p} \{ L(\bm\beta, \bm{u}) + \sum_{k = 1}^K \mathcal{P}_{\lambda_n}(\|\bm{u}_k\|)\},
\end{align} 
designed to enforce sparsity in the solution, using a tuning parameter $\lambda_n$ depending on $n$ to control the level of sparsity. The loss function is designed to capture the discrepancy between $\bX \bm\beta^*$ from $\bX \bm\beta$, which we take to be the squared error loss $L(\btheta, \bm{u})=n^{-1}\|{\bX \btheta - \bX \bm{u}}\|^2 $. In this paper, we demonstrate a variety of penalty options, including group LASSO, group SCAD, and adaptive group LASSO, representing a broad spectrum of convex and non-convex penalty functions.

\subsection{Sparse Projection-posteriors under Different Maps} \label{subsec:diffmap}
The Group LASSO is one of the most popular models for grouped regression problems in which not all groups are relevant to the study. However, if a group is deemed important, all the variables within that group are considered relevant to the response of interest. We use this group LASSO penalty $\mathcal{P}_{\lambda_n}(\|\bm{u}_k\|) = \lambda_n\sqrt{p_k}\|\bm{u}_k\|$ for $\bm{u}_k \in \R^{p_k}$ to transform a posterior sample to $\bm\beta^*$ corresponding to the tuning parameter $\lambda_n$. The induced posterior distribution of $\bm\beta^*$ is called the Group LASSO projection-posterior. 

The Group SCAD penalty extends the Smoothly Clipped Absolute Deviation (SCAD) penalty \citep{fan2001variable} to the group setting. It is a non-convex penalty function that promotes sparsity while mitigating the bias typically introduced by convex penalties such as LASSO. The group SCAD penalty $\mathcal{P}_{\lambda_n}(|t|)\}$ is given by $\lambda_n|t|\mathds{1}(|t| \leq \lambda_n) + \frac{|t|^2 - 2\tau\lambda_n|t| + \lambda_n^2}{2(\tau-1)}\mathds{1}(\lambda_n < |t| \leq \tau\lambda_n) + \frac{\lambda_n^2(\tau + 1)}{2}\mathds{1}(|t| > \tau\lambda_n),$
where \( \mathds{1}(\cdot) \) is the indicator function of a set.
for $t > 0$ and $\tau > 2$. The resulting induced posterior of $\bm\beta^*$ is called the Group SCAD projection-posterior.

While the group LASSO penalty ensures that larger groups do not unduly dominate the model just because they contain more variables, the adaptive lasso extends this concept by applying a penalty that is inversely proportional to the estimated $\ell_2$-norm of the coefficients within each group, typically derived from an initial group LASSO fit. This approach introduces an additional layer of flexibility, allowing the model to apply different penalties to different groups based on their preliminary importance. Let the group LASSO estimator of $\bm\beta$ be $\hat{\bm\beta}^\textnormal{GL} = \argmin \big\{ n^{-1} \|\bY - \bX\bm{u} \|^2 + \tilde{\lambda}_n \sum_{k = 1}^K \|\bm{u}_k\|: \bm{u}\in \R^p\big\}$
where the tuning parameter $\tilde{\lambda}_n$ should be such that the group LASSO estimator $\hat{\bm\beta}^\mathrm{GL}$ is consistent with rate $r_n$. Then, the Adaptive Group LASSO projection-posterior is based on the adaptive group LASSO penalty function, $\mathcal{P}(\|\bm{u}_k\|) = w_k \|\bm{u}_k\|$, where the weights are obtained by using the group LASSO estimator $\hat{\bm\beta}^\mathrm{GL}$, that is $w_k = \|\hat{\bm\beta}^\textnormal{GL}_k\|^{-1}$, with a thresholded weight for zero groups under the initial group LASSO fit. The induced posterior distribution of $\bm\beta^*$ is called the Group Adaptive LASSO projection-posterior.

Below, we rewrite the proposed algorithm with the three penalty choices studied across the scenarios in this paper. Although we have chosen three penalization techniques to project our initial posterior samples into the required lower-dimensional space to impose sparsity, we would like to emphasize that our proposed approach is not restricted to these three penalties. Besides these, there is another popular non-convex penalty, the minimax concave penalty (MCP) introduced in \cite{zhang2010nearly}, which tries to mitigate the bias due to shrinkage. The group MCP penalty \citep{huang2012selective} is, thus, another viable option for the sparse projection-posterior. We provide theoretical guarantees for estimation and model selection corresponding to the group MCP penalty in Supplement C.

\begin{algorithm}[H]
\DontPrintSemicolon
\caption{Sparse Projection-Posterior Sampler}
\label{alg:spp}
\KwIn{$\bX\in\R^{n\times p}$, $\bY\in\R^n$, prior scale $a_n$, penalty $P_{\lambda_n}$, tuning $\lambda_n$, draws $M$}
\KwOut{$\{\bm\beta^{*(m)}\}_{m=1}^M$}

$\hat{\bm\beta}^{\mathrm R}\leftarrow(\bX^\top\bX+a_n I_p)^{-1}\bX^\top\bY$, \quad
$\hat{\sigma}_n^2\leftarrow n^{-1}\bY^\top(I_n-H(a_n))\bY$\;

\For{$m=1,\dots,M$}{
Draw ${\sigma^*}^{2(m)}\mid\bY\sim \mathrm{InvGamma}\!\left(\frac n2,\kappa^{-1}\frac{n\hat{\sigma}_n^2}{2}\right)$,
with $\kappa=\hat{\sigma}_n^2/\tilde{\sigma}^2$\;

Draw $\bm\beta^{(m)}\sim \mathcal N_p\!\left(\hat{\bm\beta}^{\mathrm R},
{\sigma^*}^{2(m)}(\bX^\top\bX+a_n I_p)^{-1}\right)$\;

$\bm\beta^{*(m)}\leftarrow
\arg\min_{\bm u\in\R^p}\left\{
n^{-1}\|\bX\bm\beta^{(m)}-\bX\bm u\|^2+
\sum_{k=1}^K P_{\lambda_n}\!\bigl(\|\bm u_k\|_2\bigr)
\right\}$\;
}
\Return{$\{\bm\beta^{*(m)}\}_{m=1}^M$}
\end{algorithm}

\paragraph{Penalty choices $P_{\lambda_n}(\,\cdot\,)$}
Let $t=\lVert \bm u_k\rVert_{2}$ and recall the group size $p_k$.
\begin{align*}
& {\text{Group LASSO}}:
\quad
P^{\mathrm{GL}}_{\lambda_n}(t)=\lambda_n\sqrt{p_k}\,t.\\
& {\text{Group SCAD}}:
\quad \lambda_n|t|\mathds{1}(|t| \leq \lambda_n) + \frac{|t|^2 - 2\tau\lambda_n|t| + \lambda_n^2}{2(\tau-1)}\mathds{1}(\lambda_n < |t| \leq \tau\lambda_n) + \frac{\lambda_n^2(\tau + 1)}{2}\mathds{1}(|t| > \tau\lambda_n).\\
& {\text{Adaptive Group LASSO}}:
\quad
P^{\mathrm{AGL}}_{\lambda_n}(t)=\lambda_n\,w_k\,t,
\qquad
w_k=\bigl\|\hat{\bm\beta}^{\mathrm{GL}}_k\bigr\|_{2}^{-1},
\end{align*}
where $\hat{\bm\beta}^{\mathrm{GL}}$ is a preliminary group LASSO fit with its own
penalty $\tilde\lambda_n$ (chosen so that
$\hat{\bm\beta}^{\mathrm{GL}}$ is consistent at rate $r_n$). The optimization in Step (2c) was carried out in \textsf{R} using the package \texttt{grpreg}. We used \texttt{penalty = "grLasso"}, \texttt{"grSCAD"}, and \texttt{"grMCP"} for the group LASSO, group SCAD, and group MCP maps, respectively; the adaptive group LASSO map was implemented via \texttt{penalty = "grLasso"} with the \texttt{group.multiplier} option based on an initial group-LASSO fit. In all runs we set \texttt{family = "gaussian"}, \texttt{intercept = FALSE}, standardized the design matrix, and used the default convergence settings \texttt{eps = 1e-4} and \texttt{max.iter = 10000}. A single cross-validated value of \(\lambda_n\) was selected and then reused for all posterior draws.

\section{Main Results}\label{Results}
The theoretical properties of the group LASSO have been studied extensively. \cite{buhlmann2011statistics} provided oracle rates for estimation and prediction. \cite{nardi2008asymptotic} proved selection consistency using the irrepresentability condition \citep{zhao2006model, javanmard2013model} and the estimation and prediction consistency using the restricted eigenvalue condition \citep{bickel2009simultaneous}, whereas \cite{wei2010consistent} showed group selection to be consistent by making use of the sparse Riesz condition \citep{zhang2008sparsity}. 
Theoretical properties of group SCAD and group adaptive LASSO penalization techniques were derived under the high-dimensional setting by \cite{guo2015model} and \cite{wei2010consistent}, respectively. In this section, we study the main assumptions required to establish the estimation and selection consistency of our three proposed sparse projection-posterior methods, followed by the corresponding theorems. We also study a debiased projection-posterior that gives asymptotically accurate coverage.

\subsection{Main Assumptions}
This section outlines the main assumptions required for the theoretical development and justification of the proposed sparse projection-posterior methods. These assumptions ensure that the regression model behaves appropriately in sparse and high-dimensional settings and enable us to derive consistency and convergence results. Each assumption addresses a specific aspect of the model, from the properties of the design matrix to signal strength requirements, ensuring that the proposed methods achieve optimal rates of recovery, prediction, and variable selection. The conditions required for establishing the theoretical guarantees of our approach are exactly the same as their frequentist counterparts, including the order of the tuning parameter and the minimal signal strength requirements. In addition to these assumptions introduced in the frequentist penalization approaches, we only require a condition on the the prior parameter $a_n$ (\Cref{sd_assum}). Thus, our assumptions on the penalty level, design, and minimal signal are fully consistent with the frequentist high-dimensional literature for the corresponding penalties. Consequently, the projection-posterior achieves the same rates and selection guarantees while providing Bayesian uncertainty quantification, and the recommended $\lambda_n$ choices align with standard frequentist guidance.
\begin{assumption}[Bounded design condition]
\label{design}
 The predictor variables $|x_{ij}| \leq M_1$ for all $i = 1,\dots,n$, $j= 1,\dots,p$, for some $M_1 > 0$ free of $n$.
\end{assumption}

\begin{assumption}[Bounded true mean condition]
\label{mean_assum} 
The maximum of the absolute value of the expected response under the truth, $\displaystyle\max_{1\le i\le n} |\E_{\btheta^0}(Y_i)|$, is bounded.
\end{assumption}

Assumptions~\ref{design} and \ref{mean_assum} help control the complexity of the model, and are commonly assumed in the literature.

\begin{assumption}[Non-collinearity condition]
\label{sd_assum}
When $p>n$, $\mathrm{rank}(\bX) = n$ and the singular values $d_1, \dots, d_n$ of $\bX$ satisfy 
$\min\{ d_j^2: j=1,\ldots,n\} /(n a_n)\to 0$. 
\end{assumption}
The assumption avoids instability problems arising from multicollinearity by requiring that the design matrix's rank equals the sample size and its singular values remain somewhat large relative to the prior parameter $a_n$, which may be chosen as small as required.


\begin{assumption}[Compatibility condition]
\label{compat}
Let $S \subset \{1,2,\ldots,K\}$ denote the index set of active groups and let $0 < L < \infty$.
The design matrix $\bX \in \mathbb{R}^{n \times p}$ is said to satisfy the \emph{group compatibility condition}
on the cone
\small\(
\mathcal{C}(S,L)
\;=\;
\Big\{
\bm v \in \mathbb{R}^{p} :
\sum_{k \in S^{\mathrm{c}}} \sqrt{p_k}\,\|\bm v_k\|
\le
L \sum_{k \in S} \sqrt{p_k}\,\|\bm v_k\|
\Big\},
\)
with compatibility constant $\phi(S) = \phi(S,L) > 0$, if
for all $\bm v \in \mathcal{C}(S,L)$ it holds that
\small\(
\Big(
\sum_{k \in S} \sqrt{p_k}\,\|\bm v_k\|
\Big)^{2}
\;\le\;
\tfrac{\sum_{k \in S} p_k}{\phi^{2}(S)}
\cdot
\tfrac{\|\bX \bm v\|^{2}}{n}.
\)
\end{assumption}

This assumption ensures that the set of predictors satisfies a sparsity-related geometric property, needed to link the sparsity structure of the coefficients to the observed data. We also make explicit that the inequality is required only on the cone $\mathcal C(S,L)$ which is the set of directions delivered by the basic-inequality/KKT steps in our proofs. When \(p>n\), we have \(\mathrm{null}(\bm X)\neq{0}\). If there exists a nonzero \(\bm v\in \mathcal C(S,L)\cap \mathrm{null}(\bm X)\), then \(|\bm X\bm v|=0\) while \(\sum_{k\in S}\sqrt{p_k},|\bm v_k|>0\), which would force \(\phi(S,L)=0\). Assumption \ref{compat} explicitly rules out this pathology by requiring \(\phi(S,L)>0\); equivalently, it requires that the empirical Gram matrix be strictly positive on the cone \(\mathcal C(S,L)\). This is the standard compatibility/restricted-eigenvalue requirement in high-dimensional analysis: we do not ask for invertibility on \(\mathbb R^p\) (impossible when \(p>n\)), only on the restricted set of directions relevant for sparse recovery.


\begin{assumption}[Non-singularity of the Gram matrix]\label{non-singularity_condition}
Let \(\bC_{S_0}=n^{-1}\bX_{S_0}^{\mathrm{T}}\bX_{S_0}\) be the Gram matrix formed by the columns in the true active set \(S_0\). Assume \(\bC_{S_0}\) is positive definite with its smallest eigenvalue bounded away from zero, i.e.
\(
\lambda_{\min}(\bC_{S_0}) \ge M_2 > 0 ,
\)
equivalently, for all \(\bm u\in\mathbb{R}^{p_0}\),
\(
\bm u^{\mathrm{T}}\bC_{S_0}\bm u \ge M_2\|\bm u\|^2.
\)
\end{assumption}

This ensures that the relevant subset of the design matrix has eigenvalues bounded away from zero, allowing stability of estimates for the active groups and consistent model selection.

\begin{assumption}[Beta-min condition]
\label{min-beta_condition} 
The magnitude of a signal (coefficient of an active predictor) is bounded below as follows: 
    \begin{enumerate}
        \item [{\rm (A)}] (Group LASSO projection-posterior) $\displaystyle 
        {\min_{k \in S_0} \|\bm{\beta}^0_k\|_\infty} \geq \textnormal{max} \bigg \{ \frac{\sqrt{\log p_0}}{n}, \frac{\lambda_n \sqrt{s_0} p_{\textnormal{max}}^{3/2}}{\lambda_{\textnormal{min}}(\bC_{n(11)})} \bigg\}$;  
        \item [{\rm (B)}] (Group SCAD projection-posterior) $\displaystyle {\min_{k \in S_0} \|\bm{\beta}^0_k\|} \geq M_3 n^{-({1-c_3})/{2}}$ for constants $c_3, M_3 > 0$;   
        \item [{\rm (C)}] (Group adaptive LASSO projection-posterior) Let  $r_n$ be the convergence rate of an initial estimator $\hat{\bm\beta}^\mathrm{GL}$. Then 
    $ \min_{k \in S_0} \|\bm\beta_k^0\| > \max \bigg\{ p_{\textnormal{max}}^{3/4}\sqrt{\lambda_n s_0}, \frac{\sqrt{p_{\textnormal{max}} \log s_0}}{n}, \frac{p_{\textnormal{max}}^{5/2} s_0^2}{r_n p_{\textnormal{min}}}\bigg\}.$
    \end{enumerate}
\end{assumption}
The Beta-min condition is necessary to separate signals from noise. 

\begin{assumption}[Irrepresentability condition]\label{irrepresentability_condition}
    For some $0 < \nu < 1$ and for every $k \in S_0^\mathrm{c}$, we have $\|\bX_{G_k}^\mathrm{T}\bX_{S_0}(\bX_{S_0}^\mathrm{T}\bX_{S_0})^{-1}\| \leq {1-\nu}.$ 
\end{assumption}
The condition restricts the total effect of inactive groups of predictors compared to active ones and is needed to establish model selection consistency \citep{zhao2006model}. 

\subsection{Recovery and Prediction Rate}
In this section, we analyze the recovery and prediction consistency properties of the group LASSO, group SCAD, and group adaptive LASSO maps. We focus on establishing the conditions under which each posterior is consistent for estimation and demonstrates superior predictive performance in high-dimensional settings with group structure among the covariates. We study the following three properties:
\begin{align}\label{estimation 1}
     \Pi \big( \sum_{k = 1}^K \sqrt{p_k} \|\bm\beta_k^* - \bm\beta_k^0\| \geq c_1 r_{1n} \big| \bY \big) \to 0;
 \end{align} \vspace{-1.5cm}
\begin{align}\label{estimation 2}
    \Pi \big( \|\bm\beta^* - \bm\beta^0\|^2 \geq c_1^2 r_{2n} \big| \bY \big) \to 0;
\end{align} \vspace{-1.5cm}
\begin{align}\label{prediction}
     \Pi \big( n^{-1} \|\bX ( \bm\beta^* - \bm\beta^0)\|^2 \geq c_2 r_{3n} \big| \bY \big) \to 0,
\end{align} 
in probability under the true distribution for some rates $r_{1n},r_{2n},r_{3n}$. We identify the applicable rates ($r_{1n}, r_{2n}$, and $r_{3n}$) for the group LASSO, group SCAD, and group adaptive LASSO projection-posteriors. Note that the following theorems also apply to the ungrouped case by setting $p_k = 1$ for $k = 1,\dots,K$. For instance, taking $\lambda_n \asymp \sqrt{(\log p)/n}$, we obtain $r_{2n} = (s_0^2 \log p)/n$ and $r_{3n} = (s_0 \log p)/n$, which are optimal for sparse high-dimensional linear regression and are also attained by the ungrouped LASSO. 

To prove that the induced posterior from the SCAD penalized map is estimation and prediction consistent with the optimal rate, we first need to show an oracle property of this map. This requires an additional assumption about the signal strength to show that the group SCAD projection-posterior has an oracle property, as if we knew the locations of the zero coefficients in advance. Similar assumptions will later be used to show that the sparse projection-posteriors are model-selection consistent. We do not, however, need to make any assumptions on the minimum signal strength to obtain rates for the adaptive group LASSO projection-posterior. Scaling the penalty by the group LASSO estimator $\hat{\bm\beta}^\textnormal{GL}$ guarantees the same convergence rate under reduced conditions. We state the optimal recovery and prediction consistency rates for the three sparse projection-posteriors below. 

\begin{theorem}[Recovery and prediction rates] 
For $\lambda_n \geq \lambda_0$, where 
\begin{align}\label{lambda_0}
    \lambda_{0}^2(x,p) = \dfrac{8}{n} \left(1 + \sqrt{\frac{4x+4\log p}{p_{\textnormal{min}}}} + \frac{4x+4\log p}{p_{\textnormal{min}}} \right) \textnormal{ with } x> 0,
\end{align} 
the following assertions hold with posterior probability at least $1 - \textnormal{exp}(-x)$: 
\begin{enumerate}
    \item (Group LASSO Projection-posterior) Under Assumptions \ref{design}, \ref{mean_assum}, \ref{sd_assum}, \ref{compat} and \ref{non-singularity_condition}, we have for $c_1, c_2 > 0$, \eqref{estimation 1} holds with $r_{1n} = \lambda_n \sum_{k = 1}^{s_0} p_k$, \eqref{estimation 2} holds with $r_{2n} ={\lambda^2_n }\big(\sum_{k = 1}^{s_0} p_k\big)^2 / p_{\textnormal{min}}$ and \eqref{prediction} holds with $r_{3n} = \lambda_n^2 \sum_{k = 1}^{s_0} p_k$.
    \item (Group SCAD Projection-posterior) If Assumptions \ref{design}, \ref{mean_assum}, \ref{sd_assum}, \ref{compat}, \ref{non-singularity_condition} and \ref{min-beta_condition}(B) hold, then for $c_1, c_2 > 0$, \eqref{estimation 1} holds with $r_{1n} = \lambda_n \sum_{k = 1}^{s_0} p_k$, \eqref{estimation 2} holds with $r_{2n} ={\lambda^2_n }\big(\sum_{k = 1}^{s_0} p_k\big)^2 / p_{\textnormal{min}}$ and \eqref{prediction} holds with $r_{3n} = \lambda_n^2 \sum_{k = 1}^{s_0} p_k$. 
    \item (Group Adaptive LASSO Projection-posterior) 
    Let Assumptions \ref{design}, \ref{mean_assum}, \ref{sd_assum}, \ref{compat}, and \ref{non-singularity_condition} hold.   For some \(c_1, c_2 > 0\), \eqref{estimation 1} holds with \(r_{1n} = \lambda_n \sum_{k=1}^{s_0} p_k\), \eqref{estimation 2} holds with  \(r_{2n} = {\lambda_n^2 \left(\sum_{k=1}^{s_0} p_k \right)^2}/{p_{\textnormal{min}}}\), and \eqref{prediction} holds with \(r_{3n} = \lambda_n^2 \sum_{k=1}^{s_0} p_k\).
\end{enumerate}
\label{est_pred_consis}
\end{theorem}

\subsection{Model Selection Consistency}

Model selection consistency is a critical feature of sparse high-dimensional regression methods, especially when the number of predictors exceeds or is comparable to the sample size. A regression method is said to possess model selection consistency if, as the sample size increases, it accurately identifies the active groups of predictors, meaning that it selects the correct set of non-zero coefficients while exactly setting the irrelevant ones to zero. In high-dimensional settings, sparsity is key because the number of truly important variables is often much smaller than the total number of available predictors. However, one challenge of sparse regression is balancing bias and variance. For the sparse projection maps described in this paper, while aggressive penalties may shrink the components of the posterior samples too much, thus introducing bias, lighter penalties may not fully enforce sparsity. Model selection consistency ensures that, as the sample size grows, the chosen model correctly reflects the underlying sparse structure without underfitting or overfitting: $\Pi(\{\|\bm{\beta}^*_k\|>0 \text{ for all } k \in S_0, \text{and } \bm{\beta}^*_k = 0 \text{ for all } k \in S_0^\mathrm{c}|\bY) \to 1 $ 
as $n \to \infty$ in probability under the true distribution.

The group LASSO penalized estimator is model selection consistent \citep{nardi2008asymptotic} under certain assumptions, including but not restricted to the Beta-min and the irrepresentability conditions. We show here that the group LASSO projection-posterior is also model selection consistent if the prior parameter $a_n$ satisfies Assumption \ref{sd_assum}. The model selection consistency of the group SCAD penalized projection-posterior follows from the oracle property provided in Assumption \ref{oracle_SCAD}. According to the Oracle property, the minimizer of the optimization function using the group SCAD penalty results in the Oracle posterior $\bm\beta_{S_0}$ for large $n$ under certain assumptions, including a restriction on the minimum signal strength. The oracle projection is asymptotically a local minimizer under the group SCAD map, even in high-dimensional settings. Thus, for large $n$, there is a guarantee for the sparse projection $\bm\beta^*$ using the group SCAD map to select the correct model with high posterior probability. The sparse projection-posterior resulting from the group adaptive LASSO map, on the other hand, can be shown to be selection consistent under weaker conditions than the group LASSO map. We can avoid imposing the irrepresentability condition by weighting the penalty by the norm of the group LASSO estimator for each group. However, the Beta-min condition will still be required. Here, the assumption of the minimum signal strength differs from the requirements under the group LASSO and SCAD maps. 

\begin{theorem}[Model selection consistency]\label{model_sel_consis} 
The following assertions hold: 
\begin{enumerate}
    \item (Group LASSO Projection-posterior) If ${{\log (p - p_0)}/{n\lambda_n^2}} \to 0 \textnormal{ as } n \to \infty$ and Assumptions \ref{design}, \ref{mean_assum}, \ref{sd_assum}, \ref{non-singularity_condition}, \ref{min-beta_condition}(A) and \ref{irrepresentability_condition} hold, then the group LASSO projection-posterior is model selection consistent. 
    \item (Group SCAD Projection-posterior) If Assumptions \ref{design},\ref{mean_assum},\ref{sd_assum} and \ref{min-beta_condition}(B) hold, then, for $\lambda_n = o\big(n^{-{(1-c_3 + c_4)}/{2}}\big)$ and $p/(\lambda_n n^{3/2}) \to 0$, the group SCAD projection-posterior is model selection consistent.
    \item (Group Adaptive LASSO Projection-posterior) If ${p_{\textnormal{max}} \log (K - s_0)}/({\sqrt{n}\lambda_n r_n p_{\textnormal{min}}}) \to 0$ and  Assumptions \ref{design}, \ref{mean_assum}, \ref{sd_assum} and \ref{min-beta_condition}(C) hold, then the group adaptive LASSO projection-posterior is model selection consistent.
\end{enumerate}
\end{theorem}

\subsection{Uncertainty Quantification}
The sparse projection-posterior, although it concentrates around the true parameter at an optimal rate, consistently selects the correct signal groups nd automatically quantifies the associated uncertainty by virtue of it being a Bayesian method; it does not give exact frequentist coverage. To achieve targeted coverage, we need to take a further step to adjust for the penalization bias. This process, known in the literature as `debiasing,' has not previously been used in the Bayesian paradigm. Here, we debias each sparse projection-posterior sample to obtain the debiased projection-posterior samples, which we show have exact asymptotic coverage. It should be noted that no additional sampling is required at this stage. We only need to modify the sparse projection-posterior samples through a debiasing map. To define the debiased projection, we must first specify a reasonable representation of the inverse of the Gram matrix $\hat{\bm\Sigma}$. Following \cite{van2014asymptotically,honda2021biased}, define $\hat{\bm\Theta}$ that satisfies $\hat{\bm\Theta}\hat{\bm\Sigma} = \bm{I}_p$. Mimicking the construction of $\hat{\bm\Theta}$ from \cite{van2014asymptotically} and \cite{honda2021biased}, we define the group LASSO regression of the $l$-th variable in the $g$-th group on all the other groups, for $l= 1,2,\dots,p_g$, and $g = 1,2,\dots,K$, as follows: $\boldsymbol\gamma_g^{l} = \argmin_{\bm{\gamma} \in \R^{p - p_g}} \big\{n^{-1}\|\bX_{G_g}^{(l)} - \bX_{-g}\bm\gamma\|^2 + \lambda_g^l \sum_{k \neq g} \sqrt{p_k}\|\bm{\gamma}_k\| \big\},$ where $\lambda_g^l$ is the corresponding tuning parameter. Note that the vector $\bm\gamma^l_g$ can be rewritten as $\bm\gamma^l_g = ({\bm\gamma^l}^\mathrm{T}_{g,1},\dots,{\bm\gamma^l}^\mathrm{T}_{g,g - 1},{\bm\gamma^l}^\mathrm{T}_{g,g + 1},\dots,{\bm\gamma^l}^\mathrm{T}_{g,K})^\mathrm{T}$, where each $\bm\gamma_{g,k}^l \in \R^{p_k}$ for $k \in \{1,2,\dots,K\} \setminus \{g\}$. We also define the matrix $\bm\Gamma_g \in \R^{p - p_g \times p_g}$ which is composed of the group LASSO estimators by regressing all the $p_g$ variables of the $g$-th group on the rest of the $(K-1)$ groups, that is, $\bm\Gamma_g = (\bm\gamma_g^1,\bm\gamma_g^2,\dots,\bm\gamma_g^{p_g}).$ Next, let $\bm{R}_g$ denote the residual matrix obtained by regressing $\bX_{G_g}$ on $\bX_{-g}$ using the group LASSO penalty, that is, $\bm{R}_g = \bX_{G_g} - \bX_{-g}{\bm\Gamma_g} \in \R^{n \times p_g}.$
The first-order KKT conditions for these column-wise group LASSO regressions are given by $-\frac{2}{n} \bX_{-g}^\mathrm{T}(\bX_{G_g}^{(l)} - \bX_{-g} \bm\gamma^l_g) + \lambda_g^l \bm\kappa_g^l = \textbf{0} \in \R^{p - p_g},$ where $\bm\kappa_g^l = ({\bm\kappa^l}^\mathrm{T}_{g,1},\dots,{\bm\kappa^l}^\mathrm{T}_{g,g - 1},{\bm\kappa^l}^\mathrm{T}_{g,g + 1},\dots,{\bm\kappa^l}^\mathrm{T}_{g,K})^\mathrm{T} \in \R^{p - p_g}$ with $\bm\kappa_{g,k}^l \in \R^{p_k}$ for $k \neq g$ and $\|\bm\kappa_{g,k}^l\| \leq 1$ if $\|\bm\gamma_{g,k}^l\| = 0$ and $\bm\kappa_{g,k}^l = \bm\gamma_{g,k}^l/\|\bm\gamma_{g,k}^l\|$ if $\|\bm\gamma_{g,k}^l\| \neq 0$. Define the $(p - p_g) \times p_g$ matrix $\bm{K}_g = (\bm\kappa_g^1,\bm\kappa_g^2,\dots,\bm\kappa_g^{p_g})$, and express the KKT condition in the matrix form ${2}{n^{-1}}\bX_{-g}^\mathrm{T}\bm{R}_g = \bm{K}_g\bm\Lambda_g,$ where $\bm\Lambda_g \in \R^{p_g \times p_g}$ is $\textnormal{diag}(\lambda_g^1,\dots,\lambda_g^{p_g})$. Finally, following \cite{honda2021biased}, we have {\small\begin{align}\label{hat_theta}
    \hat{\bm{\Theta}}^\mathrm{T} = \begin{bmatrix}
\bm{I}_{p_1} & -\bm{\Gamma}_{2,1} & -\bm{\Gamma}_{3,1} & \dots & -\bm{\Gamma}_{K,1} \\
-\bm{\Gamma}_{1,2} & \bm{I}_{p_2} & -\bm{\Gamma}_{3,2} & \dots & -\bm{\Gamma}_{K,2} \\
-\bm{\Gamma}_{1,3} & -\bm{\Gamma}_{2,3} & \bm{I}_{p_3} & \dots & -\bm{\Gamma}_{K,3} \\
\vdots & \vdots & \vdots & \ddots & \vdots \\
-\bm{\Gamma}_{1,K} & -\bm{\Gamma}_{2,K} & -\bm{\Gamma}_{3,K} & \dots & \bm{I}_{p_K} 
\end{bmatrix} \begin{bmatrix}
\bm{T}_1^2 & \textbf{0} & \textbf{0} & \dots & \textbf{0} \\
\textbf{0} & \bm{T}_2^2 & \textbf{0} & \dots & \textbf{0} \\
\textbf{0} & \textbf{0} & \bm{T}_3^2 & \dots & \textbf{0} \\
\vdots & \vdots & \vdots & \ddots & \vdots \\
\textbf{0} & \textbf{0} & \textbf{0} & \dots & \bm{T}_K^2 
\end{bmatrix},
\end{align}} where ${\bm{T}_g}^2 = n^{-1} \bX_{G_g}^\mathrm{T}\bm{R}_g \in \R^{p_g \times p_g}$ and $\bm{\Gamma}^\mathrm{T}_{j,k} = (\bm\gamma_{j,k}^1, \bm\gamma_{j,k}^2, \dots, \bm{\gamma}_{j,k}^{p_j})^\mathrm{T} \in \R^{p_j \times p_k}$. We now define the debiased Group LASSO projection as 
\begin{align}\label{debiasedGLmap}
    \bm{\beta}^{**} = \bm{\beta}^* + \frac{1}{n}\hat{\bm{\Theta}}\bX^\mathrm{T}(\bX\bm{\beta} - \bX\bm{\beta}^*) = \bm\beta^0 + \frac{1}{n}\hat{\bm{\Theta}}\bX^\mathrm{T}\bm\eta - \bm{\Delta},
\end{align}
where $\bm{\Delta} = (\hat{\bm\Theta}\hat{\bm{\Sigma}} - \bm{I}_p)(\bm\beta^* - \bm{\beta}^0) \in \R^p$. From \eqref{hat_theta}, we have that the $(j,k)$-th block of $\hat{\bm\Theta}\hat{\bm\Sigma} - \bm{I}_p$ is given by the $(\bm{T}_j^{-2})^\mathrm{T}\bm\Lambda_j\bm{K}^\mathrm{T}_{j,k} \in \R^{p_j \times p_k}$ for $j \neq k$ and $\textbf{0} \in \R^{p_j \times p_j}$ along the diagonal when $j = k$. Note that $\bm{K}^\mathrm{T}_{j,k} = (\bm\kappa_{j,k}^1, \bm\kappa_{j,k}^2, \dots, \bm{\kappa}_{j,k}^{p_j})^\mathrm{T} \in \R^{p_j \times p_k}$. A component-wise credible set defined based on the posterior distribution of this sparse projection $\bm\beta^{**}$ can be shown to achieve accurate asymptotic coverage corresponding to the debiased group-LASSO map motivated by the debiased group-LASSO estimator \citep{honda2021biased} defined by 
{\small \begin{align}
\label{debias_estim_para}
    \hat{\bm\beta}^\textnormal{DGL} = \bm\beta^0 + n^{-1}\hat{\bm\Theta}\bX^\mathrm{T}\varepsilon - \bm\Delta^\textnormal{DGL},
\end{align}}
where $\bm\Delta^\textnormal{DGL} = (\hat{\bm\Theta}\hat{\bm{\Sigma}} - \bm{I}_p)(\hat{\bm\beta}^\textnormal{GL} - \bm{\beta}^0)$, and $\hat{\bm\beta}^\textnormal{GL}$ is the group LASSO estimator.

\begin{assumption}
\label{debias_assums} For all $j = 1,2\dots,p$, $\lambda_{\textnormal{max}}(\hat\Sigma_{j,j}) \leq C_1$  for some $C_1 > 0$, and for all $l = 1,\dots p_g$ and $g = 1,\dots,K$, $\|\boldsymbol\gamma_l^g\|  \leq C_2$  for some $C_2>0$.

\end{assumption}

\begin{theorem}[Bernstein-von Mises Theorem]\label{debiasedCLT_GL}
    If Assumptions \ref{design}, \ref{mean_assum}, \ref{sd_assum} and \ref{debias_assums} hold, we have $\|\bm\Delta_k\| = o_P(n^{-1/2})$ uniformly in $k = 1,2,\dots,K$, and $$\sup_B \Big|\Pi\big(\sigma^{-1}\sqrt{n}(\bm\beta^{**} - \bm\beta^0) \in B \big|\bY\big) - \normal_p(B;\bm{m},\hat{\bm{V}})\Big| \to 0, \textnormal{ as } n \to \infty,$$ where $\bm{m} = n^{-1/2} \hat{\bm\Theta}\bX^\mathrm{T}\bm\mu \textnormal{ and } \bm{V} = n^{-1} \hat{\bm\Theta}\bX^\mathrm{T}\bm{H}({a_n})\bX\hat{\bm\Theta}^\mathrm{T}.$
\end{theorem}
Let \( \mathcal{C}_j = [\tilde{\beta}_j - q_{j,\alpha},\ \tilde{\beta}_j + q_{j,\alpha}] \) denote the symmetric component-wise \((1 - \alpha)\)-credible interval for \(\beta_j\), for \(j = 1, \ldots, p\), where \(\tilde{\beta}_j\) is the median of the posterior distribution of \(\beta_j^{**}\), and \(q_{j,\alpha}\) is the \((1 - \alpha/2)\)-quantile of the posterior distribution of \(|\beta_j^{**} - \tilde{\beta}_j|\). The proposed credible sets will provide accurate frequentist coverage in the limit if the normal limit of the centered debiased projection-posterior matches the asymptotic normal distribution of the debiased group-LASSO estimator.

\begin{corollary}[Coverage of credible region]\label{debiased_coverage}
    If Assumptions \ref{design}, \ref{mean_assum}, \ref{sd_assum} and \ref{debias_assums} hold, then for all  $j = 1,2,\dots,p$, $\lim_{n\to \infty} \sup_{\beta_j^0 \in \R} |\prob_{\beta^0_j} \big( \beta^0_j \in \mathcal{C}_j \big) - (1 - \alpha)| = 0.$
\end{corollary}

The debiased projection described above only used two ingredients:  a sparse approximate inverse \(\hat{\bm\Theta}\) of the Gram matrix satisfying \(\|\hat{\bm\Theta}\,\hat{\bm\Sigma}-\bm I_p\|_\infty=o_P(1)\) (constructed by the nodewise group LASSO) and a penalized estimator \(\hat{\bm\beta}^{\rm init}\) for which the
remainder \(\bm\Delta^{\rm init}:=(\hat{\bm\Theta}\hat{\bm\Sigma}-\bm I_p)(\hat{\bm\beta}^{\rm init}-\bm\beta^0)\) is \(o_P(n^{-1/2})\). Hence the same debiasing formula applies verbatim when \(\hat{\bm\beta}^{\rm init}\) is the group SCAD or adaptive group LASSO (AGL) estimator.  We state the resulting Bernstein-von Mises limits in Supplement~D; the proofs follow exactly the argument of Theorem~\ref{debiasedCLT_GL}, and are
therefore omitted.

\section{Nonparametric Additive Regression}\label{sec:additive_model}
High-dimensional nonparametric additive regression under sparsity can effectively use the group LASSO formulation. The COSSO proposed by \cite{lin2006component} is a regularization technique that uses the sum of component norms as the penalty function. They also studied the convergence rate of the COSSO estimator. \cite{meier2009high} proposed a combined sparsity-smoothness penalty for high-dimensional generalized additive models, penalizing both the sparsity and the roughness. Estimation and variable selection consistency results of group LASSO and adaptive group LASSO methods under the high-dimensional additive model were established by \cite{huang2010variable}. 


 Let the response variable be $\bY \in \R^n$ and the $K$-dimensional covariate vector $(X_1,\dots,X_K)$ be collected for the $n$ subjects and stored in the matrix $\bX \in \R^{n \times K}$. Consider the model 
 \begin{align}
 \label{additive model}
    Y_i = \sum_{k = 1}^K f^0_k(X_{k,i}) + \varepsilon_i, \enskip i = 1,2,\dots,n,
\end{align}
where $X_{k,i}$ denotes the covariate information corresponding to $i$-th subject for the $k$-th variable. Here, $f^0_k(\cdot)$ is unknown for all $j = 1,2,\dots,K$ and $\varepsilon_i$ is the mean zero random noise with variance $\sigma^2$ for some $\sigma >0$. We suppose that not all of these variables are important to the response. Thus, some $f_k^0$'s are zero. The goal is to accurately and consistently identify these zero-additive components and estimate the non-zero components. 

Define $B_n$ as the number of B-spline basis functions used to expand each $f_k^0(x)$, $k = 1,2,\dots,K$. We approximate $f_k^0(X_{k,i})$ using the following B-spline expansion 
\begin{align}
\label{B-spline approximation}
    f_k(X_{k,i}) = \sum_{j = 1}^{B_n} \mathcal{B}_{k,j}(X_{k,i}) \beta_{k,j}. 
\end{align}
The group selection problem, as discussed earlier, aptly fits the goal here: to identify the set of spline coefficients that are significant in the model and discard the remaining variables. The model now becomes $\bY = \mathcal{\bm{B}} \bm\beta + \bm\varepsilon$, where $\bm\beta = (\bm\beta_1,\bm\beta_2,\dots,\bm\beta_K)^\mathrm{T} \in \R^{B_n K},$ such that $\bm\beta_k \in \R^{B_n}$ for all $k = 1,2,\dots,K$. In this model, the dimension of the coefficient vector is $B_n K = p$, say. Then, we may have $p \gg n$ and $ p$ increasing with $n$. The $i$-th row of the matrix $\mathcal{\bm{B}} \in \R^{n \times p}$ is $(\mathcal{\bm{B}}^\mathrm{T}_1(X_{1,i}),\dots,\mathcal{\bm{B}}^\mathrm{T}_K(X_{K,i}))$, where $\mathcal{\bm{B}}_k(X_{k,i}) \in \R^B$ for each $k = 1,2,\dots,K$. We write $\bm\beta^0$ as the true coefficient vector. Now, we propose to use the sparse projection-posterior method described in \Cref{Method} to get the induced posterior of $\bm\beta^*$ and define $f_k^*(X_{k,i}) = \sum_{j = 1}^{B_n} \mathcal{B}_{k,j}(X_{k,i}) \beta^*_{k,j}.$

\begin{theorem}[Consistency of Projection-Posteriors for Additive Model]\label{additve_est_consis} 
Under the additive model, the group LASSO projection-posterior satisfies the following: 
\begin{enumerate}
    \item (Recovery rate) If $\lambda_n \asymp \sqrt{\log (K B_n)/n}$ and $B_n \asymp n^{1/(1+2\alpha)}$, then, under the same Assumptions as in \Cref{est_pred_consis}, the projection-posteriors obtained by any of the three mappings satisfy $\Pi(\|f^* - f^0\|^2 \geq c_5 s_0n^{-{2\alpha}/{(2\alpha+1)}}\log p\big|\bY) \to 0 $ for $c_5 > 0$ as $n \to \infty$, in probability under the true distribution.
    \item (Model selection consistency) Under the same Assumptions as in \Cref{model_sel_consis} and for $B_n \asymp n^{1/(1+2\alpha)}$, all the three proposed projection-posteriors are selection consistent for the respective choices of $\lambda_n$: for the gruop LASSO $\lambda_n \asymp \sqrt{\log (K B_n)/n}$, for the group SCAD $\lambda_n = o\big(n^{-{(1-c_3 + c_4)}/{2}}\big)$ and $KB_n/(\lambda_n n^{3/2}) \to 0$ and for the adaptive group LASSO $\log (K - s_0)/({\sqrt{n}\lambda_n r_n}) \to 0$.
\end{enumerate}

\end{theorem}

\Cref{additve_est_consis} shows that the posterior probability that the distance between the induced posterior of the coefficients in the spline representation of the nonparametric functions and the true model converges to zero in probability. The convergence rate is influenced by three primary factors: stochastic error in estimating the nonparametric components, spline approximation error, and bias introduced by penalization. Next, we show that the posterior probability that the three projection-posteriors select all nonzero additive components converges to 1 in probability under the true distribution. In the additive model, the need to estimate smooth functions introduces an additional error source resulting from the spline approximation; that is, the modified error $\bm\varepsilon^{(2)} = \bm\varepsilon + \bm\delta$, where $\bm\delta = (\delta_1,\dots,\delta_n)^\mathrm{T}$, with $\delta_i = \sum_{k = 1}^{K} \big(f_k^0(X_{k,i}) - f_k(X_{k,i})\big)$. Since for the $j$-th component we have $\|f^0_j - f_j\| = \mathcal{O}\big(B_n^{-\alpha}\big) = \mathcal{O}\big(n^{-\alpha/(2\alpha + 1)}\big)$, with $B_n = \mathcal{O}\big(n^{1/(2\alpha + 1)}\big)$, it follows that $\|\delta_i\| = \|\sum_{j = 1}^{s_0}(f^0_j - f_j)\| \leq s_0 \|f^0_j - f_j\| = \mathcal{O}\big(s_0 n^{-\alpha/(2\alpha + 1)}\big)$. Consequently, $\|\bm\delta\| = \sqrt{\sum_{i = 1}^n \|\delta_i\|^2} = \mathcal{O}\big(\sqrt{ns_0^2n^{-2\alpha/(2\alpha + 1)}}\big) = \mathcal{O}\big({s_0}n^{1/(4\alpha + 2)}\big).$ Thus, to prove the model selection consistency result for the additive model, we need to adjust for this error and then proceed as in  \Cref{model_sel_consis}.

\section{Numerical Results}
\label{Numericals}

\subsection{Simulation Study 1: Grouped Linear Regression}\label{simu1}

\begin{figure}[htbp]
    \centering
    \includegraphics[width=0.7\linewidth]{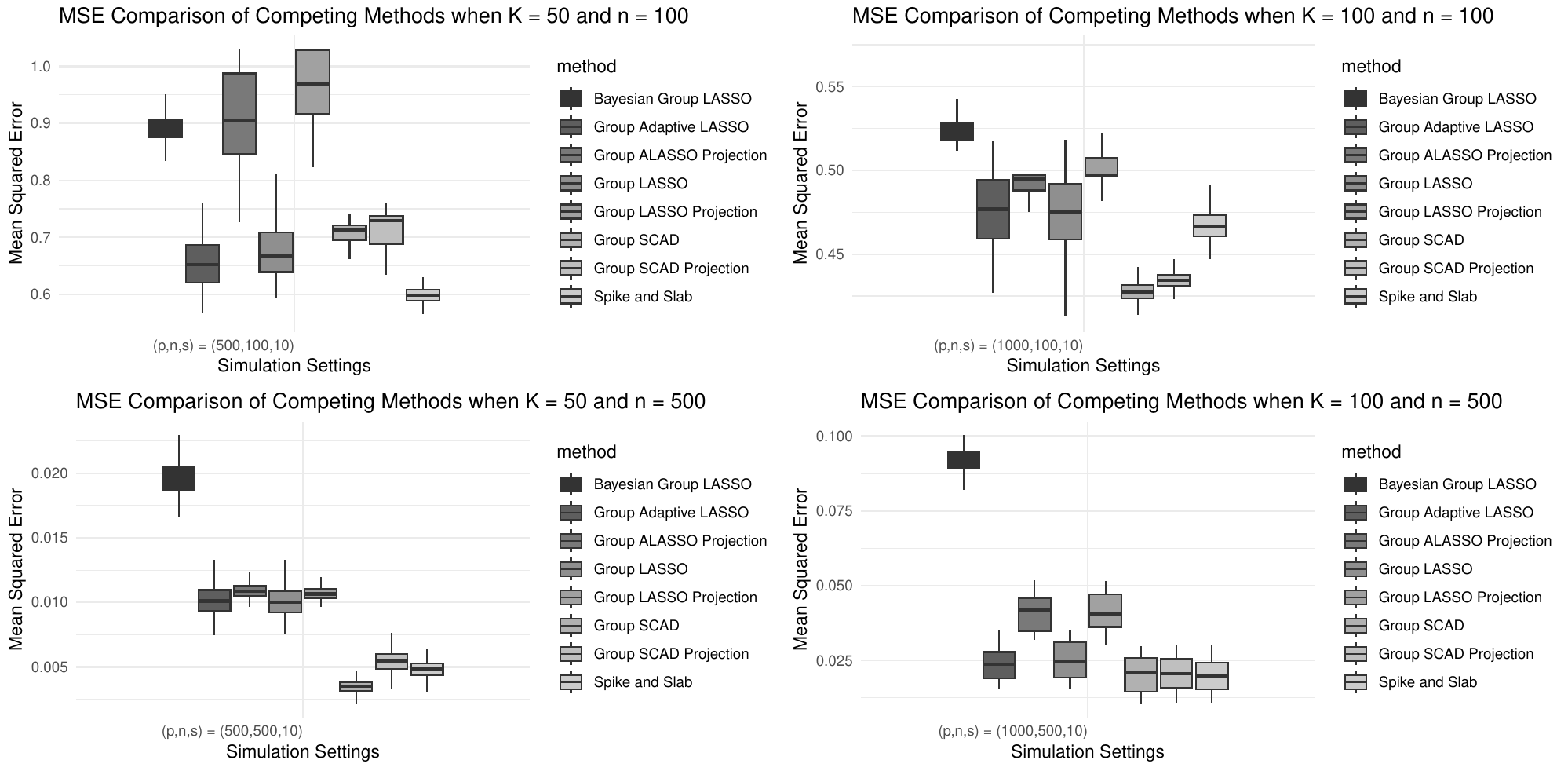}
    \caption{MSE comparisons for the four pairs of ($K,n$), namely (50,100),(50,500),(100,100) and (100,500) when $s_0 = 10$ groups are active, replicated 100 times}
    \label{fig:MSE_s10}
\end{figure}

\begin{figure}[htbp]
    \centering    \includegraphics[width=0.7\linewidth]{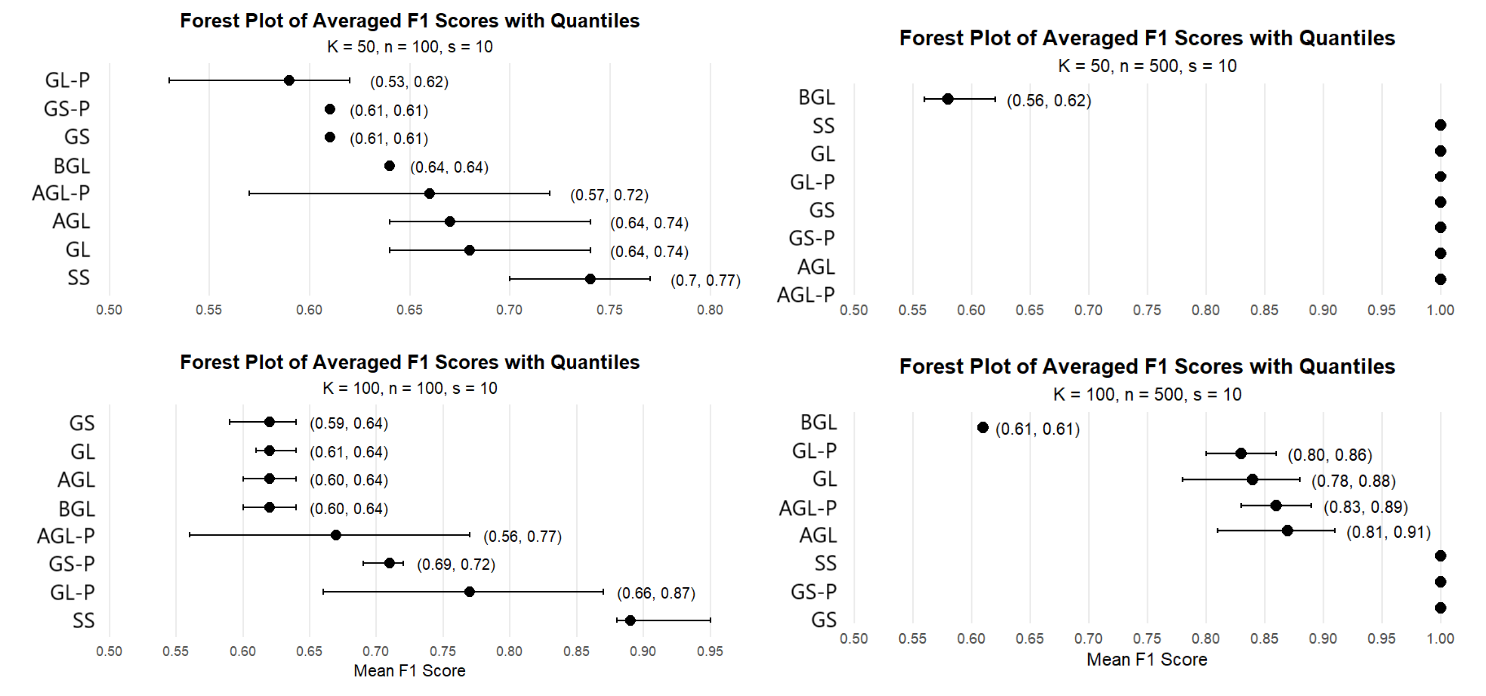}
    \caption{F1-score comparisons for the four pairs of ($K,n$), namely (50,100),(50,500),(100,100) and (100,500) when $s_0 = 10$ groups are active, replicated 100 times.}
    \label{fig:F1score_s10}
\end{figure}

\begin{figure}[htbp]
    \centering
    \includegraphics[width=\linewidth]{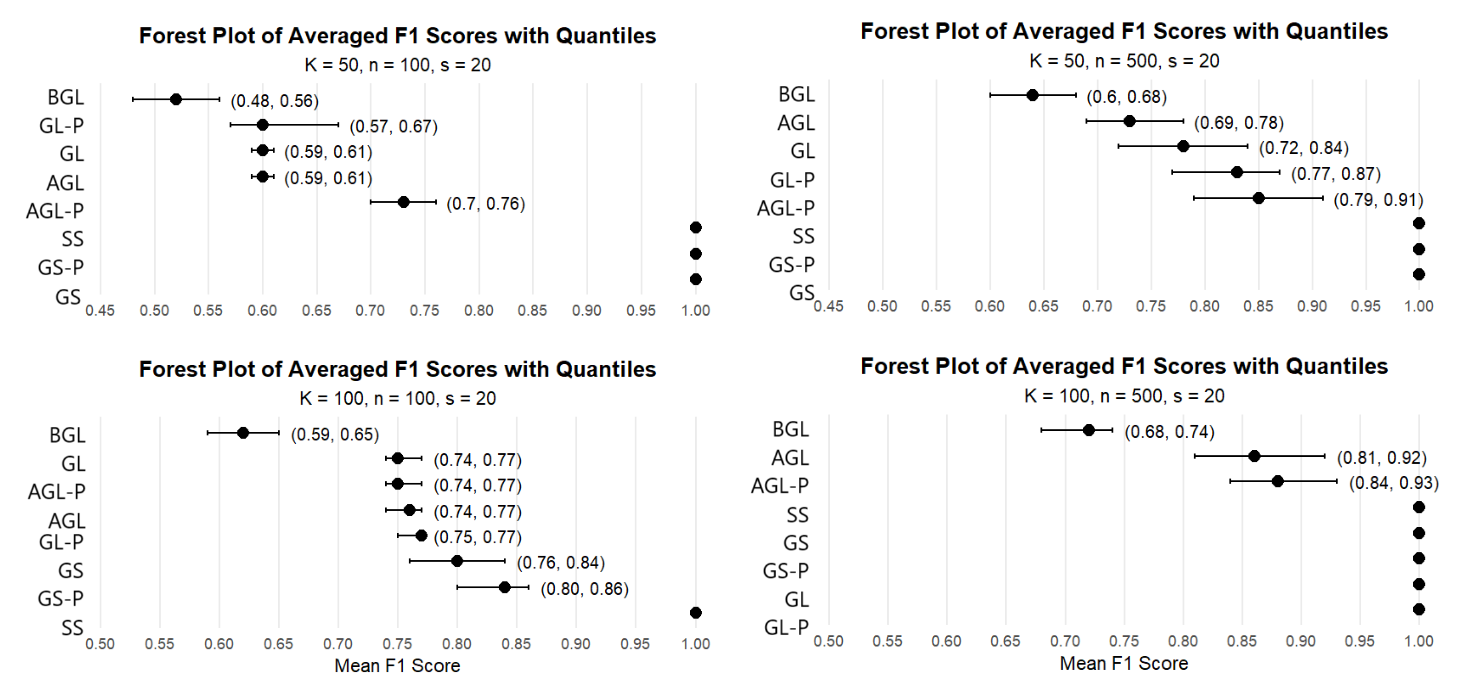}
    \caption{F1-score comparisons for the four pairs of ($K,n$), namely (50,100),(50,500),(100,100) and (100,500) when $s_0 = 20$ groups are active, replicated 100 times.}
    \label{fig:F1score_s20}
\end{figure}

For the first simulation study, we evaluated the performance of competing methods in high-dimensional grouped regression under different sparsity and dimensionality scenarios. Specifically, we considered four combinations of feature dimensions or the number of groups ($K$) and sample sizes ($n$): \((K, n) \in \{ (50, 100), (100, 100), (50, 500), (100, 500)\}\). The number of active groups, indicating nonzero coefficients, was varied across \(s_0 \in \{10, 20\}\), reflecting varying sparsity levels. Keeping an average of 10 variables per group, we get $p = 500$ when $K = 50$ and $p = 1000$ when $K = 100$, respectively, for the three ($K,n$) pairs. Eight competing methods were compared over 100 replications for each setting. The methods with which we compared our three proposed maps, namely group LASSO projection-posterior (GL-P), group SCAD projection-posterior (GS-P), and adaptive group LASSO projection-posterior (AGL-P), include group LASSO (GL), group SCAD (GS), group adaptive LASSO (AGL), debiased group LASSO, Bayesian group LASSO (BGL), and the spike-and-slab (SS). For Bayesian sparse projection-posterior methods, 5000 initial posterior samples were drawn, projected using the three proposed methods, and the median probability model (MPM) was selected as the chosen model. The Bayes estimator was selected as the posterior mean corresponding to the MPM, and the rest were set to 0. When computing the induced posterior samples using the three proposed maps, we use the penalty parameter obtained by cross-validation when regressing $\bY$ on $\bX$ under each regularization approach. For other Bayesian methods used for the purpose of comparison, we have drawn 10000 initial samples and allowed a burn in of 5000. For each replication, we recorded the Mean Squared Error (MSE) for all methods and presented the results as box plots in Figure \ref{fig:MSE_s10}, corresponding to $s = 10$. The results for $s = 20$ are similar and are displayed in \Cref{fig:MSE_s20} in the supplement. Additionally, we recorded the True Positives (TP) for each method. We measured the F1-score defined by $F_1 = ({2\times\textnormal{Precision}\times\textnormal{Recall}})/({\textnormal{Precision}+\textnormal{Recall}})$, where $\textnormal{Precision} = \textnormal{TP}/\hat{s}$ and $\textnormal{Recall} = \textnormal{TP}/s_0$ to assess the recovery of the true sparsity pattern. Here, $\hat{s}$ is the symbol used for the selected set of groups of each method. A value of $F_1$ = 1 indicates perfect precision and recall, meaning all detected signals were correct, and all actual signals were identified, whereas 0 indicates the worst performance, meaning the model did not correctly identify any signal instances. The F1-scores are visualized using forestplots (Figures \ref{fig:F1score_s10} and \ref{fig:F1score_s20} for $s = 10$ and 20, respectively) to demonstrate each method's ability to identify active groups across different scenarios accurately. This comprehensive setup allows us to evaluate the method's consistency and robustness under varying group structures and sparsity levels.


We observe from the boxplots presented in Figures \ref{fig:MSE_s10} and \ref{fig:MSE_s20} that the MSEs corresponding to all the methods reduce when the sample size increases from $n = 100$ (upper panels) to $n = 500$ (lower panels). We also notice a pattern among the penalization methods and their Bayesian projection counterparts. The MSEs for the maps using any of the three penalty functions are similar to those of the corresponding penalization methods in some cases. However, as $n$ grows, the mean squared error of BGL is maximum. GL, AGL, GL-P, and AGL-P have more or less similar characteristics. Both GS and GS-P perform better in terms of estimation error than the other two penalties and maps. The estimation accuracies of GS and GS-P are similar to those of the spike-and-slab method. Although the MSE of SS is lower, its running time is considerably higher. The run times of all these methods, averaged over 100 replications, are provided in \Cref{tab:my-timetable}.

\begin{table}[htbp]
\centering
\resizebox{0.6\linewidth}{!}{
\begin{tabular}{ccccccccc}
\hline
Setting & GL & GS & AGL & BGL & SS & GL-P & GS-P & AGL-P \\ \hline
\multirow{2}{*}{$K = 100$} & 2.7 s & 2.5 s & 3.1 s & 1.4 m & 5.8 m & 2.3 m & 3.4 m & 2.9 m \\ 
                           & (0.5 s) & (0.4 s) & (0.8 s) & (0.5 m) & (2.7 m) & (0.7 m) & (1.1 m) & (0.9 m) \\ \hline
\multirow{2}{*}{$K = 500$} & 4.3 s & 5.2 s & 4.2 s & 4.1 m & 18.8 m & 6.8 m & 8.1 m & 7.2 m \\ 
                           & (1.5 s) & (1.9 s) & (2.0 s) & (2.8 m) & (5.6 m) & (2.3 m) & (3.0 m) & (2.2 m) \\ \hline
\end{tabular}
}
\caption{Run-times averaged over 100 replicates with standard error for \( n = 100 \) and \( s = 10 \). Here s stands for seconds and m for minutes.}
\label{tab:my-timetable}
\end{table}

In Figures \ref{fig:F1score_s10} and \ref{fig:F1score_s20}, we plotted the average F1 scores for each of the 8 competing methods with their corresponding standard error bars for the for $s_0 = 10$ and $s_0 = 20$ respectively. Each of the plots contains four forestplots pertaining to the four settings, $(K,n)$ \( \in \{(50,100),(50,500),(100,100),(100,500)\}\). The F1 score, a harmonic mean of precision and recall, reflects the balance between identifying relevant groups and avoiding irrelevant ones. We observe that the F1 scores stabilize around 1 for most methods when $n$ increases from 100 to 500 for both $s_0 = 10$ and $s_0 = 20$. We also see that GS and GS-P exhibit model selection properties similar to those of the spike-and-slab method. Although the performances of GL-P and AGL-P are similar to BGL when $n = 100$, their selection accuracy exceeds that of BGL when $n$ increases to 500. The results corresponding to $s = 20$ are displayed in the supplement.

Next, we provide the average coverage of the Bayesian methods along with Debiased Group LASSO and Debiased Group LASSO Projection in Tables \ref{tab:my-table_cov_s10} and \ref{tab:my-table_cov_s20} respectively for the $s_0 = 10$ and $s_0 = 20$ setups. We used 95\% quantile-based credible regions for each Bayesian method. We averaged across all signal and noise variables to get an idea of the coverage and length patterns for each separately. Among the three sparse projections we implemented, GS-P yields accurate coverage for large samples ($n = 500$). GL-P and AGL-P fall slightly short of exact coverage. However, the debiased group LASSO map, denoted by the Debiased Projection method, achieves exact 95\% coverage with $n = 500$. The same is observed for the (non-Bayesian) debiased group LASSO confidence interval. Since the discrepancy between the results for $s_0 = 10$ and $s_0 = 20$ is minimal, we have moved \Cref{tab:my-table_cov_s20} to the supplement.

\begin{table}[htbp]
\centering
\small
\resizebox{0.75\textwidth}{!}{%
\begin{tabular}{c|cccccccc}
\toprule
Setting & &  Debiased GL & BGL & SS & GL-P & GS-P & AGL-P & Debiased Projection \\
\midrule
\multirow{4}{*}{$K=50,\,n=100$}
 & Signal Coverage & 0.91(0.11) & 0.53(0.03) & 0.81(0.02) & 0.72(0.16) & 0.89(0.13) & 0.79(0.17) & 0.91(0.05) \\
 & Signal Length   & 5.55(0.70) & 1.41(0.21) & 5.48(0.11) & 5.33(0.12) & 5.19(0.98) & 5.19(0.89) & 5.62(0.34) \\
 & Noise Coverage  & 1.00        & 1.00        & 1.00        & 0.98(0.64)  & 0.95(0.12)  & 1.00        & 1.00        \\
 & Noise Length    & 2.18(0.12) & 1.33(0.19) & 3.05(0.07) & 0.44(0.03) & 0.41(0.04) & 0.33(0.04) & 2.21(0.21) \\
\midrule
\multirow{4}{*}{$K=50,\,n=500$}
 & Signal Coverage & 0.95(0.02) & 0.87 & 0.91(0.03) & 0.93(0.09) & 0.97(0.10) & 0.93(0.09) & 0.96(0.01) \\
 & Signal Length   & 1.72(0.01) & 0.60(0.01) & 0.60(0.01) & 1.72(0.01) & 0.82(0.10) & 1.43(0.01) & 1.60(0.08) \\
 & Noise Coverage  & 1.00       & 1.00       & 1.00(0.00) & 1.00       & 1.00       & 1.00       & 1.00 \\
 & Noise Length    & 1.62(0.01) & 0.49(0.00) & 0.11(0.00) & 1.70(0.01) & 0.71(0.33) & 1.38(0.01) & 1.57(0.01) \\
\midrule
\multirow{4}{*}{$K=100,\,n=100$}
 & Signal Coverage & 0.91(0.08) & 0.40(0.01) & 0.69(0.01) & 0.65(0.01) & 0.73(0.01) & 0.72(0.06) & 0.91(0.08) \\
 & Signal Length   & 1.31(0.02) & 0.61(0.01) & 0.85(0.02) & 0.82(0.00) & 0.76(0.01) & 1.02(0.01) & 1.43(0.03) \\
 & Noise Coverage  & 1.00       & 1.00       & 1.00       & 0.96(0.01) & 1.00(0.00) & 0.96(0.01) & 1.00 \\
 & Noise Length    & 1.84(0.02) & 1.58(0.09) & 2.34(0.07) & 0.31(0.00) & 0.35(0.00) & 0.46(0.00) & 1.77(0.01) \\
\midrule
\multirow{4}{*}{$K=100,\,n=500$}
 & Signal Coverage & 0.95(0.00) & 0.80(0.01) & 0.88(0.02) & 0.93(0.00) & 0.95(0.00) & 0.94(0.00) & 0.95(0.01) \\
 & Signal Length   & 1.65(0.01) & 0.79(0.02) & 0.88(0.03) & 1.30(0.04) & 1.26(0.03) & 1.62(0.03) & 1.56(0.02) \\
 & Noise Coverage  & 1.00       & 1.00       & 1.00       & 1.00 & 1.00       & 1.00       & 1.00 \\
 & Noise Length    & 1.95(0.02) & 0.88(0.08) & 2.81(0.05) & 0.59(0.01) & 0.53(0.00) & 0.55(0.00) & 1.97(0.02) \\
\bottomrule
\end{tabular}}
\caption{Average signal and noise coverage and interval length (standard errors in parentheses) over 100 replicates with $s_0 = 10$.}
\label{tab:my-table_cov_s10}
\end{table}

\subsection{Simulation Study 2: Additive Regression}
In this simulation study, we consider an additive nonparametric regression model, as described in \eqref{additive model}. We simulate data with $n = 100$ observations and $k = 50$ predictor variables, where only $s_0 = 5$ variables are relevant to the response. These five important variables follow nonlinear relationships with the response. That is, $f_1^0(X_1) = $ \( 3\sin(X_1) \) is a sinusoidal term, $f_2^0(X_2) = $ \( 2X_2^2 \) is a quadratic term, $f_3^0(X_3) = $ \( -1.5X_3 \) is a linear term with a negative effect, $f_4^0(X_4) = $ \( \exp(X_4) \) is an exponential term, and finally $f_5^0(X_5) = $ \( \log(|X_5| + 1) \) is a logarithmic transformation, with Gaussian noise added to the response. To model the nonlinear effects across all predictors, we use a cubic B-spline basis expansion with $B_n = 8$ for each variable, yielding a transformed feature matrix in which a set of basis functions represents each variable. This new matrix, now of order $100 \times 400$, contains the expanded set of predictors, with the dimensions reflecting the number of observations and the total number of transformed features. Hence, we have $p = K \times B_n = 400$ in the model. Clearly, $\bm\beta_k \in \R^8$ for each $k = 1,2,\dots,50$. The setup allows for applying group regularization techniques to select the true active groups of variables while shrinking the coefficients of irrelevant variables to zero. Here, each of the 50 variables can be treated as a group, with the B-spline expansions of the variables forming a set of related features. Under sparsity, we aim to identify the true active variables among the 50 and recover the few truly important nonlinear relationships that affect the response, while keeping the number of selected groups minimal. Although the additive model's estimation and selection consistency results have been proven only for the group LASSO projection-posterior, we demonstrate the performance of all three proposed projection maps and five other competing methods. 

\begin{figure}[htbp]
    \centering
    \includegraphics[width=0.45\linewidth, height = 5cm]{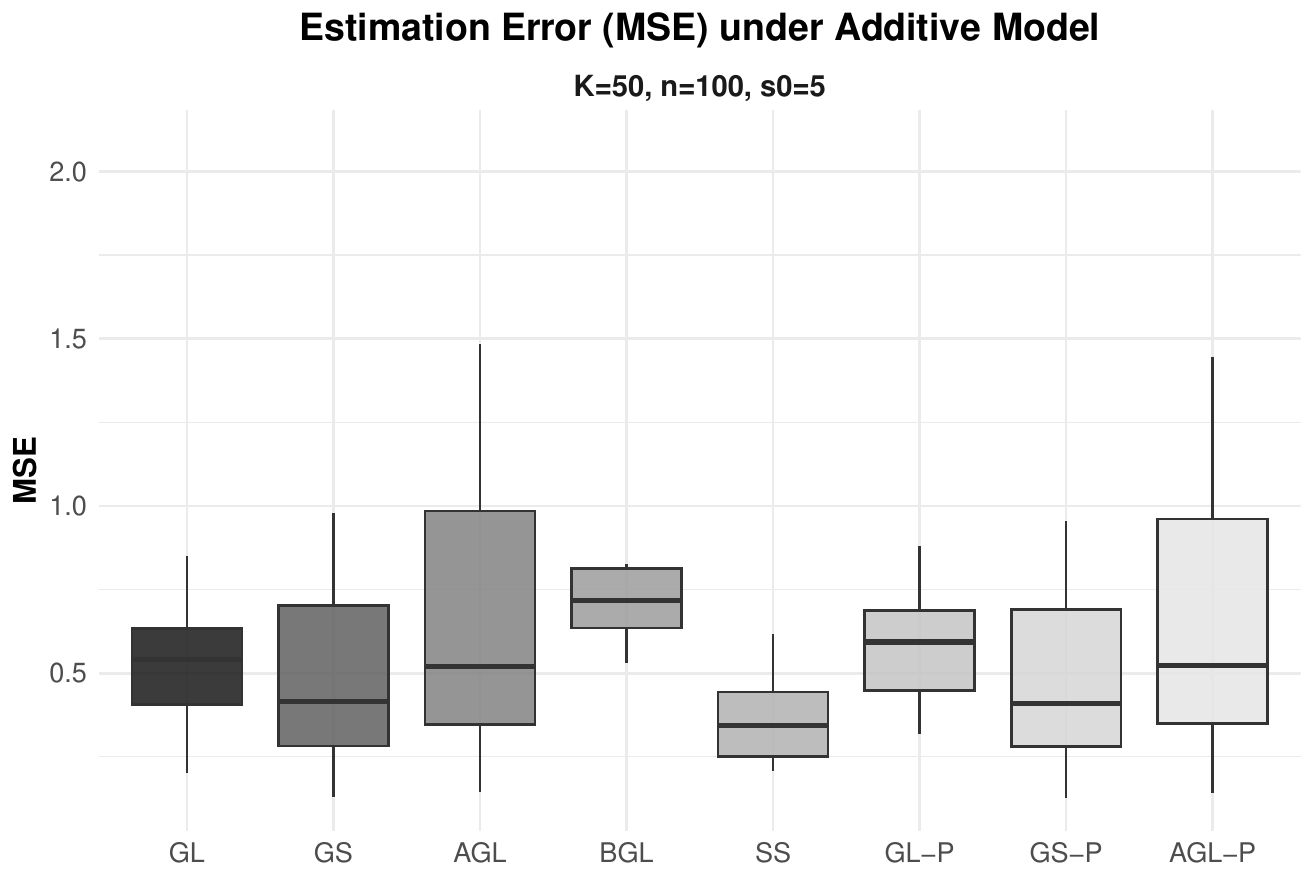}
    \includegraphics[width = 0.45\linewidth]{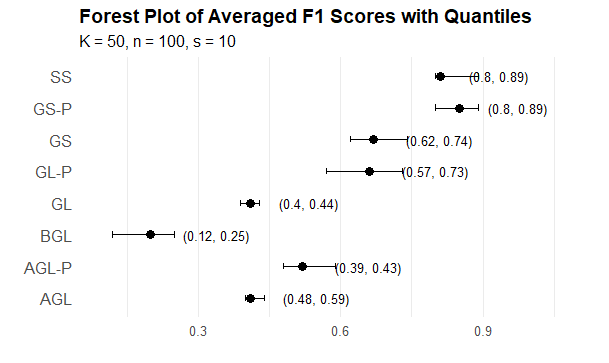}
    \caption{Boxplot of MSEs and Forestplot of F1 scores of all competing methods under the additive model with $K = 50, B_n = 8, n = 100$ and $s_0 = 5$.}
    \label{fig:additive_MSE_F1}
\end{figure}

In \Cref{fig:additive_MSE_F1}, the boxplots of the prediction MSEs of the different methods are presented in the left panel. We see that GS-P has the minimum MSE. The performances of other projection maps are again similar to those of their corresponding penalized regression techniques. On the right panel of \Cref{fig:additive_MSE_F1}, we show the forestplot of the F1 scores for the eight competing methods. Again, we observe that the F1 score of the GS-P is as good as that of spikes and slabs. Thus, the projection methods demonstrate overall successful effectuation of the asymptotic results proved in \Cref{sec:additive_model} on both estimation and selection fronts. 

\subsection{Real Data Analysis}
MRI-based volumetric measurements from Alzheimer's Disease Neuroimaging Initiative (ADNI), specifically those assessing regional brain volumes, are widely regarded as biomarkers for neurodegeneration and cognitive decline. In Alzheimer's disease (AD) research, structural changes in specific brain regions, such as the hippocampus and entorhinal cortex, are known to correlate with the progression from cognitively normal (CN) status to mild cognitive impairment (MCI) and eventually AD. By examining brain volume data, researchers can quantify these structural changes and better understand the associations between brain atrophy patterns and clinical outcomes. Using an additive model in this setting is highly appropriate, as the relationship between brain volumes and AD severity will likely be nonlinear. This approach allows for flexible modeling of complex relationships between volumetric features and cognitive outcomes, accommodating potential nonlinear effects of brain regions on disease risk. We then use the B-spline basis expansion and the group selection procedures proposed in this paper, along with existing methods, to identify the groups of features (e.g., brain regions) that are collectively relevant for predicting AD severity. In this study, we have used the Mini-Mental State Examination (MMSE) score as the indicator of Alzheimer's disease and, consequently, as the response variable of interest. This analytical framework aligns well with clinical research goals in AD, aiming to identify which brain regions and patterns of atrophy are most predictive of disease severity. By leveraging the power of grouped regularization, this study can provide insights into the neuroanatomical correlates of AD, potentially aiding in early diagnosis and targeted interventions. In this study, a total of \(K = 70\) brain regions were considered as predictors for $n = 458$ subjects, each predictor corresponding to a distinct anatomical region of interest (ROI) such as hippocampal and parahippocampal subfields, entorhinal and parahippocampal cortices, and neighboring temporal regions. Each region was modeled nonparametrically using a cubic B-spline basis with \(B_n = 8\) basis functions, resulting in a total of \(p = K \times B_n = 560\) predictors.
Thus, each brain region formed a {group} of eight correlated spline-expanded variables, collectively representing the smooth volumetric effect of that region on the response.
We note that the real data analysis uses only baseline neuroimaging-derived phenotypes (IDPs) from the ADNI database, where volumetric measures of cortical and subcortical regions serve as predictors and the MMSE score as the response. Only baseline data were analyzed to focus on cross-sectional associations between brain structure and cognitive function. Age, sex, education, and intracranial volume (ICV) were included as covariates to adjust for subject-specific confounding effects.

\begin{table}[htbp]
\centering
\resizebox{0.8\linewidth}{!}{%
\begin{tabular}{ccccccccc}
\hline
Methods & GL & GS & AGL & GL-P & GS-P & AGL-P & BGL & SS \\ \hline
\begin{tabular}[c]{@{}c@{}}Test Prediction\\ Error\end{tabular} & \begin{tabular}[c]{@{}c@{}}1.80\\ (0.36)\end{tabular} & \begin{tabular}[c]{@{}c@{}}1.14\\ (0.44)\end{tabular} & \begin{tabular}[c]{@{}c@{}}2.23\\ (0.03)\end{tabular} & \begin{tabular}[c]{@{}c@{}}1.33\\ (0.56)\end{tabular} & \begin{tabular}[c]{@{}c@{}}1.25\\ (0.47)\end{tabular} & \begin{tabular}[c]{@{}c@{}} 2.45 \\ (0.16)\end{tabular} & \begin{tabular}[c]{@{}c@{}}9.70 \\ (2.22)\end{tabular} & \begin{tabular}[c]{@{}c@{}}1.27 \\ (0.03)\end{tabular} \\ \hline
\end{tabular}%
}
\caption{Average Test Prediction Error for the different methods applied to the ADNI MRI volume data.}
\label{tab:my-table_real_MSE}
\end{table}

\Cref{tab:my-table_real_MSE} presents the average test prediction errors (with standard deviations) for different regularization and Bayesian approaches in a high-dimensional regression scenario. The dataset was split into training (70\%) and test (30\%) sets, and the reported values were averaged across 10 random splits. GS-P and GS methods achieved the lowest test errors, indicating their effectiveness in reducing prediction error under group-sparsity constraints. AGL-P and GL-P performed moderately well, balancing prediction error with model complexity. BGL had the highest test prediction error, suggesting it may not be as effective at achieving high prediction accuracy in this high-dimensional setup. Next, we discuss the brain regions (treated here as groups) selected by different methods as highly associated with the response variable (MMSE score).

The study examined a set of hippocampal and parahippocampal brain regions to identify those most associated with Alzheimer's disease severity. Specifically, the following regions were considered: Cornu Ammonis 1 (CA1), CA2, CA3, Dentate Gyrus (DG), Miscellaneous Regions (MISC), Subiculum (SUB), Entorhinal Cortex (ERC), Brodmann Area 35 (BA35), BA36, Parahippocampal Cortex (PHC), Sulcus, and a combined CA (CA) region. These regions, bilaterally present in the brain, play critical roles in memory formation, spatial navigation, and early neurodegenerative changes, particularly CA1 and ERC, which are heavily implicated in Alzheimer's pathology \citep{masurkar2018towards,rao2022hippocampus,igarashi2023entorhinal}. For simplicity, each region was abbreviated, such as Left CA1 for the left hemisphere CA1 region and similarly for other regions.

\begin{figure}
    \centering
    \includegraphics[width=\linewidth]{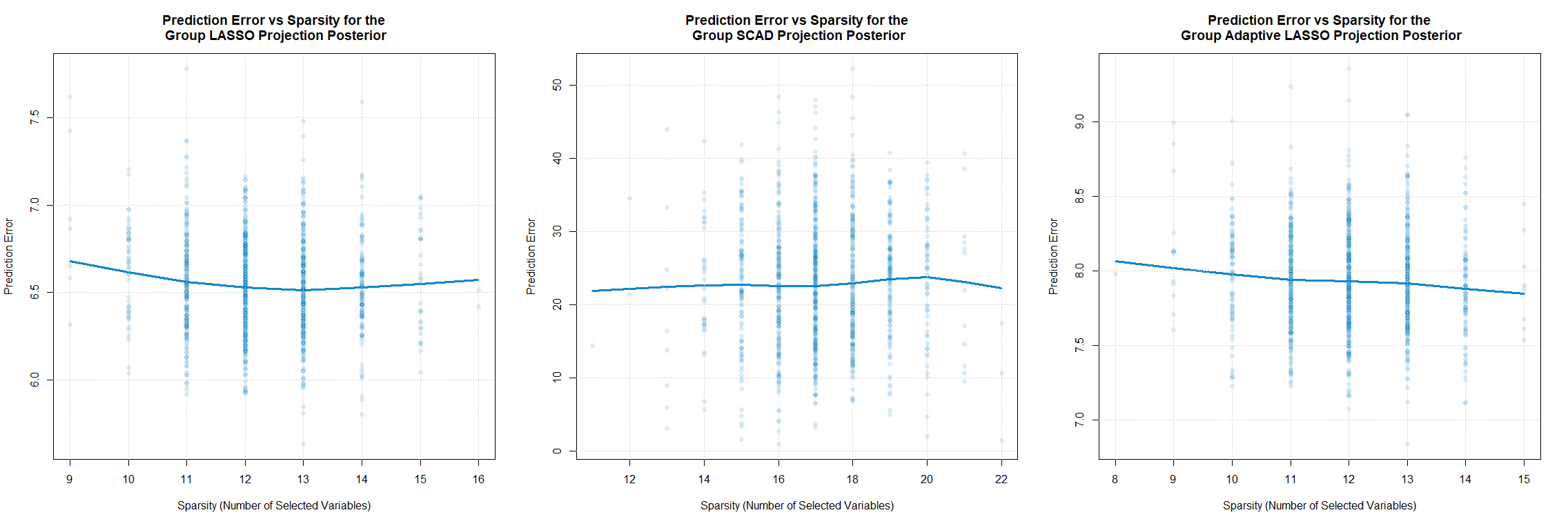}
    \caption{Plot of the distributions of the average prediction error versus the model sparsity for the three Bayesian projection-posterior methods.}
    \label{fig:pred_err_vs_sparsity}
\end{figure}

Different sparse regression methods selected varying subsets of these regions, illustrating distinct selection patterns. Group LASSO selected Left CA2, Left CA3, Left MISC, Left PHC, Left Sulcus, Right SUB, Right ERC, Right BA35, Right BA36, Right PHC, Right Sulcus, and Right CA, indicating these areas as particularly relevant to Alzheimer’s severity. AGL largely aligned with GL but included Left BA36, highlighting its potential significance. GS, on the other hand, captured a distinct subset, including Left CA2, Left CA3, Left BA35, Left PHC, Left Sulcus, Right CA1, Right SUB, Right BA35, Right PHC, and Right CA, reflecting some divergence in selection preferences for certain regions on the right hemisphere. Bayesian-inspired methods showed similar patterns. GL-P and AGL-P selected the same regions as GL, demonstrating consistency in variable selection across posterior models. However, GS-P identified Left CA2, Left CA3, Left MISC, Left BA36, Left PHC, Left Sulcus, Right SUB, Right ERC, Right PHC, and Right Sulcus, diverging slightly from GS by emphasizing a few different regions. Both SS and BGL selected the complete set of regions, suggesting a tendency towards less selectivity in high-dimensional settings. In summary, CA2, CA3, SUB, ERC, BA35, BA36, and PHC consistently emerged across models, underscoring their potential role in Alzheimer’s-related neurodegeneration. This analysis reveals that a sparse projection-posterior performs effective variable selection, comparable to traditional penalization methods, and also provides quantifiable uncertainty about the selected variables. The fact that we get the entire posterior distribution for the three sparse projection-posteriors is depicted in \Cref{fig:pred_err_vs_sparsity}. This plot shows that the in-sample prediction errors remain essentially constant for different sparse projection-posterior methods. Moreover, the density of the lines for each selected model reflects the number of models that show up with the given sparsity, thereby giving us an idea about which models appeared more frequently in the posterior.

\section{Discussion}
\label{Conclusion}

The sparse projection-posterior method is a relatively new Bayesian inference procedure with immense potential for high-dimensional models under sparsity. In this paper, we discussed several sparse projection-posterior methods for high-dimensional grouped regression. We demonstrate how different penalty functions can be used to map non-sparse posterior samples to group-sparse samples from the induced posterior. We also showed that the theoretical properties of the induced posterior depend on the chosen map. Thus, the sparse-projection-posterior method can reproduce the results of the corresponding regularization technique, with the added advantage of automatically quantifying uncertainty. The coverage properties of this approach have also been discussed. The extensive simulation studies suggest no loss in estimation and variable selection when the Bayesian regularized maps are used instead of their classical counterparts. Moreover, the proposed methods allow automatic quantification of the associated uncertainty. Compared to commonly used Bayesian methods in the grouped regression setup, we showed that all three proposed projection maps outperform the Bayesian group LASSO in terms of estimation and selection accuracy, as well as coverage properties. Although almost on par with the proposed methods, the spike-and-slab Bayesian method is computationally costly in high dimensions. In conclusion, our findings highlight the sparse projection-posterior method as a powerful alternative to standard regularization techniques and traditional Bayesian approaches. This work paves the way for a broader exploration of Bayesian approaches to sparse high-dimensional models, laying the foundation for theoretical advances and practical applications.

\section{Acknowledgement}
Data collection and sharing for the Alzheimer's Disease Neuroimaging Initiative (ADNI) is funded by the National
Institute on Aging (National Institutes of Health Grant U19 AG024904). The grantee organization is the Northern
California Institute for Research and Education.

\section{Funding}
This research is partially supported by ARO grant number 76643MA 2020-0945.

\bibliography{ref}

\newpage
\setcounter{page}{1}

\section*{Supplementary Materials}

\subsection*{Supplement A: Additional Numerical Results}

\subsubsection*{Distributed Computing}
\begin{figure}[htbp]
    \centering
    \includegraphics[width=0.7\linewidth]{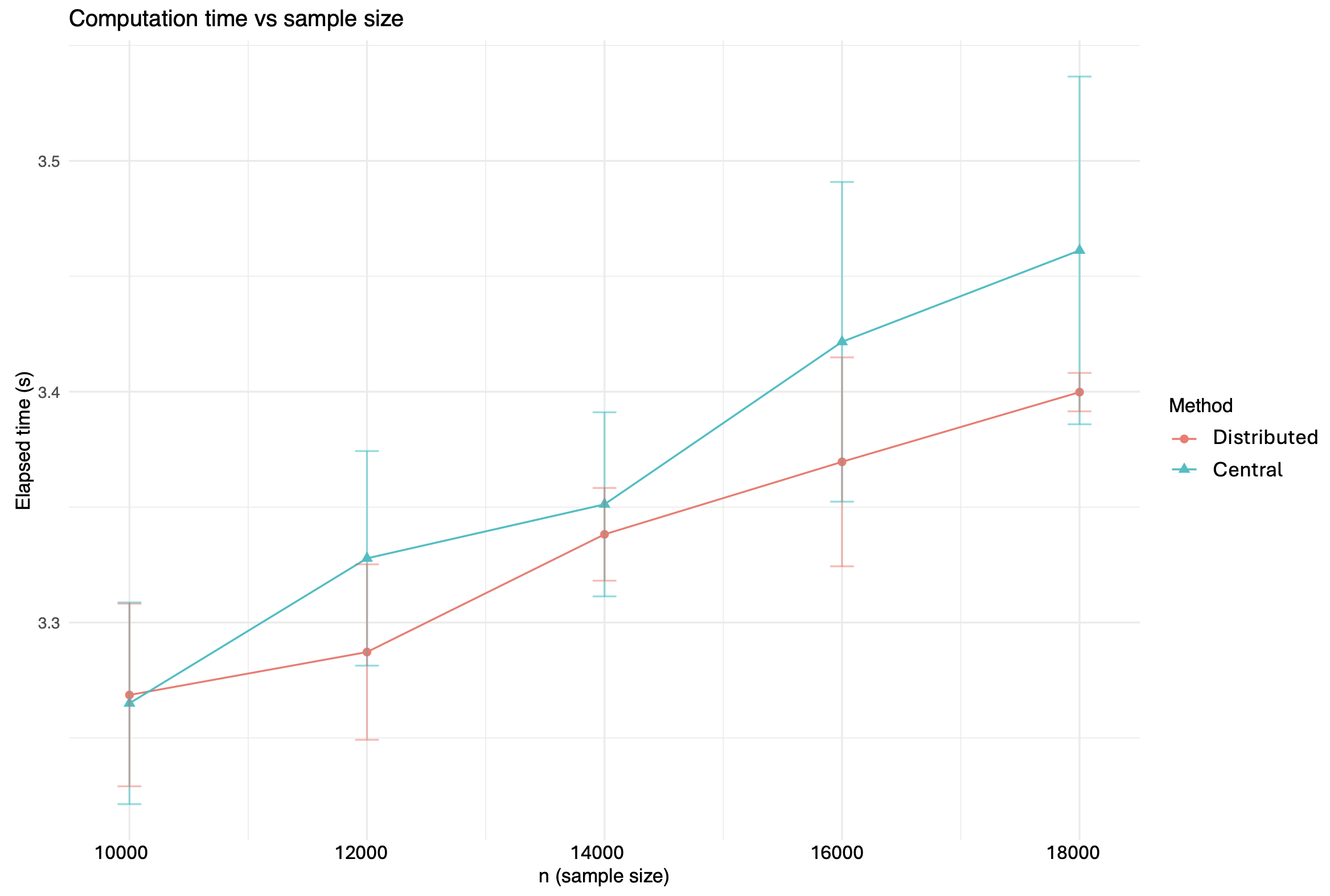}
    \caption{Computation time comparison between centralized and distributed implementations of the group LASSO projection-posterior across increasing sample sizes.}
    \label{fig:distr_runtime_gain}
\end{figure}

We compared the runtime of the proposed group LASSO projection-posterior under centralized and distributed implementations across increasing sample sizes (\(n = 10000, \ldots, 20000\)) with the dimensionality fixed at $p = 100$ and $s_0 = 5$ active groups, each of size 5. 
In the distributed version, the full dataset was partitioned into 10 shards, each processed on a separate core to compute local summary statistics 
(\(\bX^\mathrm{T}\bX,\, \bX^\mathrm{T}\bY,\, \bY^\mathrm{T}\bY\)), which were subsequently aggregated. 
As shown in Figure~\ref{fig:distr_runtime_gain}, this parallelization yielded a modest reduction in total runtime for larger \(n\), 
primarily due to the concurrent computation of local Gram matrices. 
However, the subsequent posterior sampling and projection steps, which dominate the overall computational load, incur similar costs under both centralized and distributed settings. 
Thus, while substantial speedups are not expected for the current problem size, the distributed framework offers two key advantages: firstly, it eliminates the need to access or share the complete raw data, making it naturally suited to privacy-preserving or data-federated settings; and 
secondly, it ensures that the computational time is never worse than that of the centralized alternative, with potential gains when the local computations can be efficiently parallelized. We observe similar trends for the group SCAD and adaptive group LASSO projection-posteriors; the plots are not reported to avoid repetition.

For each $n$ used in our timing study, we decomposed wall-clock time into:
(i) the {summary stage} (computing local/central $\bX^\top \bX$, $\bX^\top \bY$, $\bY^\top \bY$ and aggregating in the distributed case), and
(ii) the {sampling+projection stage} (conjugate posterior sampling of $\bm\beta$ and sparse projection for $R = 1000$ draws). It should be noted that the summary step under both distributed and central computing requires an inverse of a high-dimensional matrix. In the distributed implementation with $m$ shards, only stage~(i) differs from the centralized case; stage~(ii) should be similar as it mostly depends on the dimension rather than the sample size.

\begin{table}[h!]
\centering
\resizebox{\textwidth}{!}{%
\begin{tabular}{lcccccc}
\toprule
$n$ & Sampling{+}Projection (Central) & Sampling{+}Projection (Dist) & Summary (Central) & Summary (Dist) & Total (Central) & Total (Dist) \\
\midrule
$10{,}000$ & 2.70 & 2.70 & 0.56 & 0.57 & 3.26 & 3.27 \\
$14{,}000$ & 2.71 & 2.73 & 0.64 & 0.61 & 3.35 & 3.34 \\
$18{,}000$ & 2.74 & 2.74 & 0.72 & 0.66 & 3.46 & 3.40 \\
\bottomrule
\end{tabular}}
\caption{Runtime (seconds): decomposition of shared sampling\,+\,projection and summary stages under centralized vs.\ distributed implementations.}
\label{tab:runtime_split_full_corrected}
\end{table}

In our runs, the summary stage scales near-linearly with $n$ and benefits from distributing the load, whereas the sampling+projection stage is essentially {flat in $n$} (dominated by a single Cholesky of $\bX^\top \bX + a\bm I_p$ and $R$ projections), hence similar under centralized and distributed setups. This explains the modest but consistent gains in Figure~\ref{fig:distr_runtime_gain}.

Let $p$ and $R$ be fixed. We write
\(
T_{\text{central}}(n)\;=\; c_{\text{sum}}\,n \;+\; T_{\text{s+p}},\enskip 
T_{\text{dist}}(n;m)\;=\; \tfrac{c_{\text{sum}}}{m}\,n \;+\; T_{\text{s+p}} \;+\; T_{\text{over}},
\)
where $c_{\text{sum}}$ is the per-sample cost of forming summaries, $T_{\text{s+p}}$ is the (nearly $n$-independent) cost of sampling+projection, and $T_{\text{over}}$ is the communication and coordination overhead cost. $T_{\text{over}}$ is typically small and weakly dependent on $n$, mostly depending on the number of communication steps (one step for us) and the number of pieces to be coordinated ($m = 10$ for us). Using the empirical runtimes in Table \ref{tab:runtime_split_full_corrected}, we estimate \(T_{\text{s+p}}\approx2.7\mathrm{s}\) (essentially constant across \(n\)), and \(c_{\text{sum}}\approx4\times10^{-5}\mathrm{s}\) per sample for the centralized case.  With \(m=10\) shards and negligible \(T_{\text{over}}\), this implies a near-linear growth in total runtime for both implementations, but with a smaller slope for the distributed version.  Based on these fits, the distributed approach begins to yield visible savings beyond \(n\approx12{,}000\) (about 1-2 \%), reaching roughly 8 \% at \(n=50{,}000\), 14 \% at \(n=100{,}000\), and 20 \% at \(n=200{,}000\).  Hence, while the difference is modest for the sample sizes used in our experiments, the relative gain increases linearly with \(n\), and the distributed framework becomes substantially faster for larger datasets, while never exceeding the cost of the centralized computation.


\subsubsection*{Simulation Results for $s_0 = 20$}
Below, we provide the plots for $s_0 = 20$, keeping all other parameters required for data generation fixed as in \Cref{simu1}. We see exact similar patterns as the $s_0 = 10$ case (reported in the main paper), implying that the models are robust to varying sparsity.


\begin{table}[htbp]
\centering
\resizebox{0.9\linewidth}{!}{%
\begin{tabular}{cc|ccccccc}
\hline
\multicolumn{2}{c|}{Setting} & \begin{tabular}[c]{@{}c@{}}Debiased\\ Group LASSO\end{tabular} & \begin{tabular}[c]{@{}c@{}}Bayesian \\ Group LASSO\end{tabular} & Spike and Slab & \begin{tabular}[c]{@{}c@{}}Group ALASSO\\ Projection\end{tabular} & \begin{tabular}[c]{@{}c@{}}Group LASSO\\ Projection\end{tabular} & \begin{tabular}[c]{@{}c@{}}Group SCAD\\ Projection\end{tabular} & \begin{tabular}[c]{@{}c@{}}Debiased\\ Projection\end{tabular} \\ \hline

\multicolumn{1}{c|}{\multirow{4}{*}{\begin{tabular}[c]{@{}c@{}}$K = 50$, \\ $n = 100$\end{tabular}}} 
& \begin{tabular}[c]{@{}c@{}}Signal \\ Coverage\end{tabular} & 0.88 (0.013) & 0.372 (0.059) & 0.823 (0.01) & 0.634 (0.027) & 0.684 (0.007) & 0.69 (0.005) & 0.876 (0.021) \\
\multicolumn{1}{c|}{} & \begin{tabular}[c]{@{}c@{}}Signal\\ Length\end{tabular} & 1.826 (0.122) & 1.9 (0.295) & 2.634 (0.104) & 1.03 (0.009) & 0.95 (0.006) & 1.47 (0.24) & 1.76 (0.117) \\ \cline{2-9} 
\multicolumn{1}{c|}{} & \begin{tabular}[c]{@{}c@{}}Noise\\ Coverage\end{tabular} & 1 & 1 & 1 & 0.937 (0.016) & 0.949 (0.016) & 0.985 (0.013) & 1 \\
\multicolumn{1}{c|}{} & \begin{tabular}[c]{@{}c@{}}Noise\\ Length\end{tabular} & 1.534 (0.031) & 1.84 (0.028) & 3.91 (0.054) & 0.522 (0.007) & 0.585 (0.004) & 0.704 (0.007) & 1.62 (0.024) \\ \hline

\multicolumn{1}{c|}{\multirow{4}{*}{\begin{tabular}[c]{@{}c@{}}$K = 50$,\\ $n = 500$\end{tabular}}} 
& \begin{tabular}[c]{@{}c@{}}Signal\\ Coverage\end{tabular} & 0.952 (0.003) & 0.717 (0.017) & 0.959 (0.022) & 0.962 (0.003) & 0.963 (0.004) & 0.959 (0.001) & 0.949 (0.004) \\
\multicolumn{1}{c|}{} & \begin{tabular}[c]{@{}c@{}}Signal\\ Length\end{tabular} & 1.833 (0.009) & 1.16 (0.012) & 1.251 (0.012) & 2.03 (0.005) & 2.38 (0.007) & 1.823 (0.019) & 1.884 (0.027) \\ \cline{2-9} 
\multicolumn{1}{c|}{} & \begin{tabular}[c]{@{}c@{}}Noise\\ Coverage\end{tabular} & 1 & 0.969 (0.006) & 1 & 1 & 1 & 1 & 1 \\
\multicolumn{1}{c|}{} & \begin{tabular}[c]{@{}c@{}}Noise\\ Length\end{tabular} & 1.198 (0.03) & 0.97 (0.013) & 0.721 (0.006) & 2.01 (0.006) & 1.78 (0.007) & 1.1 (0.029) & 1.234 (0.015) \\ \hline

\multicolumn{1}{c|}{\multirow{4}{*}{\begin{tabular}[c]{@{}c@{}}$K = 100$,\\ $n = 100$\end{tabular}}} 
& \begin{tabular}[c]{@{}c@{}}Signal\\ Coverage\end{tabular} & 0.786 (0.034) & 0.178 (0.191) & 0.68 (0.059) & 0.579 (0.067) & 0.601 (0.054) & 0.632 (0.032) & 0.776 (0.03) \\
\multicolumn{1}{c|}{} & \begin{tabular}[c]{@{}c@{}}Signal\\ Length\end{tabular} & 1.207 (0.018) & 0.457 (0.002) & 0.81 (0.003) & 0.693 (0.005) & 0.782 (0.002) & 0.775 (0.005) & 1.218 (0.01) \\ \cline{2-9} 
\multicolumn{1}{c|}{} & \begin{tabular}[c]{@{}c@{}}Noise\\ Coverage\end{tabular} & 1 & 0.986 (0.002) & 1 & 1 & 1 & 1 & 0.999 (0.0001) \\
\multicolumn{1}{c|}{} & \begin{tabular}[c]{@{}c@{}}Noise\\ Length\end{tabular} & 0.987 (0.01) & 1.783 (0.012) & 1.254 (0.008) & 0.59 (0.011) & 0.578 (0.012) & 0.558 (0.013) & 1.069 (0.02) \\ \hline

\multicolumn{1}{c|}{\multirow{4}{*}{\begin{tabular}[c]{@{}c@{}}$K = 100$,\\ $n = 500$\end{tabular}}} 
& \begin{tabular}[c]{@{}c@{}}Signal\\ Coverage\end{tabular} & 0.952 (0.001) & 0.726 (0.103) & 0.923 (0.006) & 0.944 (0.003) & 0.937 (0.007) & 0.951 (0.002) & 0.953 (0.001) \\
\multicolumn{1}{c|}{} & \begin{tabular}[c]{@{}c@{}}Signal\\ Length\end{tabular} & 1.357 (0.02) & 0.659 (0.004) & 0.957 (0.023) & 1.297 (0.02) & 1.324 (0.021) & 1.109 (0.016) & 1.364 (0.015) \\ \cline{2-9} 
\multicolumn{1}{c|}{} & \begin{tabular}[c]{@{}c@{}}Noise\\ Coverage\end{tabular} & 1 & 1 & 1 & 1 & 1 & 1 & 1 \\
\multicolumn{1}{c|}{} & \begin{tabular}[c]{@{}c@{}}Noise\\ Length\end{tabular} & 0.991 (0.02) & 1.645 (0.023) & 0.791 (0.015) & 0.61 (0.008) & 0.604 (0.006) & 0.598 (0.007) & 0.927 (0.024) \\ \hline
\end{tabular}%
}
\caption{Average signal and noise coverage and length along with the standard error, averaged over 100 replicates when $s_0 = 20$. }
\label{tab:my-table_cov_s20}
\end{table}

\newpage
\subsection*{Supplement B: Proof of the Theorems}

First, we recall all notations introduced throughout the paper that will be used in the proofs.
\begin{center}
\renewcommand{\arraystretch}{0.8}
\begin{longtable}{@{}llp{9cm}@{}}
\toprule
\multicolumn{2}{@{}l}{\textbf{Symbol}} & \textbf{Description} \\
\midrule
\endfirsthead
\toprule
\multicolumn{2}{@{}l}{\textbf{Symbol}} & \textbf{Description} \\
\midrule
\endhead
\midrule
\multicolumn{3}{r}{\emph{(continued on next page)}}\\
\midrule
\endfoot
\bottomrule
\endlastfoot

\multicolumn{3}{@{}l}{\textbf{Model, data and norms}}\\
$\bm\beta, \bm\beta^0 \in \mathbb{R}^p$ && Regression coefficients and true coefficient vector\\ $\|\cdot\|, \|\cdot\|_q, \|\cdot\|_\infty$ && Euclidean/$\ell_q$/sup norms \\
$\hat{\bm\Sigma}= n^{-1}\bX^\mathrm{T} \bX$ && Gram matrix \\ $\bC_{n(11)}$ && Block of Gram on $S_0$\\[4pt]

\multicolumn{3}{@{}l}{\textbf{Group structure}}\\
$K$ && Number of groups\\
$G_1,\ldots,G_K$ && Groups (disjoint) \\
$p_k=|G_k|$ && Size of the $k$th group \\
$p=\sum_{k=1}^K p_k$; $p_{\min},p_{\max}$ && Total/min/max group size \\
$\bX_{G_k}\in\mathbb{R}^{n\times p_k}$ && Sub-design for group $k$; $X=[X_{G_1},\ldots,X_{G_K}]$ \\
$\bm\beta_k\in\mathbb{R}^{p_k}$ && Group coefficient; $\bm\beta=(\bm\beta_1^\mathrm{T},\ldots,\bm\beta_K^\mathrm{T})^\mathrm{T}$ \\
$S_0\subset\{1,\ldots,K\}$ && Index set of signal groups; $s_0=|S_0|$\\ $p_0=\sum_{k\in S_0}p_k$ && Total number of active predictors\\[4pt]

\multicolumn{3}{@{}l}{\textbf{Unrestricted (conjugate) posterior}}\\
$a_n>0$ && Prior scale; $\bm\beta|\sigma \sim \mathcal{N}_p(0, \sigma^2 a_n^{-1} I_p)$ \\
$\hat{\bm\beta}^{\,\mathrm{R}}=(\bX^\mathrm{T} \bX + a_n \bm I_p)^{-1} \bX^\mathrm{T} \bY$ && Ridge-type posterior mean \\
$\bm H(a_n)=\bX(\bX^\mathrm{T} \bX + a_n \bm I_p)^{-1}\bX^\mathrm{T}$ && Hat matrix; $\bX \bm\beta\,|\,\sigma,\bY \sim \mathcal{N}_n(\bX\hat{\bm\beta}^{\,\mathrm{R}},\,\sigma^2 \bm H(a_n))$ \\
$\bm \eta=\bX \bm\beta - \bX \bm\beta^0$ && Posterior error in the mean space\\ && $\bm\eta\,|\,\sigma,\bY \sim \mathcal{N}_n(\bm \mu,\sigma^2 \bm H(a_n))$, $\bm\mu=\bX(\hat{\bm\beta}^{\,\mathrm{R}}-\bm\beta^0)$ \\[4pt]

\multicolumn{3}{@{}l}{\textbf{Projection map (sparse projection-posterior)}}\\
$\iota:\bm\beta \mapsto \bm\beta^\ast$ && $\displaystyle \bm\beta^\ast=\arg\min_{\bm u\in\mathbb{R}^p}\Big\{L(\bm\beta,\bm u)+\sum_{k=1}^K \mathcal{P}_{\lambda_n}(\|\bm u_k\|)\Big\}$ \\
$L(\bm\beta,\bm u)=n^{-1}\|\bX \bm\beta-\bX \bm u\|^2$ && Loss function used in all maps \\
$\lambda_n$ && Sparsity tuning parameter\\
$\tilde{\lambda}_n$ && Auxiliary tuning for initial GL fit (adaptive map) \\
$\hat{\bm\beta}^{\,\mathrm{GL}}$ && Initial group-LASSO estimator \\[4pt]

\multicolumn{3}{@{}l}{\textbf{Variance calibration (immersion posterior for }\,$\sigma^2$\,\textbf{)}}\\
$\upsilon=\sigma^{-2}$ && Precision; noninformative Gamma prior; map $\upsilon\mapsto \upsilon^\ast=\kappa\,\upsilon$ \\
$\hat{\sigma}_n^2=n^{-1}\bY^\mathrm{T} (\bm I_n-\bm H(a_n))\bY$ && Residual variance \\
$\kappa=\hat{\sigma}_n^2/\tilde{\sigma}^2$ for consistent $\tilde{\sigma}^2$ && yields $\upsilon^\ast|\bY\sim \mathrm{Gamma}\!\left(\frac{n}{2},\,\kappa^{-1}\frac{n\hat{\sigma}_n^2}{2}\right)$ \\[4pt]

\multicolumn{3}{@{}l}{\textbf{Debiased projection and precision matrix surrogate}}\\
$\hat{\bm\Theta}$ && Constructed to satisfy $\hat{\bm\Theta}\hat{\bm\Sigma}\approx I_p$  \\
$\bm\Gamma_g\in\mathbb{R}^{(p-p_g)\times p_g}$ && Group-wise GL coefficients regressing $\bX_{G_g}$ on $\bX_{-g}$ \\
$\bm R_g=\bX_{G_g}-\bX_{-g}\bm\Gamma_g$ && Residual matrix; $\bm T_g^2=n^{-1}\bX_{G_g}^\mathrm{T}\bm R_g$ (block scaling) \\
$\bm \Lambda_g=\mathrm{diag}(\lambda^1_g,\ldots,\lambda^{p_g}_g)$ && KKT multipliers\\ $\bm K_g$ && Subgradient matrix from group regressions \\
$\bm\beta^{\ast\ast}=\bm\beta^\ast+\frac{1}{n}\hat{\bm\Theta}^\mathrm{T} \bX^\mathrm{T}(\bX\bm\beta-\bX\bm\beta^\ast)$ && Debiased projection \\
$\bm\Delta=(\hat{\bm\Theta}\hat{\bm\Sigma}-\bm I_p)(\bm\beta^\ast-\bm\beta^0)$ && Remainder term in debiased expansion \\[4pt]

\multicolumn{3}{@{}l}{\textbf{Rates and Constants}}\\
$r_{1n},r_{2n},r_{3n}$ && Contraction rates in $\sum_k \sqrt{p_k}\|\beta_k^\ast-\beta_k^0\|$, $\|\beta^\ast-\beta^0\|^2$, and $n^{-1}\|X(\beta^\ast-\beta^0)\|^2$ \\
$\phi(S)$ && Compatibility constant \\
$\nu$ && Irrepresentability margin (GL selection theory) \\[4pt]

\multicolumn{3}{@{}l}{\textbf{Additive model and splines}}\\
$f_k^0$ && True additive component; $Y=\sum_{k=1}^K f_k^0(X_k)+\varepsilon$ \\
$B_n$ && Number of B-spline basis per covariate \\
$\alpha>0$ && Smoothness index; typical choice $B_n\asymp n^{1/(1+2\alpha)}$ \\
$f_k(\cdot)=\sum_{j=1}^{B_n} B_{k,j}(\cdot)\,\beta_{k,j}$ && spline approximation of $f_k^0$; $f^\ast$ induced by $\beta^\ast$ \\[4pt]

\multicolumn{3}{@{}l}{\textbf{Credible sets and BvM}}\\
$C_j$ && $(1-\alpha)$ marginal credible set for $\beta_j$ \\
$\bm m,\,\hat{\bm V}$ && BvM centering and covariance for $\sqrt{n}(\beta^{\ast\ast}-\beta^0)$ \\
\end{longtable}
\end{center}

As the immersion posterior for $\sigma^2$ described in \Cref{subsec:prior-posterior} is consistent, as in \cite{pal2024bayesian}, it suffices to prove the results conditionally on given $\sigma^*$, uniformly in $\sigma^*$ in a shrinking neighborhood of the true $\sigma^0$, say, $\mathcal{U}_n$, such that $\Pi(\sigma^* \in \mathcal{U}_n|\bY) \to 1$ in probability as $n \to \infty$. Since group sparsity is a special case of the general sparse high-dimensional linear regression problem, the same map used in \cite{pal2024bayesian} is reasonable.

\begin{proof}[Proof of~\Cref{est_pred_consis}(1)]
    Since $\bm{\beta}^*$ minimizes \eqref{proj_map} under the group LASSO penalty, we can write \begin{align}\label{step1}
        \frac{1}{n} \|\bX\bm{\beta} - \bX \bm{\beta}^*\|^2 + \lambda_n\sum_{k = 1}^K \mathcal{P}( \|\bm{\beta}^*_k\| )\leq \frac{1}{n} \|\bX\bm{\beta} - \bX \bm{\beta}^0\|^2 + \lambda_n\sum_{k = 1}^K \mathcal{P}( \|\bm{\beta}^0_k\|).
    \end{align}
    Now, the left-hand side (LHS) of the above can be expressed as 
    \begin{align} 
    \label{step 2}
    \lefteqn{\frac{1}{n} \|\bm{\eta} + \bX \bm{\beta}^0 - \bX \bm{\beta}^*\|^2 + \lambda_n\sum_{k = 1}^K \mathcal{P}( \|\bm{\beta}^*_k\| ) }\nonumber \\
    &&= \frac{1}{n} \|\bX \bm{\beta}^0 - \bX \bm{\beta}^*\|^2 + \frac{1}{n} \|\bm{\eta}\|^2 - \frac{2}{n} \bm{\eta}^\mathrm{T} \bX(\bm{\beta}^* - \bm{\beta}^0) + \lambda_n\sum_{k = 1}^K \mathcal{P}( \|\bm{\beta}^*_k\| ).
\end{align}
    Then, from \eqref{step1} and \eqref{step 2}, by choosing $\lambda_n \geq 2\lambda_0$, where, $\lambda_0 \asymp \sqrt{(\log p) / n}$, we can write from \Cref{lemma_for_thm1} that
    \begin{eqnarray}
    \label{step 3}
        \lefteqn{n^{-1}\|\bX \bm{\beta}^0 - \bX \bm{\beta}^*\|^2 } \nonumber\\
       &&\leq \frac{2}{n} \bm{\eta}^\mathrm{T} \bX(\bm{\beta}^0 - \bm{\beta}^*) + \lambda_n\sum_{k = 1}^K \{\mathcal{P} (\|\bm{\beta}^0_k\|) - \mathcal{P}( \|\bm{\beta}^*_k\| )\} \\
        && \leq  \frac{2}{n} \max_{1 \leq k \leq K} \frac{\|\bm{\eta}^\mathrm{T}\bX_{G_k}\|}{\sqrt{p_k}} \sum_{k = 1}^K \sqrt{p_k}\|\bm{\beta}^0_k - \bm{\beta}^*_k\| + \lambda_n \big\{\sum_{k = 1}^{s_0} \sqrt{p_k} (\|\bm{\beta}^0_k\| - \|\bm{\beta}^*_k\|) - \sum_{k = s_0+1}^K \sqrt{p_k}\|\bm{\beta}^*_k\|\big\} \nonumber\\
        &&\leq \frac{\lambda_n}{2} \big\{ \sum_{k = 1}^{s_0} \sqrt{p_k} \|\bm{\beta}_k^0 - \bm{\beta}^*_k\| + \sum_{k = s_0+1}^K \sqrt{p_k} \|\bm{\beta}^*_k\|\big\} + \lambda_n \big\{\sum_{k = 1}^{s_0} \sqrt{p_k} \|\bm{\beta}^*_k - \bm{\beta}_k^0\| - \sum_{k = s_0+1}^K \sqrt{p_k} \|\bm{\beta}^*_k\| \big\}  \nonumber\\
        &&=\frac{3\lambda_n}{2}\sum_{k = 1}^{s_0} \sqrt{p_k} \|\bm{\beta}^*_k - \bm{\beta}_k^0\| - \frac{\lambda_n}{2}\sum_{k = s_0+1}^K \sqrt{p_k} \|\bm{\beta}^*_k\| \nonumber
    \end{eqnarray}
    This then implies by Assumption \ref{compat}, 
    \begin{align}
    \label{est_pred_proof_eq}
        \frac{1}{n} \|\bX\bm{\beta}^0 - \bX\bm{\beta}^*\|^2 + \frac{\lambda_n}{2} \sum_{k = 1}^K \sqrt{p_k} \|\bm{\beta}^*_k - \bm{\beta}_k^0\| & \leq {2\lambda_n} \sum_{k = 1}^{s_0} \sqrt{p_k} \|\bm{\beta}_k^0 - \bm{\beta}^*_k\| \nonumber\\ 
        &\leq \frac{2\lambda_n \|\bX\bm{\beta}^0 - \bX\bm{\beta}^*\|_2 \sqrt{\sum_{k = 1}^{s_0} {p_k} }}{\sqrt{n} \Phi(S_0)}.
    \end{align}
    Using $2ab \leq a^2 + b^2$ with $a = ({2\sqrt{n}})^{-1} \|\bX\bm{\beta}^0 - \bX\bm{\beta}^*\|_2$ and $b = \left(4 \lambda_n/\Phi(s_0)\right) \sqrt{\sum_{k = 1}^{s_0} p_k} $ and transposing the terms leads to 
    $ \displaystyle 
        \frac{1}{2n} \|\bX\bm{\beta}^0 - \bX\bm{\beta}^*\|^2 + \frac{\lambda_n}{2} \sum_{k = 1}^K \sqrt{p_k} \|\bm{\beta}^*_k - \bm{\beta}_k^0\| \leq \frac{16 \lambda_n^2 \sum_{k = 1}^{s_0} p_k}{\Phi^2(s_0)}.
    $ 
    \end{proof}

\begin{proof}[Proof of~\Cref{est_pred_consis}(2)]
We first need to show an oracle property of the induced posterior of $\bm\beta^*$ obtained using the group SCAD map to prove this theorem. Consider the ideal scenario where the true model is known to us. We know exactly the set of signal groups, $S_0$. In such a situation, the best posterior representation would be the conjugate normal posterior but restricted to the true model $S_0$. If $|S_0| = s_0$, then the remaining $p - s_0$ components are set to zero, and we get the oracle projection as $\bm\beta_{S_0} \in \R^p$, such that $\bm\beta_{S_0}|(S = S_0, \bY, \bX, {\sigma^*}^2) \sim \normal\big(\hat{\bm\beta}^\mathrm{R}_{S_0}, {\sigma^*}^2 \bm{H}_{S_0}(a_n)\big),$ for ${\sigma^* \in \mathcal{U}_n}$, where $\hat{\bm\beta}^\mathrm{R}_{S_0} \in \R^{s_0}$ is the ridge regression estimator with penalty $a_n^{-1}$, restricted to the true model and $\bm{H}_{S_0}(a_n)$ is defined in the same way as $\bm{H}(a_n)$, restricted to the truly active variables according to $S_0$. Let $\Pi_{S_0}(\cdot|\bY)$ be used to denote the posterior distribution of the oracle projection $\bm\beta_{S_0}$. In the following theorem, we show that when the true value of the coefficient is larger than the specified quantity in Assumption \ref{min-beta_condition}(B), this oracle posterior itself is the minimizer of the group SCAD operation, that is, $\bm\beta_{S_0} = \iota(\bm\beta_{S_0})$, under the group SCAD penalty. This is possible because the SCAD penalty flattens out for large values of the coefficient, thereby ceasing to contribute to the minimization of the function. Thus, under a known structure, the group SCAD map returns the conjugate posterior distribution under a strong signal.

\begin{lemma}[Oracle property of group SCAD projection-posterior]\label{oracle_SCAD}
    Under Assumptions \ref{design}, \ref{mean_assum}, \ref{sd_assum} and \ref{min-beta_condition}(B), if $\lambda_n = o\big(n^{-{(1-c_3 + c_4)}/{2}}\big)$ and $p/(\lambda_n n^{3/2}) \to 0$, then $|\Pi(z|\bY) - \Pi_{S_0}(z|\bY)| \to 0$ as $n \to \infty$.
\end{lemma}
Then, choosing $\lambda_n \geq \lambda_0$ and following the technique laid out in \eqref{step1}, \eqref{step 2} and \eqref{step 3} in the proof of \Cref{est_pred_consis}(1), for the group SCAD penalty $n^{-1}\|\bX \bm{\beta}^0 - \bX \bm{\beta}^*\|^2 \leq 2n^{-1} \bm{\eta}^\mathrm{T} \bX(\bm{\beta}^0 - \bm{\beta}^*) + \lambda_n\sum_{k = 1}^K \{\mathcal{P} (\|\bm{\beta}^0_k\|) - \mathcal{P}( \|\bm{\beta}^*_k\| )\}$, which can be estimated by
    \begin{align*}
       &\frac{2}{n} \max_{1 \leq k \leq K} \frac{\|\bm\eta^\mathrm{T}\bX_{G_k}\|}{\sqrt{p_k}} \sum_{k = 1}^K \sqrt{p_k} \|\bm\beta^0_k - \bm\beta_k^*\| + \lambda_n \big\{ \sum_{k = 1}^{s_0} \left( \mathcal{P}(\|\bm\beta^0_k\|) - \mathcal{P}(\|\bm\beta^*_k\|)\right) - \sum_{k = s_0+1}^K \mathcal{P}(\|\bm\beta^*_k\|)\big\}\\
       \leq & {\lambda_n} \big\{ \sum_{k = 1}^{s_0} \sqrt{p_k} \|\bm\beta^0_k - \bm\beta_k^*\| + \sum_{k = s_0 + 1}^K \sqrt{p_k} \|\bm\beta^*_k\|\big\} + \frac{1}{2}(\tau + 1)s_0 \lambda_n^2 - \lambda_n \sum_{k = s_0+1}^K \mathcal{P}(\|\bm\beta^*_k\|)\\
       =& \lambda_n \sum_{k = 1}^{s_0} \sqrt{p_k} \|\bm\beta^0_k - \bm\beta^*_k\| + \frac{1}{2}(\tau+ 1)s_0\lambda_n^2, 
    \end{align*}
    by \cite{huang2007asymptotic} and \Cref{oracle_SCAD}. Then, by Assumption \ref{compat}, we can write \begin{align*}
        n^{-1}\|\bX\bm\beta^0 - \bm\beta^*\|^2 + \frac{\lambda_n}{2}\sum_{k = 1}^{s_0} \sqrt{p_k} \|\bm\beta^0_k - \bm\beta^*_k\| &\leq \frac{3\lambda_n}{2}\sum_{k = 1}^{s_0} \sqrt{p_k} \|\bm\beta^0_k - \bm\beta^*_k\| + \mathcal{O}(s_0\lambda_n^2)\\
        &\leq \frac{2\lambda_n\|\bX\bm\beta^0 - \bX\bm\beta^*\| \sqrt{\sum_{k = 1}^{s_0}p_k}}{\sqrt{n}\Phi(S_0)} + \mathcal{O}(s_0\lambda_n^2).
    \end{align*}
    The rest follows as in the proof of \Cref{est_pred_consis}(1), after \eqref{est_pred_proof_eq}.
\end{proof}

\begin{proof}[Proof of~\Cref{est_pred_consis}(3)]
    Let $\hat{S}$ be the set of variables selected by the group LASSO penalty. Since we weigh the adaptive LASSO map by the inverse of the group norms of $\hat{\bm\beta}^\textnormal{GL}$, the variables not included in $\hat{S}$ will receive $\infty$ penalty and will hence not appear in $S^*$ too, where $S^*$ refers to the set of variables selected by the sparse-projection (adaptive group LASSO) map. Thus, we may only restrict ourselves to $\hat{S}$ and claim that $\bm\beta^*_{\hat{S}}$ minimizes \eqref{proj_map} under the group adaptive LASSO penalty pertaining only to the variables in $\hat{S}$. That is,
    \begin{align*}
        \frac{1}{n}\|\bX\bm\beta - \bX_{\hat{S}}\bm\beta^*_{\hat{S}}\|^2 + \lambda_n \sum_{k \in \hat{S}} \frac{\sqrt{p_k}}{\|\hat{\bm\beta}_k\|}\|\bm\beta^*_k\| \leq \frac{1}{n}\|\bX\bm\beta - \bX_{\hat{S}}\bm\beta^0_{\hat{S}}\|^2 + \lambda_n \sum_{k \in \hat{S}} \frac{\sqrt{p_k}}{\|\hat{\bm\beta}_k\|}\|\bm\beta^0_k\|.
    \end{align*}
    Then, we can rewrite the above as \begin{align}
        0 &\geq \frac{1}{n}\|\bX\bm\beta - \bX_{\hat{S}}\bm\beta^0_{\hat{S}} + \bX_{\hat{S}}\bm\beta^0_{\hat{S}} - \bX_{\hat{S}}\bm\beta^*_{\hat{S}}\|^2 - \frac{1}{n}\|\bX\bm\beta - \bX_{\hat{S}}\bm\beta^0_{\hat{S}}\|^2 - \lambda_n \sum_{k \in \hat{S}} \frac{\sqrt{p_k}}{\|\hat{\bm\beta}_k\|}\|\bm\beta^*_k - \bm\beta^0_k\| \nonumber\\
        & = \frac{1}{n}\|\bX_{\hat{S}}(\bm\beta^*_{\hat{S}} - \bm\beta^0_{\hat{S}})\|^2 - \frac{2}{n}\bm\eta^\mathrm{T}\bX_{\hat{S}}(\bm\beta^*_{\hat{S}} - \bm\beta^0_{\hat{S}}) - \lambda_n \sum_{k \in \hat{S}} \frac{\sqrt{p_k}}{\|\hat{\bm\beta}_k\|}\|\bm\beta^*_{\hat{S}} - \bm\beta^0_{\hat{S}}\|\nonumber
    \end{align} 
    Now, it has been shown in \cite{wei2010consistent} that there exists a positive constant $\psi$, such that $P(\min_{k \in S} \|\hat{\bm\beta}_k\| \geq \psi \min_{k \in S} \|\bm\beta^0_k\|) \to 1$ as $n \to \infty$. Consequently, choosing $\lambda_n \geq \lambda_0$, by \Cref{lemma_for_thm1}, $n^{-1} \|\bX(\bm\beta^* - \bm\beta^0)\|^2$ can be written as
    \begin{align*}
         n^{-1} \|\bX_{\hat{S}}(\bm\beta^*_{\hat{S}} - \bm\beta^0_{\hat{S}})\|^2 &
        \leq \frac{2}{n}\bm\eta^\mathrm{T}\bX_{\hat{S}}(\bm\beta^*_{\hat{S}} - \bm\beta^0_{\hat{S}}) + \lambda_n \sum_{k \in \hat{S}} \frac{\sqrt{p_k}}{\|\hat{\bm\beta}_k\|}\|\bm\beta^*_{\hat{S}} - \bm\beta^0_{\hat{S}}\|\\
        & \leq \frac{2}{n} \max_{1 \leq k \in \hat{S}} \frac{\|\bm\eta^\mathrm{T}\bX_{G_k}\|}{\sqrt{p_k}} \sum_{k \in \hat{S}} \sqrt{p_k} \|\bm\beta^0_k - \bm\beta_k^*\| + \frac{\lambda_n}{\psi\min_{k \in S_0}\|\bm\beta_k^0\|} \sum_{k \in \hat{S}} \sqrt{p_k} \|\bm\beta^0_k - \bm\beta_k^*\|\\
        & \leq \lambda_n \Big(1 + \frac{1}{\psi\min_{k \in S_0}\|\bm\beta_k^0\|}\Big) \sum_{k \in \hat{S}} \sqrt{p_k} \|\bm\beta^0_k - \bm\beta_k^*\|
    \end{align*}

    Then, using the Compatibility condition in Assumption \ref{compat}, we have,

    \begin{align*}
         n^{-1} \|\bX(\bm\beta^* - \bm\beta^0)\|^2 & \leq  \lambda_n \Big(1 + \frac{1}{\psi\min_{k \in S_0}\|\bm\beta_k^0\|}\Big) \frac{\|\bX\beta^* - \bX\beta^0\| \sqrt{\sum_{k = 1}^{s_0}p_k}}{\sqrt{n}\Phi(s_0) }
    \end{align*}

    Assuming that $\min\{\|\bm\beta^0_k\|: k \in \hat{S}\} $ is larger than some positive constant and that the group sizes do not vary with $n$, we can show the following bounds holds using the inequality $2ab \leq a^2 + b^2$ and tracing the same line of argument undertaken in the proof of \Cref{est_pred_consis}(1) succeeding \eqref{est_pred_proof_eq}, we have $n^{-1}\|\bX(\bm\beta^* - \bm\beta^0)\|^2 = \mathcal{O}_P({s_0\lambda_n^2}) \textnormal{ and } \sum_{k = 1}^K {\sqrt{p_k}} \|\bm\beta^*_k - \bm\beta^0_k\| = \mathcal{O}_P({s_0\lambda_n}).$
\end{proof}

\begin{proof}[Proof of~\Cref{model_sel_consis}(1)]
    Defining $\bm\xi_{S} = \textnormal{vec}(\sqrt{p_k} \bm{z}_k, k \in S) \in \R^{\sum_{k \in S}p_k}$, we can see that the theorem will be proved if we can show that the following two hold, \begin{align}\label{first_part_model_sel_proof}
        \sup_{\sigma^* \in \mathcal{U}_n}\Pi\big(\|\bm\beta^*_{S_0} - \bm\beta^0_{S_0}\|_\infty < \min_{k \in S_0} \|\bm\beta_k^0\|_\infty\big| \bY,\sigma^*\big) \to 1 \text{ in probability as } n \to \infty,
    \end{align}
    \vspace{-0.5cm}\begin{align}\label{second_part_model_sel_proof}
        \sup_{\sigma^* \in \mathcal{U}_n}\Pi\big( p_k^{-1/2} \|\bm\xi_k\| < 1 \enskip \textnormal{ for } k \in S_0^\mathrm{c} \big| \bY,\sigma^* \big) \to 1 \text{ in probability as } n \to \infty,
    \end{align} From the first-order subgradient KKT condition of the group LASSO optimization problem \begin{align*}
    & - \frac{2}{n}\bX_{G_k}^\mathrm{T}\bX(\bm\beta^* - \bm\beta) + \frac{2}{n} \bX_{G_k}^\mathrm{T}\bm\eta = \lambda_n \sqrt{p_k} \bm{z}_k,
\end{align*} where $\bm{z}_k \in \R^{p_k}$ is such that $\|\bm{z}_k\| \leq 1$ when $\bm\beta^*_k = 0$ and $\bm{z}_k = {\bm\beta^*_k}/{\|\bm\beta^*_k\|_2}$ when $\bm\beta^*_k \neq 0$, and defining $\bC_{n(11)} = n^{-1} \bX_{S_0}^\mathrm{T} \bX_{S_0}$ and $\bC_{n(21)} = n^{-1} \bX_{S_0^\mathrm{c}}^\mathrm{T} \bX_{S_0}$, we can write
    \begin{align}\label{active_part_of_KKT}
        \bm\beta^*_{S_0} = \bm\beta^0_{S_0} + \bC_{n(11)}^{-1} \bigg(\frac{1}{n} \bX_{S_0}^\mathrm{T}\bm\eta - \lambda_n \bm\xi_{S_0}\bigg),    \end{align} \vspace{-0.5cm}
 \begin{align}\label{noise_part_of_KKT}
        \lambda_n\bm\xi_{S_0^\mathrm{c}} = \frac{1}{n}\bX_{S_0^\mathrm{c}}^\mathrm{T}\boldeta + \bC_{n(21)}\bC_{n(11)}^{-1}\bigg(\lambda_n \bm\xi_{S_0} - \frac{1}{n}\bX_{S_0}^\mathrm{T}\boldeta \bigg).
    \end{align}
    We will prove \eqref{first_part_model_sel_proof} using the condition in \eqref{active_part_of_KKT}, and similarly, we will show \eqref{second_part_model_sel_proof} to be true using \eqref{noise_part_of_KKT} in \Cref{lemma_for_groupLASSO_model_sel} stated below.
\end{proof}

\begin{lemma}\label{lemma_for_groupLASSO_model_sel} Under Assumptions \ref{design}, \ref{mean_assum}, \ref{sd_assum}, \ref{non-singularity_condition}, \ref{min-beta_condition}(A) and \ref{irrepresentability_condition},
        $$\Pi\big(\|\bm\beta^*_{S_0} - \bm\beta^0_{S_0}\|_\infty < \min_{k \in S_0} \|\bm\beta_k^0\|_\infty\big| \bY,\sigma^*\big) \to 1 \mbox{ and }  \Pi\big( p_k^{-1/2} \|\bm\xi_k\| < 1 \enskip \textnormal{ for } k \in S_0^\mathrm{c} \big| \bY,\sigma^* \big) \to 1$$ 
        in probability uniformly in ${\sigma^* \in \mathcal{U}_n}$ as $n \to \infty$.
\end{lemma}

\begin{proof}[Proof of~\Cref{model_sel_consis}(2)]
    The proof of this theorem follows from \Cref{oracle_SCAD}, which shows that the group SCAD map leads to the oracle projection $\bm\beta_{S_0}$.
\end{proof}

\begin{proof}[Proof of~\Cref{model_sel_consis}(3)]
     With $\bm{u}_1 = \bm{u} - \bm\beta^0$, we can write 
    \begin{align*}
        \bm\beta^* & = \argmin_{\bm{u}} \bigg\{\frac{1}{n} \|\bX\bm\beta - \bX\bm\beta^0 + \bX\bm\beta^0 - \bX\bm{u}\|^2 + \lambda_n \sum_{k = 1}^K \frac{\sqrt{p_k}}{\|\hat{\bm\beta}_k\|}\|\bm{u}_k\|\bigg\}\\
        & = \argmin_{\bm{u}_1} \bigg\{ \frac{1}{n}\|\bm\eta - \bX\bm{u}_1\|^2 + \lambda_n \sum_{k = 1}^K\frac{\sqrt{p_k}}{\|\hat{\bm\beta}_k\|} \|\bm{u}_1 + \bm\beta^0_k\|\bigg\},
    \end{align*} Defining $\bC_n = n^{-1}\bX^\mathrm{T}\bX$ and $\bZ_n = n^{-1/2}\bX^\mathrm{T}\bm\eta$, by the KKT conditions for $k \in S_0$,
    \begingroup
    \renewcommand{\theequation}{\alph{equation}}
    \setcounter{equation}{0}  
        \begin{align}\label{KKT1}
        \frac{1}{\sqrt{n}}\big( \bC_{n(11)}(\sqrt{n}\bm{u}_{1(1)}) - \bZ_{n(1)}\big)_k = -\lambda_n \frac{(\bm{u}_{1k} + \bm\beta^0_k)\sqrt{p_k}}{\|\bm{u}_{1k} + \bm\beta^0_k\| \|\hat{\bm\beta}_k\|} \textnormal{ and } \|\bm{u}_{1k} \| \leq \|\bm\beta^0_k\|
    \end{align}
    and for $k \in S_0^\mathrm{c}$, \begin{align}\label{KKT2}
        -\lambda_n \frac{\sqrt{np_k}}{\|\hat{\bm\beta}_k\|} \bm{1}_{p_k} \leq \big( \bC_{n(21)}(\sqrt{n}\bm{u}_{1(1)}) - \bZ_{n(2)}\big)_k \leq \lambda_n \frac{\sqrt{np_k}}{\|\hat{\bm\beta}_k\|} \bm{1}_{p_k}.
    \end{align}
    \endgroup
 Recall that $p_0 = \sum_{k = 1}^{s_0} p_k$ is the total number of active variables in the model. Then, defining the vector $\bm{M}_{S_0} \in \R^{p_0}$ such that $\bm{M}_{S_0} = (\bm{M}_1,\dots,\bm{M}_{s_0})^\mathrm{T}$, where $$\bm{M}_k = -\lambda_n {(\bm{u}_{1k} + \bm\beta^0_k)\sqrt{p_k}}/({\|\bm{u}_{1k} + \bm\beta^0_k\| \|\hat{\bm\beta}_k\|}).$$
 Here, the event in \eqref{KKT1} may be represented as  
 \begin{align*}
     A=& \big\{\|(\bC_{n(11)}^{-1}\bZ_{n(1)} - \sqrt{n}\bC_{n(11)}^{-1}\bm{M}_{S_0})_k\| = \sqrt{n} \|\bm{u}_{1k}\| \enskip \textnormal{ for } k \in S_0\big\} \cap \{\|\bm{u}_{1k}\| \leq \|\bm{\beta}^0_k\| \enskip \textnormal{ for } k \in S_0\}\\
     = & \big\{\|(\bC_{n(11)}^{-1}\bZ_{n(1)} - \sqrt{n}\bC_{n(11)}^{-1}\bm{M}_{S_0})_k\| \leq \sqrt{n}\|\bm\beta^0_k\| \enskip \textnormal{ for } k \in S_0\big\}
\end{align*}
Similarly, defining $\bm{P}_{S_0} = \bX_{S_0}(\bX_{S_0}^\mathrm{T}\bX_{S_0})^{-1}\bX_{S_0}^\mathrm{T}$, the event in \eqref{KKT2} can be simplified to 
\begin{align*}
    B=& \big\{\|(\bC_{n(21)}\big[\bC^{-1}_{n(11)}\bZ_{n(1)} - \sqrt{n}\bC^{-1}_{n(11)}\bm{M}_{S_0}\big] - \bZ_{n(2)})_k \| \leq \lambda_n ({\sqrt{np_k}}/{\|\hat{\bm\beta}_k\|} )\|\bm{1}_{p_k}\| \big\}\\
    = & \big\{ \|n^{-1/2} \bX_{S_0^\mathrm{c}}^\mathrm{T}\bm{P}_{S_0}\boldeta - \sqrt{n}\bC_{n(21)}\bC^{-1}_{n(11)}\bm{M}_{S_0} - n^{-1/2}\bX_{S_0^\mathrm{c}}^\mathrm{T}\boldeta\| \leq \lambda_n ({\sqrt{np_k}}/{\|\hat{\bm\beta}_k\|} )\|\bm{1}_{p_k}\| \big\}\\
    \subseteq & \big\{n^{-1/2}\|\bX_{S_0^\mathrm{c}}^\mathrm{T}(\bm{I}_n - \bm{P}_{S_0})\boldeta \| \leq \lambda_n ({\sqrt{np_k}}/{\|\hat{\bm\beta}_k\|} )\|\bm{1}_{p_k}\| - \sqrt{n} \|\bC_{n(21)}\bC^{-1}_{n(11)}\bm{M}_{S_0}\|\big\}.
\end{align*}
Consequently, for $\sigma^* \in \mathcal{U}_n$, \begin{align*}
    \Pi(\|\bm\beta^*_k\| \neq 0 \textnormal{ for } k \in S_0, \|\bm\beta^*_k\| = 0 \textnormal{ for } k \in S_0^\mathrm{c}|\bY) \geq \Pi(A \cap B|\bY) \geq 1 - \Pi(A^\mathrm{c}|\bY) - \Pi(B^\mathrm{c}|\bY).
\end{align*}
The theorem then follows from \Cref{lemma_A_groupAdap} and \Cref{lemma_B_groupAdap}.
\end{proof}

\begin{lemma}\label{lemma_A_groupAdap} Under Assumptions \ref{design}, \ref{mean_assum}, \ref{sd_assum} and \ref{min-beta_condition}(C), $\sup_{\sigma^* \in \mathcal{U}_n}\Pi(A^\mathrm{c}|\bY,\sigma^*) \to 0$ as $n \to \infty$.\end{lemma}

\begin{lemma}\label{lemma_B_groupAdap} Under Assumptions \ref{design}, \ref{mean_assum}, \ref{sd_assum} and \ref{min-beta_condition}(C), $\sup_{\sigma^* \in \mathcal{U}_n}\Pi(B^\mathrm{c}|\bY,\sigma^*) \to 0$ as $n \to \infty$.\end{lemma}

\begin{proof}[Proof of~\Cref{debiasedCLT_GL}]  
    To prove the asymptotic normality of the debiased LASSO projection $\bm\beta^{**}$ cemtered at the truth $\bm\beta^0$, we only need to control $\bm\Delta$, since we already know that \begin{align*}
        & \sup_{\sigma^* \in \mathcal{U}_n}\E(n^{-1}\hat{\bm\Theta} \bX^\mathrm{T} \bm\eta | \bY,\sigma^*) = n^{-1} \hat{\bm\Theta} \bX^\mathrm{T}\bm\mu, \nonumber \\
        & \sup_{\sigma^* \in \mathcal{U}_n}\textnormal{var}(n^{-1}\hat{\bm\Theta} \bX^\mathrm{T} \bm\eta | \bY,\sigma^*) = n^{-1} {\sigma^*}^2 \hat{\bm\Theta} \bX^\mathrm{T} \bm{H}(a_n) \bX \hat{\bm\Theta}^\mathrm{T}.
    \end{align*} 
    We also know that the $j$-th sub-vector of the $p$-dimensional vector $\bm{\Delta}$ can be written as $$\bm\Delta_j = (\bm{T}_j^{-2})^\mathrm{T} \bm{\Lambda}_j \sum_{k \neq j} \bm{K}_{j,k}^\mathrm{T} (\bm\beta^{*}_k - \bm{\beta}^0_k) \in \R^{p_j}.$$ Then, by the construction of $\hat{\bm\Theta}$, we have for $j = 1,2,\dots,K$,  \begin{align*}
        \|\bm\Delta_j\| \leq \mathcal{O}\bigg(\sqrt{{\big(\max_{1 \leq k \leq K}p_k^2\big) \log n}/{n}} \bigg) \sum_{k = 1}^K \sqrt{p_k} \|\bm\beta^*_k - \bm\beta^0_k \| \leq C\lambda_0s_0\sqrt{{\big(\max_{1 \leq k \leq K}p_k^3\big) \log n}/{n}}.
    \end{align*} 
    in view of \Cref{lemma_for_Delta} and \Cref{est_pred_consis}(1). 
\end{proof}

\begin{proof}[Proof of~\Cref{debiased_coverage}]
First, we show that the means of the two limiting normal distributions are the same. That is, we need to show that the mean of the random variable $\sqrt{n}(\bm\beta^{**} - \hat{\bm\beta}^\textnormal{DGL}) = \sqrt{n}(\bm{m} + \bm\beta^0 - \hat{\bm\beta}^\textnormal{DGL})$ vanishes as $n$ grows to $\infty$. From \eqref{debias_estim_para}, we can write the above quantity as 
\begin{align*}
     \|\sqrt{n}(\bm{m} - \hat{\bm\beta}^\textnormal{DGL} + \bm{\beta}^0)\| =& \sqrt{n} \| \hat{\bm\Omega}^{-1/2} n^{-1}\hat{\bm\Theta}\bX^\mathrm{T}(\bm\mu - \bm\varepsilon) + \bm\Delta^\textnormal{DGL} \|\\
     = & \sqrt{n} \|n^{-1} \hat{\bm\Theta}\bX^\mathrm{T}(\bX\hat{\bm\beta}^\mathrm{R} - \bX\bm\beta^0  - \bY + \bX\bm\beta^0) + \bm\Delta^\textnormal{DGL} \|\\
     \leq & n^{-1/2} \|\bY - \bX\hat{\bm\beta}^\mathrm{R}\| \cdot \| \hat{\bm\Theta}\bX^\mathrm{T} \|_{\textnormal{op}} + \sqrt{n}\|\bm\Delta^\textnormal{DGL}\|,
\end{align*} 
which is $o_P(1)$ because the first term is a product of two quantities, $n^{-1/2} \|\bY - \bX\hat{\bm\beta}^\mathrm{R}\|$ and $\| \hat{\bm\Theta}\bX^\mathrm{T} \|_{\textnormal{op}}$, of which the first vanishes by \Cref{lemma_ridge_conv} and the second can be shown to be bounded using the Proposition 5 in \cite{honda2021biased}, which says that $n^{-1}\hat{\bm\Theta}\bX^\mathrm{T}\bX \hat{\bm\Theta}^\mathrm{T} \to \bm\Theta$ blockwise in probability. The second term $\sqrt{n}\|\bm\Delta^\textnormal{DGL}\|$ has been proven to be $o(1)$ in Proposition 3 of \cite{honda2021biased}. For completeness and ease of understanding, both these propositions have been restated in \Cref{lemma_for_hat_Theta} and \Cref{lemma_for_Delta}, respectively. 
Finally, since 
    $$\bigg|\frac{\hat{\bm\Theta}\bX^\mathrm{T}\bm{H}({a_n})\bX\hat{\bm\Theta}^\mathrm{T}}{\hat{\bm\Theta}\bX^\mathrm{T}\bX\hat{\bm\Theta}^\mathrm{T}} - 1\bigg| = \bigg|\frac{\hat{\bm\Theta}\bX^\mathrm{T}\big(\bm{I}_p - \bm{H}({a_n})\big)\bX\hat{\bm\Theta}^\mathrm{T}}{\hat{\bm\Theta}\bX^\mathrm{T}\bX\hat{\bm\Theta}^\mathrm{T}}\bigg| \leq \text{tr}\left(\bm{I}_p - \bm{H}(a_n)\right) = o(1),$$
     the variances of the two asymptotic normal distributions match as $n \to \infty$.
\end{proof}

\begin{proof}[Proof of~\Cref{additve_est_consis}(1)]
    Recalling that $f_k(X_{k,i}) = \sum_{j = 1}^{B_n} \mathcal{B}_{k,j}(X_{k,i})\beta_{k,j}$ and the corresponding map-induced version $f^*_k(X_{k,i}) = \sum_{j = 1}^{B_n} \mathcal{B}_{k,j}(X_{k,i})\beta^*_{k,j}$, we can write \begin{align*}
        \|f^* - f^0\|^2  \leq 2(n^{-1}\|\mathcal{\bm{B}}\bm\beta^* - \mathcal{\bm{B}}\bm\beta^0\|^2 + \|\mathcal{\bm{B}}\bm\beta^0 - f_0\|^2)
        = \mathcal{O}(s_0 B_n \lambda_n^2) + \mathcal{O}(s_0 B_n^{-2\alpha}),
    \end{align*} 
    where the first bound follows from theorem \Cref{est_pred_consis}(1) and the second bound is due to approximation by B-spline basis expansion of $f_0(\cdot).$ Then, for $B_n \asymp n^{1/(2\alpha + 1)}$ and $\lambda_n \asymp \sqrt{(\log p)/n}$, we have 
    $\|f^* - f^0\|^2 = \mathcal{O}(s_0 n^{-{2\alpha}/({2\alpha + 1})} \log p).$
\end{proof}

\begin{proof}[Proof of~\Cref{additve_est_consis}(2)]
    We note that, although the true underlying data-generating model is a nonparametric additive model, we use a B-spline basis expansion followed by a group-sparsity operation to construct the model. Hence, we need to account not only for random error but also for the approximation error introduced by misspecifying the model as a B-spline regression. Thus, besides $\bm\varepsilon$, the  model error now also includes $\bm\delta = (\delta_1,\dots,\delta_n)^\mathrm{T}$, where $\delta_i = \sum_{k = 1}^{K} \big(f_k^0(X_{k,i}) - f_k(X_{k,i})\big)$. We know that $\|\bm\delta\| = \mathcal{O}\big({s_0}n^{1/(4\alpha + 2)}\big).$ The theorem then readily follows from \Cref{model_sel_consis}(1), replacing $\bm\varepsilon$ by $\bm\varepsilon^{(2)} = \bm\varepsilon + \bm\delta$.
\end{proof}

\subsection*{Proofs of the Lemmas}

 \begin{lemma}\label{lemma_bound_Fn}
    Under Assumption \ref{non-singularity_condition}, we have that \begin{align}\label{Fn}
        F_n = \frac{(\boldeta - \bm{\mu})^{\mathrm{T}} \bX_{G_k} \bX^\mathrm{T}_{G_k}(\boldeta - \bm{\mu})}{(\boldeta - \bm{\mu})^{\mathrm{T}} \bX_{G_k} (\bX_{G_k}^\mathrm{T} \bm{H}(a_n) \bX_{G_k})^{-1} \bX_{G_k}^\mathrm{T} (\boldeta - \bm{\mu})} = \mathcal{O}(1).
    \end{align}
 \end{lemma}

 \begin{proof} It is easy to see that,
     \begin{align*}
         0\le F_n \leq \frac{\|(\bm\eta - \bm\mu)^\mathrm{T}\bX_{G_k}\|^2 \lambda_{\textnormal{max}}(\bX_{G_k}^\mathrm{T}\bm{H}(a_n)\bX_{G_k})}{\|(\bm\eta - \bm\mu)^\mathrm{T}\bX_{G_k}\|^2}
         \leq \lambda_{\textnormal{max}}(\bm{H}(a_n))\cdot \lambda_{\textnormal{max}}(\bX_{G_k}^\mathrm{T}\bX_{G_k})
     \end{align*} is bounded.
 \end{proof}

 \begin{lemma}\label{lemma_for_thm1}
     For all $x > 0$ and for $$\lambda_{0}^2(x) = \dfrac{8}{n} \left(1 + \sqrt{\frac{4x+4\log p}{p_{\textnormal{min}}}} + \frac{4x+4\log p}{p_{\textnormal{min}}} \right),$$ we have, $$\sup_{\sigma^* \in \mathcal{U}_n}\Pi\Big(\big\{ \max_{1 \leq k \leq K} \frac{\|\bm{\eta}^\mathrm{T}\bX_{G_k}\|^2}{np_k} \leq \frac{\lambda_{0}^2(x)}{4}\big\}\Big|\bY,\sigma^*\Big) \geq 1 - \textnormal{exp}(-x).$$
 \end{lemma}

 \begin{proof}
     From the posterior distribution of $\bX\bm\beta$, we can write $$\bX_{G_k}^\mathrm{T}(\bm\eta - \bm\mu)|(\bY,\sigma^*) \sim \normal_{p_k}(\textbf{0}_{p_k}, {\sigma^*}^2 \bX_{G_k}^\mathrm{T}\bm{H}(a_n)\bX_{G_k}).$$ Now consider the singular value decomposition of the design matrix given by $\bX = \bm{U}\bm{D} \bm{V}^\mathrm{T}$, where the columns of $\bm{U}$ are the left singular vectors and $\bm{V}^\mathrm{T}$ has rows that are the right singular vectors, such that $\bm{U}^\mathrm{T}\bm{U} = \bm{I}_n$ and $\bm{V}^\mathrm{T}\bm{V} = \bm{I}_p$. Then, we have $$\bm{I}_{n} - \bm{H}(a_n)= \bm{U}\left(\bm{I}_n - \bm{D}(\bm{D}\bm{D}^\mathrm{T} + a_n\bm{I}_n)^{-1}\bm{D}\right)\bm{U}^\mathrm{T},$$ and consequently $\bm{I}_{n} - \bm{H}(a_n)$ is non-negative definite. Now, note that,
\begin{align*}
    \max_{1 \leq k \leq K} \| \boldsymbol \mu^{\mathrm{T}} \bX_{G_k} \| & = \max_{1 \leq k \leq K} \| {\bX_{G_k}}^{\mathrm{T}} \bm{H}(a_n) \big( \bX\btheta^0 + \bm \varepsilon \big) - {\bX_{G_k}}^{\mathrm{T}} \bX \btheta^0 \|\\
    & = \max_k \|{\bX_{G_k}}^{\mathrm{T}} \bm{H}(a_n) \bm \varepsilon - {\bX_{G_k}}^{\mathrm{T}} \big( \bm{I}_n - \bm{H}(a_n)\big) \bX \btheta^0 \|\\
    & \leq \max_k \| {\bX_{G_k}}^{\mathrm{T}} \bm{H}(a_n) \boldsymbol \varepsilon \| + \max_k \| {\bX_{G_k}}^{\mathrm{T}} \big( \bm{I}_n - \bm{H}(a_n)\big) \bX \btheta^0 \|\\
    & \leq \max_k \big[ \max_{1 \leq i \leq n} |\varepsilon_i| \|{\bX_{G_k}}^{\mathrm{T}} \bm{H}(a_n)\| + \max_{1 \leq i \leq n} |\E(Y_i)| \| {\bX_{G_k}}^{\mathrm{T}} \big( \bm{I}_n - \bm{H}(a_n) \big)\| \big]\\
    & = \mathcal{O}\big(K \sqrt{ \log n}\big).
\end{align*}
Then, defining $\chi^2_k = (\boldeta - \bm{\mu})^{\mathrm{T}} \bX_{G_k} (\bX_{G_k}^\mathrm{T} \bm{H}(a_n) \bX_{G_k})^{-1} \bX_{G_k}^\mathrm{T} (\boldeta - \bm{\mu})$, and noting that the posterior distribution of $\chi^2_k$ given data follows a $\chi^2$-distribution with ${p_k}$ degrees of freedom, we get by Cauchy-Schwarz inequality, uniformly in $\sigma^*\in \mathcal{U}_n$, 
\begin{align*}
    &\sup_{\sigma^* \in \mathcal{U}_n}\Pi \Big( \big\{ \max_{1 \leq k \leq K}  \frac{\| \boldeta^{\mathrm{T}} \bX_{G_k} \|^2}{np_k} \leq \lambda^2_{0}(x)/4 \big\} \big| \bY, \sigma^* \Big)\\
    \geq & \sup_{\sigma^* \in \mathcal{U}_n}\Pi \Big( \big\{ \max_{1 \leq k \leq K}  \frac{\| (\boldeta - \bm{\mu})^{\mathrm{T}} \bX_{G_k} \|^2}{np_k} + \max_{1 \leq k \leq K} \frac{\|\bm{\mu}^\mathrm{T}\bX_{G_k}\|^2}{np_k} \leq \frac{\lambda^2_{0}(x)}{8} \big\} \big| \bY, \sigma^* \Big) \\
    \geq & 1 - \sup_{\sigma^* \in \mathcal{U}_n}\Pi \Big( \big\{ \max_{1 \leq k \leq K}  \frac{\chi^2_k}{np_k} \cdot F_n + \max_{1 \leq k \leq K} \frac{\|\bm{\mu}^\mathrm{T}\bX_{G_k}\|^2}{np_k} > \frac{\lambda^2_{0}(x)}{8} \big\} \big| \bY, \sigma^* \Big) \\
    \geq & 1 - \sup_{\sigma^* \in \mathcal{U}_n}\Pi \Big( \max_{1 \leq k \leq K} \chi^2_k \cdot F_n + \max_{1 \leq k \leq K}{\|\bm{\mu}^\mathrm{T}\bX_{G_k}\|^2} > p_k \bigg(1 + \sqrt{\frac{4x+4\log p}{p_{\text{min}}}} + \frac{4x+4\log p}{p_{\text{min}}} \bigg) \bigg| \bY, \sigma^* \Big)\\
    \geq & 1 - \text{exp}(-x),
\end{align*}
because $F_n$, introduced in \eqref{Fn}, is bounded by \Cref{lemma_bound_Fn}, $\max_{1 \leq k \leq K} {\|\bm{\mu}^\mathrm{T}\bX_{G_k}\|^2}$ has been shown to be $\mathcal{O}\big(K^2{{\log n}}\big)$ and $\prob\big(\chi_k^2 \geq p_k(1+a)\big) \leq \text{exp}\left(-{p_k}\big(a - \log (1+a)\big)/2\right)$ by \cite{wallace1959bounds}. Then, taking $a = \sqrt{{4x}/{p_k}} + {4x}/{p_k},$ and using the estimate $a - \log(1+a) \geq a^2/2(1+a) \geq 2x/p_k$, it follows as in \cite{buhlmann2011statistics} that $\sup_{\sigma^* \in \mathcal{U}_n} \Pi \big( \chi^2_k \geq p_k(1+a) \big| \bY,\sigma^* \big) \leq e^{-x}$. 
 \end{proof}

 \begin{lemma}[Lemma 1 of \cite{pal2024bayesian}]
\label{lemma_an}
Under Assumption \ref{sd_assum},
$$\displaystyle \max_{j=1,\ldots,n} \big |1-\frac{d_j^2}{d_j^2 + a_n} \big|
=\max_{j=1,\ldots,n} \big |\frac{a_n}{d_j^2 + a_n} \big|=o(n^{-1}),$$ where $d_1,\ldots,d_n$ are the singular values of $\bX$. 
\end{lemma}

\begin{lemma}[Lemma 5 of \cite{pal2024bayesian}]\label{lemma_ridge_conv}
    Under Assumption \ref{sd_assum}, $n^{-1/2} \|\bY - \bX \hat{\boldsymbol \theta}^\mathrm{R} \|_1 = o_P(1).$
\end{lemma}

\begin{lemma}[Proposition 3 of \cite{honda2021biased}]\label{lemma_for_hat_Theta}
    Let $m$ be a fixed positive integer, and $\{j_1,\dots,j_m\} \subset \{1,2,\dots,K\}$. If Assumption \ref{debias_assums} holds, then, uniformly in $\{j_1,\dots,j_m\} $, as $n \to \infty,$  we have that $\max(|\lambda_{\text{min}}\big(\bm{D})|, |\lambda_{\text{max}}|(\bm{D})\big) \to 0$ in true probability, where $\hat{\bm{\Omega}} = n^{-1} \hat{\bm\Theta}\bX^\mathrm{T}\bX\hat{\bm\Theta}^\mathrm{T}$ and $$\bm{D} = \begin{bmatrix}
        \hat{\bm\Omega}_{j_1,j_1} & \dots & \hat{\bm\Omega}_{j_1,j_m} \\
\vdots & \ddots & \vdots \\
\hat{\bm\Omega}_{j_m,j_1} & \dots & \hat{\bm\Omega}_{j_m,j_m}
    \end{bmatrix} -
        \begin{bmatrix}
\bm\Theta_{j_1,j_1} & \dots & \bm\Theta_{j_1,j_m} \\
\vdots & \ddots & \vdots \\
\bm\Theta_{j_m,j_1} & \dots & \bm\Theta_{j_m,j_m} 
\end{bmatrix}.$$
\end{lemma}

\begin{lemma}[Proposition 5 of \cite{honda2021biased}]\label{lemma_for_Delta}
    Under Assumption \ref{debias_assums} we have $$\|\bm\Delta^\text{DGL}_j\| < C \max_{1\leq k \leq K} p_k^2 \cdot \frac{s_0 \log n}{n}$$ uniformly in $j = 1,2,\dots,K$, with probability tending to 1 for sufficiently large $C$.
\end{lemma}

\begin{proof}[Proof of~\Cref{oracle_SCAD}]
We need to show that $\bm\beta_{S_0}$ is the minimizer of \eqref{proj_map} under the group SCAD penalty. Writing $Q = n^{-1}\|\bX\bm{\beta}_{S_0} - \bX\bm{u}\|^2 + \sum_{k = 1}^K \mathcal{P}_{\lambda_n}(\|\bm{u}_k\|),$ for the $k$-th group, $$\frac{\partial Q}{\partial \bm{u}_k} = -\frac{1}{n}\bX_{G_k}^\mathrm{T}(\bX\bm\beta_{S_0} - \bX\bm{u}) + \frac{P'_{\lambda_n}(\|\bm{u}_k\|)\bm{u}_k}{\|\bm{u}_k\|},$$ where $P_{\lambda_n}'(t) = \lambda_n \big( \mathbbm{1}(t < \lambda_n) + \lambda^{-1}_n(\tau - 1)^{-1}(\tau\lambda_n - t)_+\mathbbm{1}(t > \lambda_n)\big).$ Now, following \cite{guo2015model}, for the oracle posterior $\bm\beta_{S_0}$ to be the minimizer of $Q$, the following second-order sufficiency conditions should be satisfied with high posterior probability:
    \begin{enumerate}
        \item $V_k(\bm\beta_{S_0}) = 0, \|\bm\beta_{S_0,k}\| > \tau\lambda_n \textnormal{ for } k \leq s_0$,
        \item $\|V_k(\bm\beta_{S_0})\| < \lambda_n, \|\bm\beta_{S_0,k}\| < \lambda_n \textnormal{ for } k > s_0$,
    \end{enumerate} assuming, for the convenience of notation, that the first $s_0$ groups are the active groups. The first condition holds if $\Pi(\|\bm\beta_{S_0,k}\|>\tau\lambda_n|\bY) \to 1$ as $n \to \infty$ for $k \leq s_0.$ By the triangle inequality we have $\|\bm\beta_{S_0,k}\| \geq \|\bm\beta^0_k\| - \|\bm\beta_{S_0,k} - \bm\beta^0_k\|$. 
 
    Then by invoking Assumption \ref{min-beta_condition}(B) and recalling that $\lambda_n = o\big(n^{-{(1-c_3 + c_4)}/{2}}\big)$, we observe that for any $\epsilon > 0$, the first condition boils down to
    \begin{align*}
        &\sup_{\sigma^* \in \mathcal{U}_n}\Pi\big(\cup_{k \leq s_0} \big\{\sqrt{n} \|\bm\beta_{S_0,k} - \bm\beta^0_k\| \geq \epsilon n^{c_3/2}\big\}|\bY,\sigma^*\big) \leq \sum_{k = 1}^{s_0} \frac{\sup_{\sigma^* \in \mathcal{U}_n}\E\big(\sqrt{n} \|\bm\beta_{S_0,k} - \bm\beta^0_k\| \big|\bY,\sigma^*\big)}{\epsilon n^{c_3/2}}, 
    \end{align*}
and can be bounded by ${\epsilon}^{-1} n^{-c_3/2}\sum_{k = 1}^{s_0} \big\{\E\big(\sqrt{n} \|\bm\beta_{S_0,k} - \hat{\bm\beta}^\mathrm{R}_{S_0;k}\| \big|\bY\big) + \E\big(\sqrt{n} \|\hat{\bm\beta}^\mathrm{R}_{S_0;k} - \bm\beta^0_k\| \big|\bY\big) \big\} = o(1)$. 

To prove the second condition, we only need $\sup_{\sigma^* \in \mathcal{U}_n} \Pi(\max_{k > s_0}\|V_k(\bm\beta_{S_0,k}\| < \lambda_n \big|\bY,\sigma^*) \to 1$ as $n \to \infty$, the rest is trivial for $k > s_0$. First, we denote $\bm\beta_{S_0,(1)}$ as the vector of the components of the oracle projection $\bm\beta_{S_0}$ pertaining to the $s_0$ active groups only. Then, $\bm\beta_{S_0,(1)} \in \R^{\sum_{k \leq s_0}p_k}$. Noting that $\bm\beta_{S_0,(1)}|\bY,\sigma^* \sim \normal \big((\bX_{(1)}^\mathrm{T}\bX_{(1)} + a_n \bm{I}_n)^{-1} \bX_{(1)}^\mathrm{T}\bY, {\sigma^*}^2 \bm{H}_{(1)}(a_n)\big)$, where $\bm{H}_{(1)}(a_n) = \bX_{(1)}(\bX_{(1)}^\mathrm{T}\bX_{(1)} + a_n \bm{I}_n)^{-1} \bX_{(1)}^\mathrm{T}$, by Assumption \ref{sd_assum}, we estimate the probability by
    \begin{align*}
        & \sum_{k > s_0} \sup_{\sigma^* \in \mathcal{U}_n}\Pi(n^{-1/2}\|\bX_{G_k}^\mathrm{T}(\bX\bm\beta - \bX_{(1)}\bm\beta_{S_0,(1)})\| > \sqrt{n}\lambda_n \big| \bY,\sigma^*)\\
        = & (p-s_0) \mathcal{O}\bigg(\frac{1}{\sqrt{n}\lambda_n\|\bX^\mathrm{T}_{G_k}[\bm{H}(a_n) + \bm{H}_{(1)}(a_n)]\bX_{G_k}\|\sigma^2} \bigg)\\
        \leq & (p-s_0) \mathcal{O}\bigg(\frac{1}{\sqrt{n}\lambda_n\|\bX^\mathrm{T}_{G_k}\bX_{G_k}\|\sigma^2} \bigg),
    \end{align*} 
    by the non-negative definiteness of $\bm{I}_{n} - \bm{H}(a_n)$. We know $\bX\bm\beta - \bX_{(1)}\bm\beta_{S_0,(1)} \big| (\bY,\sigma^*) \sim \normal_n\big(\bm{0},[\bm{H}(a_n) + \bm{H}_{(1)}(a_n)]{\sigma^*}^2\big)$. Hence, the assertion follows using the lower bound for $\lambda_n$ from \Cref{lemma_an}.
\end{proof}

\begin{proof}[Proof of Lemma~\ref{lemma_for_groupLASSO_model_sel}]
Using Markov inequality, we have,
    \begin{eqnarray*}
        \lefteqn{\sup_{\sigma^* \in \mathcal{U}_n}\Pi\big(\|\bm\beta^*_{S_0} - \bm\beta^0_{S_0}\|_\infty > \min_{k \in S_0} \|\bm\beta_k^0\|_\infty\big| \bY,\sigma^*\big)} \\
         && \le\frac{\sup_{\sigma^* \in \mathcal{U}_n}\E\big(\|\bm\beta^*_{S_0} - \bm\beta^0_{S_0}\|_\infty \big| \bY,\sigma^*\big)}{\min_{k \in S_0} \|\bm\beta_k^0\|_\infty}\\
        && \le \frac{1}{\alpha}  \bigg[ \sup_{\sigma^* \in \mathcal{U}_n}\E\big( \|n^{-1}\bC_{n(11)}^{-1}\bX^\mathrm{T}_{S_0^\mathrm{c}} \bm\eta\|_\infty \big| \bY,\sigma^* \big) + \big\|{\lambda_n} \bC_{n(11)}^{-1}\bm\xi_{S_0} \big\|_{\infty}\bigg]\\
         &&= \text{I} + \text{II}, \text{ say}.
    \end{eqnarray*}

    Now, \begin{align*}
        \text{I} & = \frac{1}{\min_{k \in S_0} \|\bm\beta_k^0\|_\infty} \sup_{\sigma^* \in \mathcal{U}_n}\E\big( \|n^{-1}\bC_{n(11)}^{-1}\bX^\mathrm{T}_{S_0^\mathrm{c}} (\bm\eta - \bm\mu) + n^{-1} \bC_{n(11)}^{-1} \bX^\mathrm{T}_{S_0}\bm\mu \|_\infty \big| \bY,\sigma^* \big)\\
        & \leq \frac{1}{\min_{k \in S_0}\|\bm\beta_k^0\|_\infty} \big[\sup_{\sigma^* \in \mathcal{U}_n}\E\big( \|n^{-1}\bC_{n(11)}^{-1}\bX^\mathrm{T}_{S_0^\mathrm{c}} (\bm\eta - \bm\mu)\|_\infty \big| \bY,\sigma^* \big) + \|n^{-1} \bC_{n(11)}^{-1} \bX^\mathrm{T}_{S^\mathrm{c}_0}\bX (\hat{\bm\beta}^\mathrm{R} - \bm\beta^0)\|_\infty \big] \\
        & = \frac{1}{\min_{k \in S_0} \|\bm\beta_k^0\|_\infty} \big[ \mathcal{O}\big(\sqrt{\log p_0}/n\big) + \frac{1}{n} \|\bC_{n(11)}^{-1}\bX_{S_0^\mathrm{c}}^\mathrm{T} \bX \big((\bX^\mathrm{T}\bX + a_n \mathrm{I}_p)^{-1}\bX^\mathrm{T} (\bX\beta^0 + \bm{\varepsilon}) - \bm\beta^0\big) \|_\infty  \big]\\ 
        & \leq \frac{1}{\min_{k \in S_0} \|\bm\beta_k^0\|_\infty} \big[\mathcal{O}\big({\sqrt{\log p_0}}/{n }\big) + \frac{1}{n} \big( \big\|\bC_{n(11)}^{-1}\bX_{S_0^\mathrm{c}}^\mathrm{T} \big(\bm{I}_n - \bm{H}(a_n) \big)\bX\bm\beta^0 \big\|_\infty + \|\bC_{n(11)}^{-1}\bX^\mathrm{T}_{S_0^\mathrm{c}}\bm{H}(a_n)\bm\varepsilon\|_\infty \big) \big]\\
        & = \frac{1}{\min_{k \in S_0} \|\bm\beta_k^0\|_\infty} \mathcal{O}\big(\sqrt{\log p_0}/n\big),
    \end{align*}
    since we have that $\|\bX^\mathrm{T}_{S_0^\mathrm{c}}\bm{H}(a_n)\bm\varepsilon\|_\infty = \max_{1 \leq i \leq n} |\varepsilon_i| \|\bX^\mathrm{T}_{S_0^\mathrm{c}}\bm{H}(a_n)\|_\infty$ and $\|\bX_{S_0^\mathrm{c}}^\mathrm{T} \big(\bm{I}_n - \bm{H}(a_n) \big)\bX\bm\beta^0 \|_\infty = \max_{1 \leq i \leq n} |\E(Y_i)| \|\bX_{S_0^\mathrm{c}}^\mathrm{T} \big(\bm{I}_n - \bm{H}(a_n)\|_\infty$. Moreover, following \cite{nardi2008asymptotic}, we can bound II by \begin{align*}
    & \lambda_n \frac{\sqrt{\sum_{k = 1}^{s_0}p_k}}{\lambda_{\textnormal{min}}(\bC_{n(11)})}\|\bm\xi_{S_0}\|_{\infty}
      \leq  (\lambda_n \sqrt{s_0} p_{\textnormal{max}}^{3/2})/\lambda_{\textnormal{min}}(\bC_{n(11)})
    \end{align*} using the $\ell_{\infty}$-$\ell_2$ norm inequality. This proves \eqref{first_part_model_sel_proof} by Assumption \ref{min-beta_condition}(A). Next, we consider \eqref{second_part_model_sel_proof} and express \eqref{noise_part_of_KKT} as $\bm\xi_{S_0^\mathrm{c}} = \bm{W}_1 + \lambda_n^{-1} \bm{W}_2,$ where $$\bm{W}_1 = n^{-1}\bX_{S_0^\mathrm{c}}^\mathrm{T}\bX_{S_0}\bC_{n(11)}^{-1}\bm\xi_{S_0} \text{ and } \bm{W}_2 = n^{-1}\bX_{S_0^\mathrm{c}}^\mathrm{T} (\bm{I}_n - \bX_{S_0^\mathrm{c}}\bC_{n(11)}^{-1}\bX_{S_0^\mathrm{c}}^\mathrm{T}) \bm\eta.$$ For simplicity, we assume that the first $s_0$ groups are active and the remaining $(K-s_0)$ groups are noise. Then, supposing $\bm{W}_1 = (\bm{W}_{1, s_0 + 1}^\mathrm{T},\dots,\bm{W}_{1, K}^\mathrm{T})^\mathrm{T}$ and $\bm{W}_2 = (\bm{W}_{2, s_0 + 1}^\mathrm{T},\dots,\bm{W}_{2, K}^\mathrm{T})^\mathrm{T}$ , and using $\|\bm{W}_{2,k}\| \leq \sqrt{p_k} \|\bm{W}_{2,k}\|_\infty$, we can write for some $k \in S_0^\mathrm{c}$, following from Assumption \ref{irrepresentability_condition} and using Markov's inequality, \begin{align*}
        \sup_{\sigma^* \in \mathcal{U}_n}\Pi\big(p_k^{-1/2} \|\bm\xi_k\| > 1 \big| \bY,\sigma^* \big) & \leq \sup_{\sigma^* \in \mathcal{U}_n}\Pi\big( p_k^{-1/2} \|\bm{W}_{1,k}\| + \lambda_n^{-1} \|\bm{W}_{2,k}\|_\infty > 1 \big| \bY,\sigma^* \big)\\
        & \leq \sup_{\sigma^* \in \mathcal{U}_n}\Pi\big(\lambda_n^{-1} \|\bm{W}_{2,k}\|_\infty > 1 - 1 + \nu \big|\bY,\sigma^*)\\ 
        & \leq \frac{\sup_{\sigma^* \in \mathcal{U}_n}\E\big( \|\bm{W}_{2,k}\|_\infty \big| \bY,\sigma^* \big)}{\lambda_n\nu}\\
        & \leq \frac{1}{\lambda_n \nu} \sup_{\sigma^* \in \mathcal{U}_n}\E \big[ \| n^{-1} \bX_k^\mathrm{T} (\bm{I}_n - \bX_{S_0}\bm\Sigma^{-1}_0 \bX_{S_0}^\mathrm{T})(\bm\eta - \bm\mu)\|_\infty \big| \bY,\sigma^* \big] \\
        & \qquad \qquad \qquad + \|n^{-1}\bX_k^\mathrm{T} (\bm{I}_n - \bX_{S_0}\bm\Sigma^{-1}_0 \bX_{S_0}^\mathrm{T})\bX(\hat{\bm\beta}^\mathrm{R} - \bm\beta^0)\|_\infty \\
        & = \mathcal{O}\big(\sqrt{{\log (p - p_0)}/{n\lambda^2_n}}\big),
    \end{align*}
    where the first part of the inequality follows, by Assumption \ref{sd_assum}, as 
    \begin{align*}
        \sup_{\sigma^* \in \mathcal{U}_n}\text{var}(\bm{W}_{2,k}| \bY) & = \text{var}\big( n^{-1} \bX_k^\mathrm{T} (\bm{I}_n - \bX_{S_0}\bm\Sigma^{-1}_0 \bX_{S_0}^\mathrm{T})\bm\eta \big| \bY,\sigma^* \big) \\
        & = {\sigma^*}^2 n^{-2} \bX_k^\mathrm{T} (\bm{I}_n - \bX_{S_0}\bm\Sigma^{-1}_0 \bX_{S_0}^\mathrm{T}) \bm{H}(a_n) (\bm{I}_n - \bX_{S_0}\bm\Sigma^{-1}_0 \bX_{S_0}^\mathrm{T})^\mathrm{T} \bX_k\\
        & \leq {\sigma^*}^2 n^{-2} \|\bX_k^\mathrm{T} \bm{H}(a_n) \bX_k\|\\
        & \leq {\sigma^*}^2 n^{-2} \|\bX_k\|^2.
    \end{align*}

    Thus, for any $k \in S_0^\mathrm{c}$, $  \sup\{\Pi\big(p_k^{-1/2} \|\bm\xi_k\| > 1 \big| \bY,\sigma^* \big):
    \sigma^* \in \mathcal{U}_n\} \to 0$ since $\log (p - p_0)/n\lambda_n^2 \to 0$ as $n \to \infty$.
\end{proof}

\begin{proof}[Proof of Lemma~\ref{lemma_A_groupAdap}]
Define the event $R_1 = \{\|\hat{\bm\beta}_k\| \geq \psi \|\bm\beta_k^0\| \enskip \textnormal{ for } k \in S_0\}$. From \cite{wei2010consistent}, we know $P(\min_{k \in S_0} \|\hat{\bm\beta}_k\| \geq \psi\min_{k \in S_0} \|\bm\beta_k^0\|) \to 1$ as $n \to \infty,$ where the inverse of $\|\hat{\bm\beta}_k\|$ is used as the weight in the group adaptive LASSO projection map, and hence needs to be controlled. Thus, \begin{align*}
    & \sup_{\sigma^* \in \mathcal{U}_n}\Pi(A^\mathrm{c}|\bY,\sigma^*) \\
    \leq & \sup_{\sigma^* \in \mathcal{U}_n}\Pi(A^\mathrm{c}\cap R_1|\bY,\sigma^*) + P(R_1^\mathrm{c})\\
    \leq & \sup_{\sigma^* \in \mathcal{U}_n}\Pi\big(\big\{\|(\bC_{n(11)}^{-1}\bZ_{n(1)})_k\| + \sqrt{n}\|(\bC_{n(11)}^{-1}\bm{M}_{S_0})_k\| > \sqrt{n}\|\bm\beta^0_k\| \enskip \textnormal{ for } k \in S_0\big\} \\
    & \qquad \qquad \qquad \qquad \qquad \qquad \qquad \qquad \cap \{\|\hat{\bm\beta}_k\| \geq \psi \|\bm\beta_k^0\| \enskip \textnormal{ for } k \in S_0\}\big|\bY,\sigma^*\big) + o(1)\\
    \leq & \sup_{\sigma^* \in \mathcal{U}_n}\Pi\big(\big\{\max_{k \in S_0}\|(\bC_{n(11)}^{-1}\bZ_{n(1)})_k\| > \min_{k \in S_0} \big(\sqrt{n}\|\bm\beta^0_k\| - \sqrt{n}\|(\bC_{n(11)}^{-1}\bm{M}_{S_0})_k\|\big)\big\}\\
    & \qquad \qquad \qquad \qquad \qquad \qquad \qquad \qquad\cap \{\|\hat{\bm\beta}_k\| \geq \psi \|\bm\beta_k^0\| \enskip \textnormal{ for } k \in S_0\}\big|\bY,\sigma^*\big) + o(1)\\
    \leq & \sup_{\sigma^* \in \mathcal{U}_n}\Pi\Big(\Big\{\frac{1}{\sqrt{n}}\max_{k \in S_0}\|(\bC_{n(11)}^{-1}\bX_{(1)}^\mathrm{T}\boldeta)_k\| > \sqrt{n}\min_{k \in S_0}\|\bm\beta^0_k\| - \frac{s_0\lambda_n\sqrt{n} \lambda^{-1}_{\textnormal{min}}(\bC_{n(11)})p_{\textnormal{max}}^{3/2}}{\psi \min_{k \in S_0} \|\bm\beta^0_k\|}\Big\} \Big| \bY,\sigma^*\Big) + o(1),
\end{align*} since \begin{align*}
    \min_{k \in S_0} \big(\sqrt{n}\|\bm\beta^0_k\| - \sqrt{n}\|(\bC_{n(11)}^{-1}\bm{M}_{S_0})_k\|\big) & \geq \sqrt{n}\min_{k \in S_0} \|\bm\beta^0_k\| - \sqrt{n}\max_{k \in S_0}\|(\bC_{n(11)}^{-1}\bm{M}_{S_0})_k\|\\
    & \geq \sqrt{n}\min_{k \in S_0} \|\bm\beta^0_k\| - \frac{\sqrt{n}p_0}{\lambda_{\textnormal{min}}(\bC_{n(11)})}\max_{k \in S_0}\|\bm{M}_k\|\\
    & \geq \sqrt{n}\min_{k \in S_0} \|\bm\beta^0_k\| - \frac{\sqrt{n p_{\textnormal{max}}} \lambda_n s_0 p_{\textnormal{max}}}{\lambda_{\textnormal{min}}(\bC_{n(11)})\psi\min_{k \in S_0}\|\bm\beta^0_k\|}\\
    & = v_1, \textnormal{ say.}
\end{align*}
Under Assumption \ref{min-beta_condition}(C), we can bound $v_1$ as follows:
\begin{align*}
    v_1 &  = \sqrt{n}\min_{k \in S_0} \|\bm\beta_k^0\| - \frac{\sqrt{n} s_0 \lambda_n (p_{\textnormal{max}})^{3/2}}{\lambda_{\textnormal{min}}(\bC_{n(11)}) \psi \min_{k \in S_0} \|\bm\beta_k^0\|}\\
    & = \sqrt{n}\min_{k \in S_0} \|\bm\beta_k^0\| + o(\sqrt{n}\min_{k \in S_0} \|\bm\beta_k^0\|).
\end{align*}
Then, using Markov's inequality followed by a maximal inequality, we will have \begin{align*}
    \sup_{\sigma^* \in \mathcal{U}_n}\Pi(A^\mathrm{c}|\bY,\sigma^*) & \leq \frac{1}{v_1}\sup_{\sigma^* \in \mathcal{U}_n}\E\big(n^{-1/2}\max_{k \in S_0}\|(\bC_{n(11)}^{-1}\bX_{(1)}^\mathrm{T}\boldeta)_k\| \big| \bY,\sigma^* \big)\\
    & \leq \frac{1}{v_1\sqrt{n}} \big\{\sup_{\sigma^* \in \mathcal{U}_n}\E\big(\max_{k \in S_0}\|\big(\bC_{n(11)}^{-1}\bX_{(1)}^\mathrm{T}(\boldeta - \bm{\mu})\big)_k\| \big| \bY,\sigma^* \big) + \max_{k \in S_0} \|\bC_{n(11)}^{-1}\bX_{(1)}^\mathrm{T}\bm{\mu}\|\big\}\\
    & \leq K \frac{\sqrt{p_{\textnormal{max}} \log s_0}}{v_1\sqrt{n}}\to 0, 
\end{align*}
 by Assumption \ref{min-beta_condition}(C).
\end{proof}

\begin{proof}[Proof of Lemma~\ref{lemma_B_groupAdap}]
We first define the event $R_2 = \{\|\hat{\bm\beta}_k\|^{-1} > r_n\},$ where $r_n$ is the rate of convergence of the group LASSO estimator $\hat{\bm\beta} = (\hat{\bm\beta}^\mathrm{T}_1,\dots,\hat{\bm\beta}^\mathrm{T}_K)^\mathrm{T}$. Then,
\begin{align*}
    &\sup_{\sigma^* \in \mathcal{U}_n}\Pi(B^\mathrm{c}|\bY,\sigma^*)\\
    \leq & \sup_{\sigma^* \in \mathcal{U}_n}\Pi(B^\mathrm{c}\cap R_2|\bY,\sigma^*) + P(R_2^\mathrm{c})\\
    \leq & \sup_{\sigma^* \in \mathcal{U}_n}\Pi\Big(\Big\{\frac{1}{\sqrt{n}}\|(\bX_{S_0^\mathrm{c}}^\mathrm{T}(\bm{I}_n - \bm{P}_{S_0})\boldeta)_k\| \geq \lambda_n r_n{\sqrt{n p_k}}\|\bm{1}_{p_k}\| - \sqrt{n}\|\bC_{n(21)}\bC_{n(11)}^{-1}\bm{M}_{S_0}\|\Big\} \big| \bY,\sigma^*\Big) + o(1)\\
    \leq & \sup_{\sigma^* \in \mathcal{U}_n}\Pi\Big(\Big\{\frac{1}{\sqrt{n}}\|(\bX_{S_0^\mathrm{c}}^\mathrm{T}(\bm{I}_n - \bm{P}_{S_0})\boldeta)_k\| \geq v_2\Big\}\Big|\bY,\sigma^*\Big) + o(1), 
\end{align*} where 
\begin{align*}
    v_2 = & \lambda_n r_n{\sqrt{n p_k}}\|\bm{1}_{p_k}\| - \sqrt{n}\|\bC_{n(21)}\bC_{n(11)}^{-1}\bm{M}_{S_0}\|\\
    \geq & \sqrt{n} \lambda_n r_n p_{\textnormal{min}} - \frac{\sqrt{n} p_0^2 \lambda_n \sqrt{p_{\textnormal{max}}}}{\lambda_{\textnormal{min}}(\bC_{n(11)})\psi \min_{k \in S_0}\|\bm\beta^0_k\|}\\
    \geq & \sqrt{n} \lambda_n r_n p_{\textnormal{min}} - \frac{\sqrt{n} \lambda_n s_0^2 ({p_{\textnormal{max}}})^{5/2}}{\lambda_{\textnormal{min}}(\bC_{n(11)})\psi \min_{k \in S_0}\|\bm\beta^0_k\|}\\
    = & \sqrt{n} \lambda_n r_n p_{\textnormal{min}} \big(1 
    + o(1)\big) 
\end{align*} 
 by Assumption \ref{min-beta_condition}(C). 
 
Finally, using Markov's inequality and recalling that ${p_{\textnormal{max}} \log (K - s_0)}/({\sqrt{n}\lambda_n r_n p_{\textnormal{min}}}) \to 0$ as $n \to \infty$, we can conclude that $\sup\{\Pi(B^\mathrm{c}|\bY,\sigma^*): \sigma^* \in \mathcal{U}_n\} \to 0.$
\end{proof}

\subsection*{Supplement C: Sparse Projection-posterior based on Group MCP penalty}

Our existing proof for the group LASSO estimator already provides the appropriate scaffold: the argument proceeds through the basic inequality, stochastic score control (Lemma~\ref{lemma_for_thm1}), the cone condition induced by the penalty, the compatibility condition, and the resulting estimation and prediction rates. To extend the argument to the \emph{group MCP} estimator, we only need to modify the steps involving the penalty algebra. All other ingredients, including Lemma~\ref{lemma_for_thm1} and Lemma~\ref{lemma_bound_Fn}, remain unchanged.

For a group \(k\) with Euclidean norm \(t = \|\bm{\beta}_k\|_2\), the group MCP penalty with parameters \((\lambda, \gamma)\), where \(\gamma > 1\), is defined as
\[
P_{\lambda,\gamma}(t)
=
\begin{cases}
\lambda t - \dfrac{t^2}{2\gamma}, & 0 \le t \le \gamma \lambda,\\
\dfrac{1}{2}\gamma \lambda^2, & t > \gamma \lambda.
\end{cases}
\]
Its group-level derivative is
\[
P'_{\lambda,\gamma}(t) =
\begin{cases}
\lambda - \dfrac{t}{\gamma}, & 0 \le t \le \gamma \lambda,\\
0, & t > \gamma \lambda.
\end{cases}
\]
Two key properties will be repeatedly used:  
\begin{itemize}  
\item 
For any \(u,v \in \mathbb{R}^{p_k}\) with \(t_u = \|u\|_2\) and \(t_v = \|v\|_2\),
\[
P_{\lambda,\gamma}(t_u) - P_{\lambda,\gamma}(t_v)
\le
\lambda (t_u - t_v)
- \frac{1}{2\gamma}(t_u - t_v)_+^2\le P_{\lambda,\gamma}(t_u) - P_{\lambda,\gamma}(t_v) \le \lambda (t_u - t_v);\] 
\item if \(t_u, t_v \ge \gamma \lambda\), then 
\[ P_{\lambda,\gamma}(t_u) = P_{\lambda,\gamma}(t_v) = \tfrac{1}{2}\gamma \lambda^2,\] so in this region, the penalty exerts no shrinkage. 
\end{itemize} 
These properties imply a locally stronger curvature control than the group LASSO and complete bias elimination for sufficiently large groups. The fundamental basic inequality remains identical to the group LASSO case:
\[
\frac{1}{n}\|\bX \bm{\beta}^0 - \bX \bm{\beta}^*\|_2^2
\le
\frac{2}{n}\boldsymbol{\eta}^\mathrm{T} \bX(\bm{\beta}^0 - \bm{\beta}^*)
+ \sum_{k=1}^K \left\{ P_{\lambda,\gamma}(\|\bm{\beta}_k^0\|_2)
- P_{\lambda,\gamma}(\|\bm{\beta}_k^*\|_2) \right\}.
\]
The stochastic term is controlled by Lemma~\ref{lemma_for_thm1} with the same choice of \(\lambda \ge \lambda_0(x)\). The only modification concerns the penalty difference. Where the group LASSO argument uses
\(
\sum_{k=1}^K \lambda \sqrt{p_k}
\big( \|\bm{\beta}_k^0\|_2 - \|\bm{\beta}_k^*\|_2 \big),
\)
we replace it with the group MCP bound
\[
P_{\lambda,\gamma}(\|\bm{\beta}_k^0\|_2)
- P_{\lambda,\gamma}(\|\bm{\beta}_k^*\|_2)
\le
\lambda\big( \|\bm{\beta}_k^0\|_2 - \|\bm{\beta}_k^*\|_2 \big)
- \frac{1}{2\gamma}\big( \|\bm{\beta}_k^*\|_2 - \|\bm{\beta}_k^0\|_2 \big)_+^2.
\]
Summing over all groups yields
\[
\sum_{k=1}^K
\Big\{ P_{\lambda,\gamma}(\|\bm{\beta}_k^0\|_2)
- P_{\lambda,\gamma}(\|\bm{\beta}_k^*\|_2) \Big\}
\le
\lambda
\sum_{k=1}^K
\sqrt{p_k}
\big( \|\bm{\beta}_k^0\|_2 - \|\bm{\beta}_k^*\|_2 \big)
-
\frac{1}{2\gamma}
\sum_{k=1}^K
\big( \|\bm{\beta}_k^*\|_2 - \|\bm{\beta}_k^0\|_2 \big)_+^2.
\]
This preserves the decomposable structure used to establish the cone condition, but with an additional nonnegative curvature term that sharpens the inequality and reduces estimation bias. Following the same steps as in the group LASSO case, we obtain
\[
\frac{1}{n}\|\bX \Delta\|_2^2
+
\lambda \sum_{k \notin S_0} \sqrt{p_k}\|\bm{\beta}_k^*\|_2
+
\frac{1}{2\gamma}\sum_{k=1}^K
\big( \|\bm{\beta}_k^*\|_2 - \|\bm{\beta}_k^0\|_2 \big)_+^2
\le
\frac{3\lambda}{2}\sum_{k\in S_0}\sqrt{p_k}\|\Delta_k\|_2,
\]
where \(\Delta_k = \bm{\beta}_k^* - \bm{\beta}_k^0\).
Dropping the nonnegative curvature term yields the usual inequality
\(
\sum_{k \notin S_0} \sqrt{p_k}\|\bm{\beta}_k^*\|_2
\le (3/2)
\sum_{k\in S_0} \sqrt{p_k}\|\Delta_k\|_2,
\)
while retaining it leads to a strictly tighter constant. Under Assumption~\ref{compat}, we have
\[
\sum_{k\in S_0}\sqrt{p_k}\|\Delta_k\|_2
\le
\frac{\sqrt{\sum_{k\in S_0}p_k}}{\phi(S_0)}
\frac{\|\bX\Delta\|_2}{\sqrt{n}}.
\]
Substituting into the basic inequality and applying \(2ab \le a^2+b^2\) yields
\[
\frac{1}{2n}\|\bX\Delta\|_2^2
+\frac{\lambda}{2}\sum_{k=1}^K \sqrt{p_k}\|\Delta_k\|_2
\lesssim
\frac{\lambda^2}{\phi^2(S_0)}\sum_{k\in S_0}p_k.
\]
Hence, the estimation and prediction rates match those obtained under the group LASSO penalty, with improved constants due to the concavity of the MCP penalty.

For exact support recovery, we impose a groupwise \(\beta\)-min condition:
\[
\min_{k\in S_0} \|\bm{\beta}_k^0\|_2
\ge (1+\delta)\gamma \lambda,
\]
together with the weighted group irrepresentability condition.
This ensures that the active groups lie within the flat region of the MCP penalty (no shrinkage), while inactive groups are shrunk to zero via the KKT conditions. This is the MCP analogue of Assumption~\ref{min-beta_condition}(B) used for SCAD.

Thus, the overall structure of the proof remains unchanged, and the group MCP yields identical convergence rates with reduced estimation bias.

\subsection*{Supplement D: Debiased Projection-posterior for group SCAD and adaptive group LASSO}

\begin{assumption}[GS and AGL conditions]\label{ass:GS-AGL}
Assume the design and regularity conditions used in Theorem~\ref{debiasedCLT_GL} hold.
In addition:

\begin{enumerate}
\item[\textnormal{(GS)}] Let \(\hat{\bm\beta}^{\rm GS}\) be any stationary point of the group SCAD objective.  Suppose there is a beta-min constant \(c_0>0\) such that \(\min_{k\in S_0}\|\beta^0_k\|\ge c_0\), and \(\hat{\bm\beta}^{\rm GS}\) is selection consistent and \(\|\hat{\bm\beta}^{\rm GS}-\bm\beta^0\|_2=\mathcal{O}_P(\lambda_n \sum_{k = 1}^{s_0}p_k/\sqrt{p_\textnormal{min}})\) with $\lambda_n$ as defined in \Cref{est_pred_consis}(2).
\item[\textnormal{(AGL)}] Let \(\hat{\bm\beta}^{\rm AGL}\) be the adaptive group LASSO estimator with weights \(w_k=\|\tilde\beta_k\|^{-\gamma}\) (\(\gamma>0\)). Assume the initial estimator \(\tilde\beta\) is zero-consistent: for some rate \(r_n\to\infty\), \(\max_{k\in S_0^c}\|\tilde\beta_k\|=\mathcal{O}_P(r_n^{-1})\) and
\(\min_{k\in S_0}\|\tilde\beta_k\|\ge c_0\) w.h.p., so that \(w_k=\mathcal{O}_P(r_n^\gamma)\) for \(k\in S_0^c\) and \(w_k=\mathcal{O}_P(1)\) for \(k\in S_0\).  With tuning satisfying the condition in \Cref{est_pred_consis}(3), the estimator is selection consistent and \(\|\hat{\bm\beta}^{\rm AGL}-\bm\beta^0\|_2=\mathcal{O}_P(\lambda_n \sum_{k = 1}^{s_0}p_k/\sqrt{p_\textnormal{min}})\).
\end{enumerate}
\end{assumption}

\noindent Define the remainder terms
\[
\bm\Delta^{\rm GS}=(\hat{\bm\Theta}\hat{\bm\Sigma}-I_p)(\hat{\bm\beta}^{\rm GS}-\bm\beta^0),
\;
\bm\Delta^{\rm AGL}=(\hat{\bm\Theta}\hat{\bm\Sigma}-I_p)(\hat{\bm\beta}^{\rm AGL}-\bm\beta^0),
\]
and the corresponding debiased projections
\[
{\bm\beta}^{**}_{\,\textnormal{DGS}} = \bm\beta^0 + \frac{1}{n}\hat{\bm\Theta}^\mathrm{T}\bX^{\mathrm{T}}\bm\eta - \bm\Delta^{\rm GS},\;
{\bm\beta}^{**}_{\,\textnormal{DAGL}} = \bm\beta^0 + \frac{1}{n}\hat{\bm\Theta}^\mathrm{T}\bX^{\mathrm{T}}\bm\eta - \bm\Delta^{\rm AGL}.
\]

\begin{theorem}[BvM for debiased GS/AGL projection-posteriors]
\label{thm:UQ-GS-AGL}
Suppose the assumptions of Theorem~\ref{debiasedCLT_GL} on the construction of \(\hat{\bm\Theta}\) and the posterior covariance \(\bm H(a_n)\) hold, and in addition Assumption~\ref{ass:GS-AGL} holds.

\smallskip
\noindent\textnormal{(i) \textbf{Group SCAD.}}
We have \(\|\bm\Delta^{\rm GS}\|_\infty=o_P(n^{-1/2})\) and
\[
\sup_{B}\Big|
\Pi\!\big(\sigma^{-1}\sqrt{n}\,({\bm\beta}^{**}_{\,\textnormal{DGS}}-\bm\beta^0)\in B\,\big|\,\bY\big)
- \mathcal{N}_p\!\big(B;\,\bm m,\bm V\big)\Big|\;\to\;0,
\]
where \(\bm m=n^{-1/2}\hat{\bm\Theta}^{\!\mathrm{T}}\bX^{\mathrm{T}}\bm\mu\) and
\(\bm V=n^{-1}\hat{\bm\Theta}^{\!\mathrm{T}}\bm H(a_n)\hat{\bm\Theta}\) are same as in \Cref{debiasedCLT_GL}.

\smallskip
\noindent\textnormal{(ii) \textbf{Adaptive group LASSO.}}
We have \(\|\bm\Delta^{\rm AGL}\|_\infty=o_P(n^{-1/2})\) and
\[
\sup_{B}\Big|
\Pi\!\big(\sigma^{-1}\sqrt{n}\,({\bm\beta}^{**}_{\,\textnormal{DAGL}}-\bm\beta^0)\in B\,\big|\,\bY\big)
- \mathcal{N}_p\!\big(B;\,\bm m, \bm V\big)\Big|\;\to\;0,
\]
with the same Gaussian limit \(\mathcal{N}_p(\cdot;\bm m,\bm V)\) as in part
\textnormal{(i)} and Theorem~\ref{debiasedCLT_GL}.
\end{theorem}

\begin{corollary}[Coverage]
Under the conditions of Theorem~\ref{thm:UQ-GS-AGL}, component-wise credible intervals and ellipsoidal credible sets obtained from the GS or AGL debiased projection-posteriors have asymptotically exact
frequentist coverage.
\end{corollary}
The proofs are identical to those of Theorem~\ref{debiasedCLT_GL} and \Cref{debiased_coverage} and hence, we skip them to avoid repetition.
\end{document}